# Pathfinding of Digital False Information Diffusion by Conformal Field Theory and Feynman's Green's Function: Multiple Analysis Considering Stress Energy Tensor and Symmetry of Virasoro Operators

Yasuko Kawahata †

Faculty of Sociology, Department of Media Sociology, Rikkyo University, 3-34-1 Nishi-Ikebukuro,Toshima-ku, Tokyo, 171-8501, JAPAN.

ykawahata@rikkyo.ac.jp

**Abstract:** This study delves into the application of Conformal Field Theory (CFT) to understand information diffusion within digital media and its broader social implications. Focusing on the digital native generation, vulnerable to misinformation and data privacy issues, the research highlights the urgency of addressing psychological health in the digital realm. By leveraging CFT's analytical power, particularly in tracing information flows and assessing the impact of misinformation, the study offers novel insights into the dynamics of digital communication. It proposes a model for predicting and analyzing the spread of biased content, emphasizing the need for improved information literacy and personal data protection. This paper explores the concept of infinitesimal conformal transformations using the Virasoro operator, forming a conformal transformation group. It considers relaxing isometric transformations, making conformal transformations possible. The Virasoro operator, comprising an infinite number of elements, constitutes a Lie group and Lie algebra, forming a complex polynomial on a circle. These mathematical concepts underlie conformal field theory. It discusses the application of these ideas in assessing risk scores for misinformation spread in digital environments, considering factors like propagation speed and user behavior. In the examination of multiple analytical methods, the approach to quantitatively analyze cases that are likely to expose the risk of spreading filter bubbles such as negative information, rumors, and fake news is to incorporate a two-dimensional conformal field theory approach using the Ward-Takahashi identity and path integral with trace anomalies and metric are introduced. Trace anomalies and metrics were introduced. A trace anomaly refers to the spread of information in an unexpected way, usually indicating anomalous behavior or hidden dynamics in the system. Metric variation hypothesized and analyzed how diffusion characteristics, such as the rate and pattern of information diffusion, vary over time.

## 1. Introduction

With the rapidly evolving digital environment, online health, especially psychological health, has become an important concern. With the increasing use of the Internet and social media, the spread of misinformation and the impact of negative information have become serious issues. In this environment, the digital native generation, especially those with limited information literacy, is at risk of being influenced by misinformation and malicious content. Challenges include modeling the complex real-world data and addressing privacy concerns, requiring interdisciplinary approaches and further research.Finally, we performed a multiple analysis considering the symmetry of the stress energy tensor and the Vilasoro operator, each of which was subjected to random number generation to verify the results and provide quantitative and qualitative discussion to scrutinize the method.

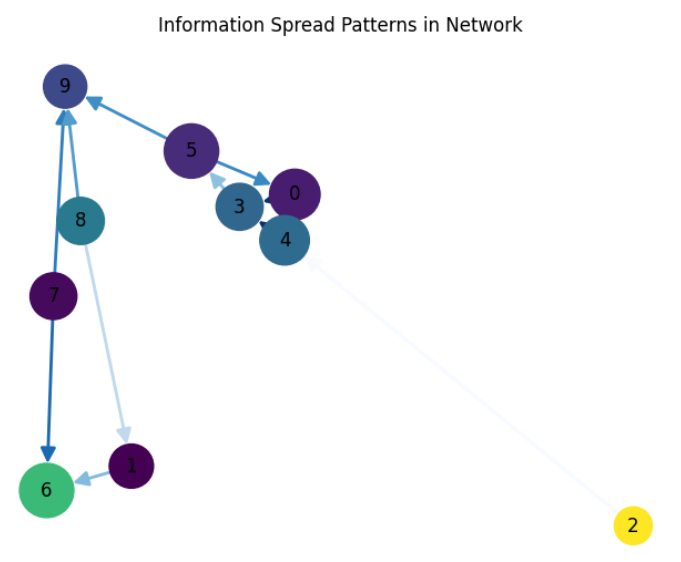

Fig. 1: Example:Information Spread Patterns in Network

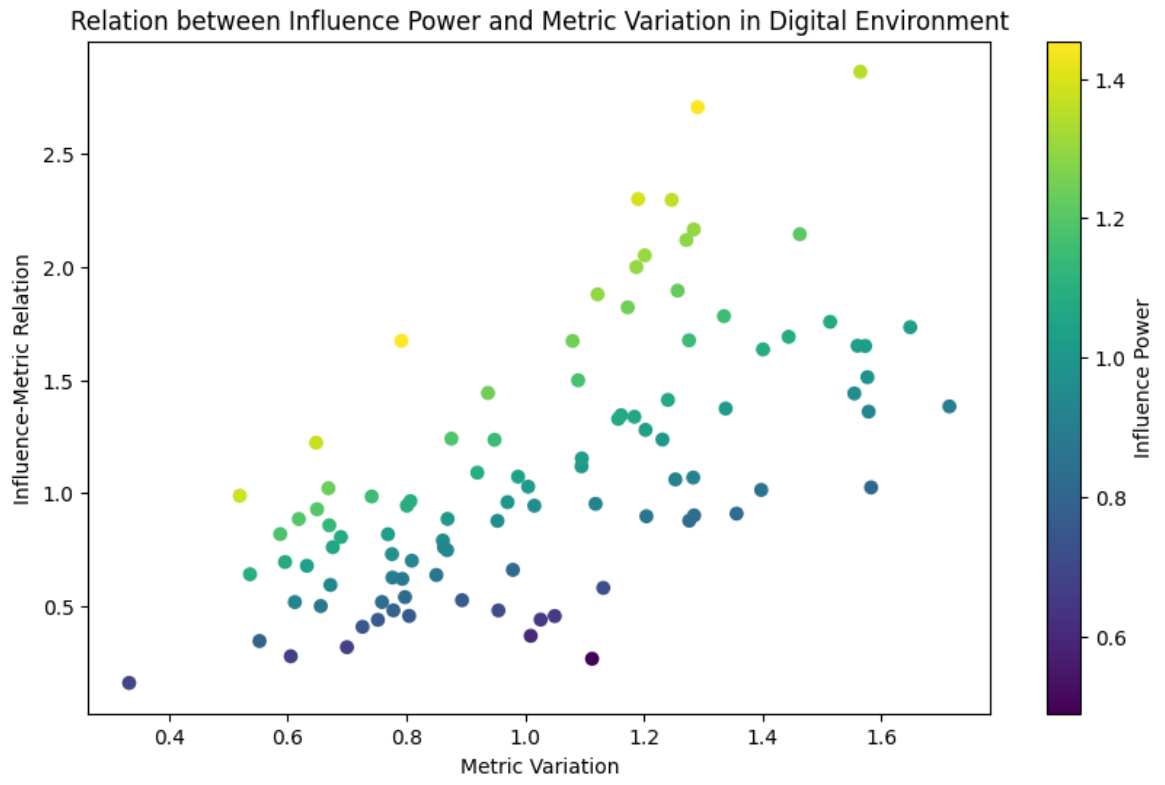

Fig. 2: Example:Relation between Influence Power and Metric Variation in Digital Environment

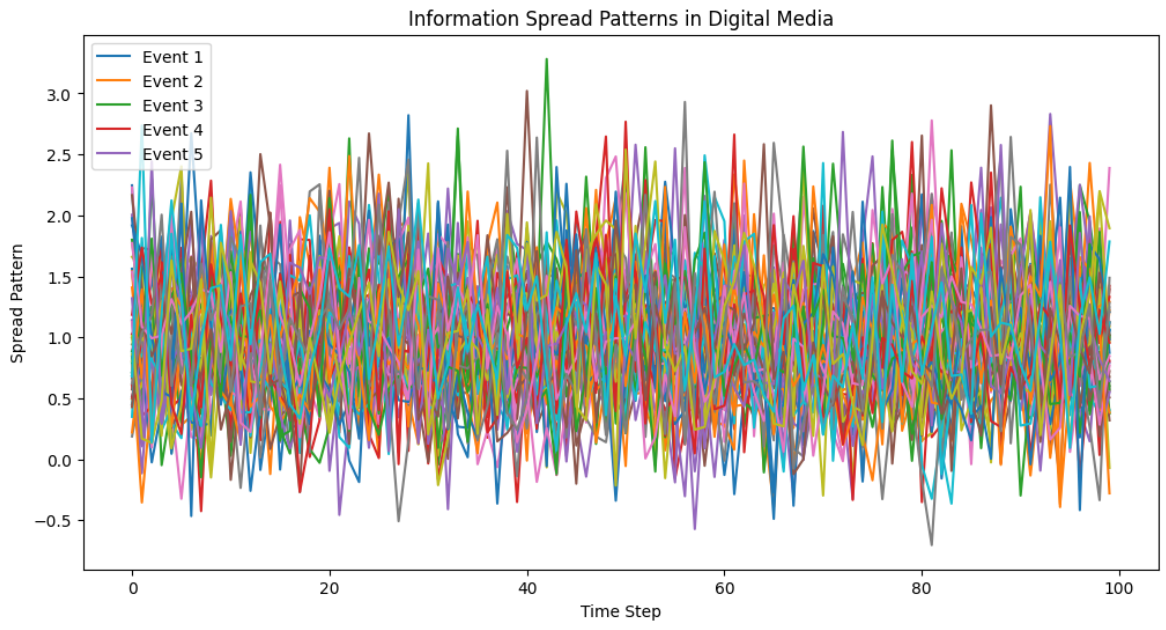

Fig. 3: Example:Information Spread Patterns in Network

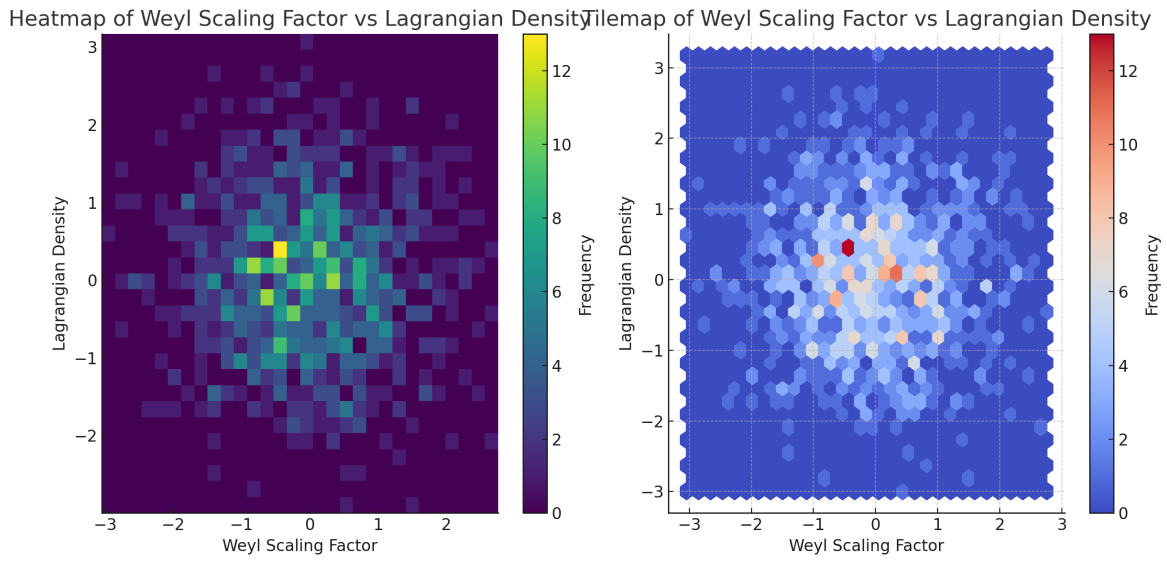

Fig. 4: Example:Information Spread Patterns in Digital Media

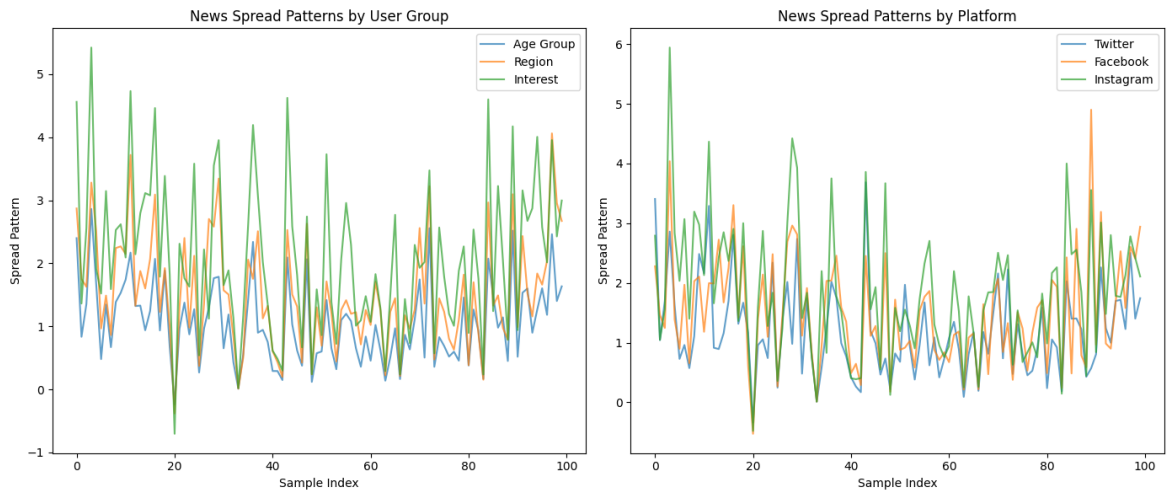

Fig. 5: Example:News Spread Patterns by Platform and News Spread Patterns by User Group

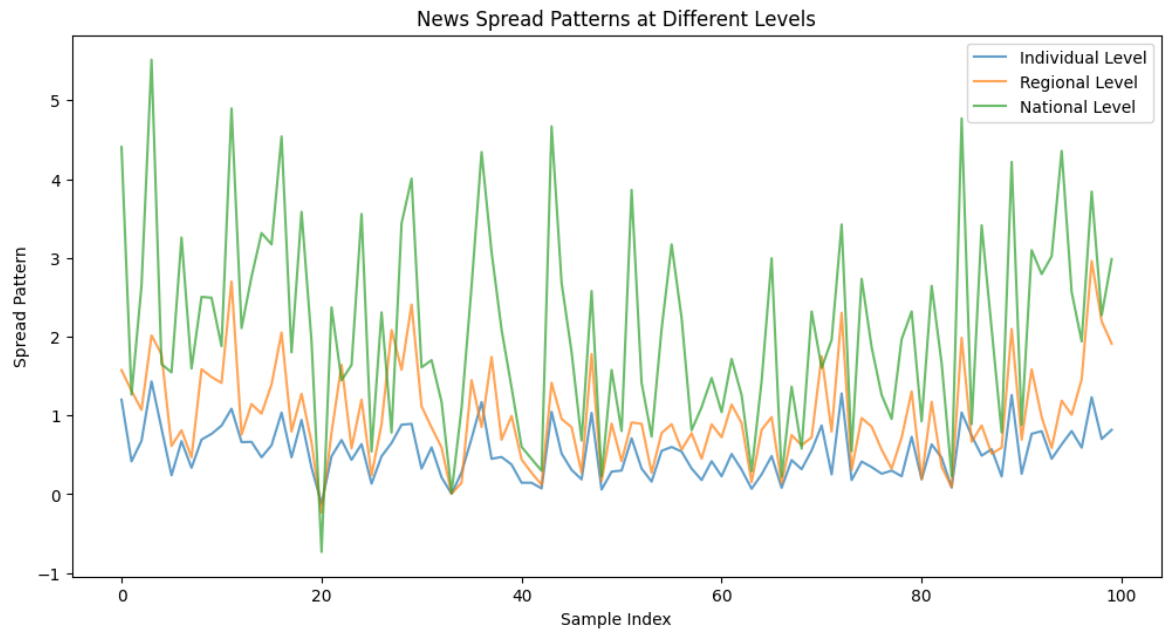

Fig. 6: Example:News Spread Patterns at Different Levels

As shown in the examples in Fig. 1-3, the theory of applied dynamics provides opportunities for hypothesizing and reasoning about various real-world physical phenomena with applications to social physics, such as opinion dynamics. This example is the result of the hypothesis of time series diffusion of social events, trace anomaly coefficients, and diffusion patterns of influence on user groups, which is introduced in this paper, and is applied to the diffusion of rumors by hypothesizing the diffusion network and its strength, path integrals, and diffusion of information per agent and per person, By hypothesizing the spread of rumors, fake news and other misinformation, and the risk of viral spread, it can be used to predict the spread of viruses, and to control superspreaders.

This paper also introduces the stress energy tensor, and introduces not only path integrals but also time-delayed diffusion path integrals with the Green's function. (However, since this is still a difficult discussion, this paper is limited to the introduction and hypotheses.)

Fig. 5 and 6 show opinion dynamics, News Spread Patterns at Different Levels, News Spread Patterns by Platform and News Spread Patterns by User Group. The simulation example hypothesizes the characteristics of topic spread characteristics for a single topic by Platform and News Spread Patterns by User Group. (However, in this paper, we limit ourselves to the hypotheses.) In Fig.7, the number of individuals in a social network is specified to simulate the propagation of opinions and external influences (e.g., news events and media influences), and the temporal variation of the dynamic order is examined to discuss the filter bubble and vacuum expectation, or the probability of excluding highly diffuse information such as incorrect information. and others were also introduced. (However, also limited to a hypothesis in this paper.)

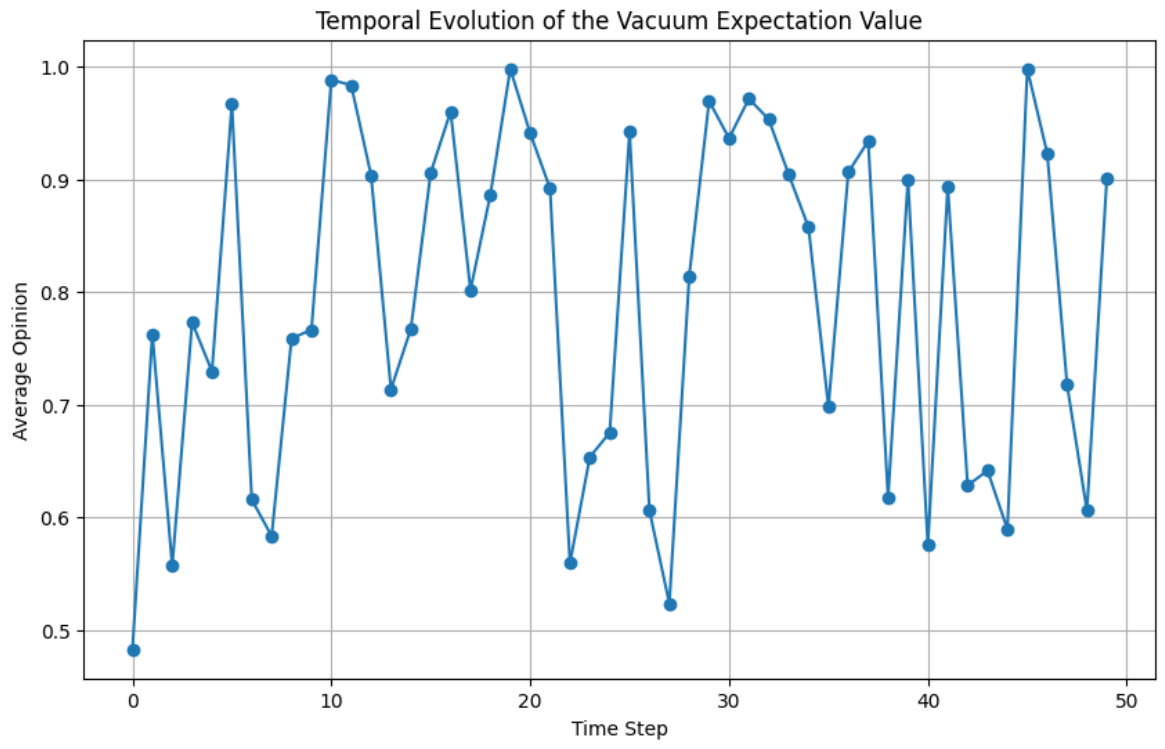

Fig. 7: Example:Temporal Evolution of the Vacuum Expectation Value

In this study, we apply Conformal Field Theory (CFT) to analyze the pattern of information diffusion on digital media and its social impact. Conformal field theory is a theoretical framework for describing physical systems with special symmetries such as scale invariance and angle conservation in physics. This theory opens new avenues to mathematically understand how misinformation and negative content diffuse in social networks and affect individuals and communities.

This paper also introduces the stress energy tensor, and introduces not only path integrals but also time-delayed diffusion path integrals with the Green's function. (However, since this is still a difficult discussion, this paper is limited to the introduction and hypotheses.)

In particular, by considering the path integral of information diffusion, it is possible to trace the flow of information in different time frames and communities. This allows for a detailed analysis of the process from the initial appearance of misinformation or negative information to its spread and its impact on individual opinions and behavior.

In addition, this study will focus specifically on the digital native generation, highlighting its importance in terms of personal data protection. With digital media deeply rooted in our daily lives, protecting personal information and improving information literacy are essential to protect ourselves from misinformation and negative influences. This approach, based on conformal field theory, is expected to contribute to policymaking and the design of educational programs for the healthy use of digital media.

## Research Background: Digital Environment and Health Diffusion and Impact of the Digital Environment

In recent years, the rapid evolution and diffusion of digital technology has facilitated the acquisition, exchange, and dissemination of information. The rise of social media, in particular, has revolutionized the way people communicate. These digital platforms promote diversity of opinion and have a significant impact on shaping the attitudes of individuals and communities.

## Digital Native Generation and Information Literacy

The digital native generation has been surrounded by digital technology since birth. However, they do not necessarily have a high level of information literacy, or the ability to evaluate information and think critically. The digital environment, where misinformation and biased information can spread easily, poses a particular risk to this generation.

## Personal Data Protection and Privacy Challenges

Online activities are supported by the accumulation and analysis of personal data. The importance of personal data protection is heightened by the risk of data breaches and privacy violations. Ensuring user data protection and privacy is an important issue for the health of the digital society.

## Health Implications

The digital environment affects psychological and social health. Particularly among young people, social media use has been reported to increase stress, anxiety, and loneliness. To promote a healthy digital environment, it is essential to curb the spread of misinformation and improve information literacy.

## Importance of simulations based on conformal field theory

## Modeling Information Diffusion Patterns

Conformal field theory provides a powerful tool for understanding and modeling information diffusion dynamics. The theory helps to analyze in detail the flow of information and its impact in the digital environment. In terms of risk assessment of misinformation and negative information Through simulations, it is possible to gain a deeper understanding of the impact of misinformation and negative information, especially on the digital native generation. This can contribute to the development of information literacy education and awareness programs. In terms of privacy protection and the promotion of a healthy digital environment, this model can be used to simulate the use of personal data and its impact, highlighting the importance of privacy protection. This approach aims to contribute to the promotion of a healthy digital environment. To understand the utility of applying conformal field theory (CFT) to the dynamics of information diffusion in the digital environment, one must first understand the basic concepts of conformal field theory and how it is suited to model information flow and interaction.

## Basic Concepts of Conformal Field Theory

Conformal symmetry and conformal field theory are theories that have properties that are invariant to scale transformations that preserve local angles. This property means that the physical system is scale independent, meaning that the same laws apply to small and large scale phases.

## Primary Fields and Correlation Functions

In conformal field theory, there are basic elements called primary fields, and the interactions between them are described using correlation functions. The correlation function indicates the strength of the interaction between different primary fields.

## Application to information diffusion models

## Information Sources and Primary Fields

Each source of information in the digital environment (e.g., news articles, social media posts, etc.) can be modeled as a primary field. This makes it possible to analyze the interaction between information sources and the strength of their influence. 2.

## Correlation Functions and Propagation of Information

The correlation function between sources of information indicates how much one source of information influences another and how diffuse that information is. Through this analysis, we can understand the dynamics of information propagation patterns and influence. 3.

## Conformal Invariance and Scale Invariance

Conformal invariance in the diffusion of information suggests that similar patterns exist at different scales (e.g., individual, community, and national levels). For example, it implies that the laws of information propagation within a small community are applicable to large social networks.

## Specific formulas and applications

## Mathematical formulas for correlation functions

An example of a concrete correlation function is the two-point correlation function: $G(x, y) = \langle \Phi(x)\Phi(y) \rangle$. Here, $\Phi(x)$ and $\Phi(y)$ represent different information sources, and $G(x, y)$ indicates the strength of the interaction between these information sources.

## Modeling the propagation of information

For example, to model how a particular news event (primary field $\Phi(x)$) spreads on social media (primary field $\Phi(y)$

The correlation function $G(x, y)$ is used to model how a particular news event (primary field $\Phi(x)$) spreads on social media (primary field $\Phi(y)$). This analysis allows us to understand how widely a particular news item is likely to propagate, and what characteristics make it more likely to spread more widely.

Again, as we will assume, the application of conformal field theory to models of information diffusion provides a new approach to understanding the complex dynamics of information flow in the digital environment and to analyze and control the spread of information based on misinformation and bias. However, further research is needed to conduct detailed analysis based on actual data and to bridge the gap between theoretical models and reality.

Based on the aforementioned, we further introduce several conformal field theory-based theories. This paper will be based on network analysis to simulation of social negative news diffusion analysis based on conformal field theory, but we assume that in reality it is a complex task to apply physics theory to social science problems. For example, the process of computing dynamic ordinals using the Ward-Takahashi identity or Feynman's Green's function may provide deep insights into patterns of information diffusion. Such simulations can be very useful in understanding and counteracting patterns of fake news and misinformation diffusion. However, real-world social networks are so complex that it is difficult for theoretical models to fully capture real-world situations. Therefore, such simulations should be interpreted in combination with real data.

## Ward-Takahashi Identity and Feynman's Green's Function

### Introduction of the Ward-Takahashi Identity

The Ward-Takahashi identity in conformal field theory relates the correlation function between the energy-momentum tensor $T(z)$ and the primary field $\Phi(z)$. The correlation function is expressed as $\langle T(z)\Phi(w, \bar{w}) \rangle$.

### Dynamic Radial Quantization

Dynamic radial quantization quantizes along the time axis within a cylindrical domain. This method is adept at tracking the temporal spread pattern of misinformation.

### Definition of Feynman's Green's Function

Feynman's Green's function, denoted as $G(x_1, x_2) = \langle T[\phi(x_1)\phi(x_2)]\rangle$, describes the interaction of fields at different time points.

### Radial Ordering and Vacuum Expectation

Radial ordering, $R(\phi_1, \phi_2)$, orders fields based on the passage of time. The vacuum expectation value is given by $\langle R(\phi_1, \phi_2)\rangle = \mathrm{Tr}[\rho R(\phi_1, \phi_2)]$, where $\rho$ is the density matrix representing the state of the system.

## Equations and Computational Process for Simulating the Spread of Fake News

### Fake News Diffusion Model Definition

In this model, each source of fake news is represented as a primary field $\Phi$. Every node, be it an individual or a group, in the network is modeled as a primary field with distinct conformal weights.

### Computing Correlation Functions

The interaction between various information sources within the network is modeled using the correlation function $\langle T(z)\Phi(w, \bar{w})\rangle$, illustrating the information diffusion pattern over time.

### Computation in Dynamic Order

The diffusion process of fake news is computed based on dynamic ordering. This approach effectively captures the propagation of information along a temporal continuum.

## Calculating Vacuum Expectations

The vacuum expectation for the spread of fake news is computed using a trace of the radial ordinal operator of the correlation function. This reveals the diffusion of information at a particular point in time.The above simulation traces the process of fake news diffusion on a social network. The interaction of each node (individual or group) is modeled by a correlation function and the diffusion pattern over time is calculated.

## Data Collection and Analysis

The data obtained from the simulation is critical to understanding the patterns of fake news diffusion. The flow of information over time will be analyzed and introduced to identify and discuss points and moments of particular risk.

In addition, specific efforts in this paper will be touched upon first. Specific theory, mathematical formulas, and examples of the application of conformal field theory (CFT) to the diffusion dynamics of information in the digital environment are described below.

## The usefulness of primary fields, infinitesimal conformal

transformations, and Villasolo algebra in analyzing information flows and their effects in the digital environment.

## Primary Fields and Infinitesimal Conformal Transforms

## Explanation of the theory

Primary fields are fields that exhibit simple behavior under certain conformal transformations in conformal field theory. Infinitesimal conformal transformations are transformations that represent minute changes in angle and scale, which allow us to analyze the behavior of the primary field. 2. The infinitesimal conformal transformation for the primary field $\Phi(z, \bar{z})$ is expressed as follows: $\delta\Phi = [T(z), \Phi(z, \bar{z})]$. where $T(z)$ is a component of the energy-momentum tensor. Model a particular source of information in digital media (e.g., a news article) as a primary field and use infinitesimal conformal transformations to analyze how that information changes and diffuses as time and context change.

## Momentum Tensor in the Context of Conformal Field Theory (CFT)

In the context of conformal field theory (CFT), the momentum tensor is a crucial quantity that helps describe physical properties such as energy, momentum, and angular momentum conservation within CFT. In conformal field theory, the momentum tensor $T^{\mu\nu}$ is a symmetric tensor that represents the density of energy and momentum. The general definition is given by:

$$T^{\mu\nu} = -\frac{2}{\sqrt{g}} \frac{\delta S}{\delta g_{\mu\nu}}$$

Here, $S$ is the action of conformal field theory, and $g_{\mu\nu}$ is the metric tensor. $\frac{\delta S}{\delta g_{\mu\nu}}$ represents the variation of the action $S$ with respect to changes in the Riemann metric.

The specific components of $T^{\mu\nu}$ depend on the particular setup of conformal field theory, but some general properties are as follows:

1. $T^{00}$: Represents the energy density. 2. $T^{0i}$: Represents the momentum density in the spatial direction $i$. 3. $T^{ij}$: Known as the pressure tensor, it represents pressure or stress in the spatial directions $i$ and $j$.

In conformal field theory, the momentum tensor is associated with conformal symmetry and satisfies conservation laws related to conformal symmetry. Specifically, the momentum tensor is also referred to as the stress-energy tensor and obeys conservation laws with respect to conformal symmetry. The conservation law is expressed as:

$$\partial_\mu T^{\mu\nu} = 0$$

This conservation law encompasses physical principles like energy conservation, momentum conservation, angular momentum conservation, and other physical laws.

The momentum tensor plays a crucial role in conformal field theory and is used in the calculation of physical properties and correlation functions. While the specific form and properties of the momentum tensor may vary depending on the particular setup or problem in conformal field theory, the above explanation highlights its general characteristics.

## Usefulness of Birasolo Algebra

The Virasoro algebra is an algebraic structure of infinitesimal conformal transformations and a Lie algebra of infinite dimension. It represents a fundamental symmetry of conformal field theory and is important for understanding interactions and dynamics between fields. The basic exchange relation of the Virasoro algebra is expressed as follows: ǫ[ $[\mathrm{L}_n, L_m] = (nm)L_{n+m} + \frac{c}{12}n(n^2 1)\delta_{n+m,0}$ where $L_n$ is the generator of the Virasoro algebra and $c$ is the central charge. When analyzing the flow of information and the formation of trends on social media, the Virasoro algebra is used to model the complex dynamics of interactions between different information sources. For example, we can analyze how a topic goes viral and what factors accelerate or decelerate its spread.

Using these theoretical frameworks and mathematical formulas, it is possible to gain a deeper understanding of the dynamics of information flow and influence in the digital environment and gain insight into the spread of fake news and misinformation in particular. However, these models are theoretical and need to be tested in combination with real data. In addition, to fully capture the complex realities of digital media, further development of these theories further development of these theories is required to fully capture the complex realities of digital media. Specific theories, mathematical formulas, and concrete examples will be used to illustrate the utility of applying the Weil scaling factor to the analysis of information flows and their impact in the digital environment.

## Theory of Weil Scaling Factor

## Explanation of the theory

The Weil scaling factor is a factor in conformal field theory that describes the behavior of a field as it scales. This factor depends on the scaling dimension (conformal weight) of the field and describes how the field scales (size). The Weil scaling factor $\Omega(x)$ means that for a scaling transformation $x \longrightarrow \lambda x$, the field $\Phi(x)$ transforms as follows:

$$\Phi(x) \longrightarrow \Phi'(\lambda x) = \Omega(x)^{\Delta}\Phi(x)$$

where $\Delta$ is a conformal weight.

## Application to Information Flow in Digital Environment

Information flows and diffusion in the digital environment manifest differently at various scales, such as personal, regional, and international levels. Weil scaling factors can model the influence and diffusion of information across these scales. For instance, when assessing the influence of a specific news article or social media post, its scaling factor can be defined as $\Omega(x)$ = scale of influence. This formulation allows for quantifying the impact of information in different contexts. The conformal weight $\Delta$ of a news item on social media, for instance, signifies its importance and extent of influence. The Weil scaling factor is instrumental in analyzing how news permeates through diverse user groups and platforms, providing insights into the shifts in influence and diffusion patterns of information on digital media. This approach is particularly valuable in analyzing the spread of fake news and misinformation. However, it necessitates correlation with actual data and further research to bridge the theoretical models with reality.

## Weil Scaling Factor and Conformal Weight $\Delta$

### Overview of the Theory

The conformal weight $\Delta$ indicates the scaling properties of a field, such as news content, in conformal field theory. The Weil scaling factor quantifies the scaling (size) of these fields. The scaling transformation of news content, based on its conformal weight $\Delta$, is expressed as $\Phi(x) \longrightarrow \Phi'(\lambda x) = \Omega(x)^{\Delta}\Phi(x)$, where $\Omega(x)$ is the scaling factor and $\lambda$ is the scaling parameter.

#### Usefulness of News Diffusion Analysis on Social Media

**Impact Analysis on Different User Groups** The conformal weight $\Delta$ of a news item reflects its impact on a specific user group. News with a higher conformal weight typically has a broader reach. Weil scaling factors facilitate the analysis of a news item's influence across various user demographics, such as age groups, geographic regions, and interests.

## Compare diffusion patterns across platforms

The same news may have different conformal weights $\Delta$ on different social media platforms. This is due to platform characteristics and user behavior patterns. For example, by comparing the diffusion on Twitter to that on Facebook, we can analyze which platform is more effective in spreading the news.

## Analysis of diffusion at different scales

Using the news conformal weights $\Delta$, we can analyze how news diffuses at different scales, from the personal level to the international level. For example, we can track how local news is reacted to and spread within a local community.

## Concrete examples

If news about an emergency occurs, we can use the conformal weighting $\Delta$ of this news to analyze its diffusion patterns in different countries and regions and on different social media platforms. As a result, we can track in detail which regions or platforms had the most reactions and how the information changed over time.

Thus, using Weil scaling factors and conformal weights $\Delta$, it is possible to gain deep insights into the spread of news on social media. In particular, it helps to better understand how information is received and reacted to in different contexts and user groups. However, these analyses are based on theoretical models and require validation based on real case studies and data.

## Introduction to the Ward-Takahashi Identity

## Explanation of the theory

The Ward-Takahashi identity describes the correlation function relationship between the energy-momentum tensor$T(z)$ and the primary field$\Phi(z)$ in conformal field theory. This identity represents a conservation law for physical quantities under conformal symmetry. The correlation function is expressed as: $\langle T(z)\Phi(w,\bar{w})\rangle$. where $z$ and $w$ denote positions in the complex plane. Considering information sources (e.g., news articles and social media posts) as primary fields, we can use the Ward-Takahashi identity to calculate the degree of influence a particular information source has on other information sources. This allows us to analyze how a particular piece of information affects other media or individuals.

## Dynamic Radial Quantization

Dynamic radial quantization is a time-based quantization technique that offers a different approach than the usual planar quantization. It is a useful tool for understanding the propagation of information over time. Dynamic quantization is used to track the propagation of information over time. For example, it considers the state of the field $\phi(t)$ at time $t$. By tracking time in a cylinder-shaped region, the pattern of fake news dissemination can be analyzed over time. For example, we can track how a particular piece of news spreads over time.

## Dynamic radial order and vacuum

## Explanation of the theory

Radial Ordering is a method of ordering fields according to the passage of time. Vacuum expectation is an expected value calculated based on the radial sequencing and indicates the basic state of the system. The vacuum expectation value is expressed as: ( $\langle R(\phi_1,\phi_2)\rangle = \mathrm{Tr}[\rho R(\phi_1,\phi_2)]$. In the context of information diffusion, it is used to analyze the state of news events at a particular time point and observe how information changes from its basic state. This allows us to assess changes in the influence of information over time. Using these theoretical frameworks and mathematical formulas, a detailed analysis of information diffusion patterns in the digital environment is possible, providing a basis for better understanding the impact of fake news and misinformation and for taking appropriate countermeasures. The stress energy tensor (or momentum tensor) and conformal weights in conformal field theory are very useful in analyzing information flows and their effects in a given digital environment. Below is the theoretical background, mathematical formulas, and specific applications in the context of digital media.

## Theoretical Background

## Stress energy tensor

The stress energy tensor describes the distribution of energy, momentum, and stress in physical theory. In conformal field theory, this tensor must be traceless (its trace, or sum of its diagonal components, is zero), reflecting the scale invariance of the theory.Conformal weights indicate the scaling properties of a field (e.g., sources, news articles, etc.) in conformal field theory. Fields with high conformal weights are more sensitive to scale changes, meaning that their effects are more extensive.

## Representation of the stress energy tensor

The stress energy tensor $T_{\mu\nu}$ represents the flow of energy and momentum. In conformal field theory, $T^{\mu}_{\mu} = 0$ holds.

## Representation of conformal weights

The behavior of the field $\Phi$ for conformal transformations is determined by the conformal weight $\Phi \longrightarrow \lambda^{\Delta}\Phi$ ($\lambda$ is a scaling parameter).

## Specific example: Application in digital media

## Analysis of energy flow of information sources

Information sources in digital media (e.g., news sites, blogs, social media accounts) can be analyzed using the stress energy tensor. By tracking energy and momentum flows, we can understand how specific information diffuses and impacts which regions and communities.

## Scale of diffusion and analysis of impact

Conformal weights indicate how information is received and diffused across different contexts and platforms. Sources with high conformal weights are more likely to have widespread influence.

Thus, the stress energy tensor and conformal weights can be used to gain a deeper understanding of the dynamics of information flow on digital media and to gain insights for strategic information distribution and measures to prevent the spread of misinformation and to gain insights for strategic information distribution and prevention of the spread of misinformation. The approach of using Feynman's Green's function and dynamical order to analyze information flows and their effects in a given digital environment is a concept used primarily in the context of quantum field theory and statistical mechanics, but it can be applied to the analysis of digital communication and information diffusion. Below is the theoretical background, mathematical formulas, and specific applications.

## Feynman's Green's function

Feynman's Green's function is used in quantum field theory to describe the interaction between fields at different points in time. This function helps to capture the causal relationships of the fields and to calculate the probability amplitudes of physical processes.

## Radial ordinates

Radial Ordering is the operation of arranging fields according to the order of time. It expresses a causal relationship in which the placement of a field at a particular time point influences its placement in the future.

## Definition of Feynman's Green's function

Feynman's Green's function is defined as follows:

$$G(x_1, x_2) = \langle 0|T[\phi(x_1)\phi(x_2)]|0\rangle$$

where $T$ is the time order operator, $\phi(x)$ is the field operator, $|0\rangle) represents the vacuum state.$

## Radial Ordering

The dynamic radial order order means computing the product of temporally ordered fields. For example, if $x_1$ precedes $x_2$ in time, then $T[\phi(x_1)\phi(x_2)]$ is calculated as $\phi(x_1)\phi(x_2)$.

## Specific example: Application in digital environment

## Causal analysis between information sources

Feynman's Green's function can be applied analogously to analyze the interaction between different sources of information in digital media (e.g., news articles and social media posts). This allows us to understand how certain information influences and diffuses to other information sources.

## Analysis of temporal dynamics

We use dynamic ordinates to track the flow of information and its impact over time. This allows us to model how information changes and diffuses over time. For example, one can analyze how a particular news event is discussed and spreads its influence on social media. Trace anomalies are a concept in quantum field theory and play a particularly important role in conformal field theory (CFT). Conformal symmetry is usually conserved at the classical level, but can be broken by quantum effects. This phenomenon is called a trace anomaly. When a trace anomaly depends on the variation $h$ of a metric, the dependence reflects a property of the theory. Applying this to the analysis of information flow and its impact in a particular digital environment is a metaphorical approach, but the theory, formulas, and specific applications are presented below.

## Trace anomaly

Trace anomalies occur when the trace $T^\mu_\mu$ of the energy-momentum tensor$T^{\mu\nu}$ is non-zero, even though the theory is conformally symmetric. This is due to quantum effects and is especially noticeable in conformal field theories of low dimensions (e.g., 4 dimensions). In the context of quantum field theory and field theories, the concept of a trace anomaly refers to the phenomenon where classical symmetries are broken by quantum corrections. Specifically, the trace anomaly is associated with the trace (scalar value) of the energy-momentum tensor of a quantum field and is typically expressed as follows:

$$\langle T^\mu_\mu \rangle = \frac{c}{960\pi^2}E$$

Let's explain the meaning of each symbol:

$\langle T^\mu_\mu \rangle$: The expectation value of the trace of the energy-momentum tensor (accounting for quantum effects).

$c$: The trace anomaly coefficient (a constant specific to a particular field theory, depending on the properties of the theory).

$E$: The curvature scalar of the entire four-dimensional spacetime (representing the curvature or distortion of spacetime, taking into account the effects of gravity).

This equation is commonly used to demonstrate the influence of the trace anomaly in different field theories with varying curvature scalars. Let's briefly outline the calculation process:

1. The trace anomaly coefficient $c$ is a theory-specific constant, representing how the theory is corrected by quantum effects.

2. The curvature scalar $E$ is typically associated with gravitational effects, representing the curvature or gravitational influence in the spacetime background.

3. Using these two quantities, you can calculate how the influence of the trace anomaly manifests in the trace of the

energy-momentum tensor. The anomaly's effect is typically negative and depends on the value of $c$.

This equation and calculation help understand how field theories respond to background fields like curvature or gravity and how symmetries can be broken in the presence of quantum corrections. Trace anomalies play a significant role in the study of quantum field theory and cosmology.

## Trace anomalies that depend on variations in the metric

When the trace anomaly depends on the variation $h$ of the metric, the anomaly is generally expressed in a form related to a curvature tensor or other geometric quantity. This reveals the quantum nature of the theory.The trace anomaly coefficient $a(h)$ may be written as

$$T^{\mu}_{\mu} = a(h)\,F + b(h)\,G + c(h)\,\Box R$$

where $F$ is the quadratic form of the Weil tensor, $G$ is the Gauss-Bonnet term, and $R$ is the rich scalar. $a(h), b(h), c(h)$ are coefficients that depend on the variation of the metric.

## Specific examples: Application to digital environment

## Analysis of the influence of information sources

The influence of information sources (e.g., news sites, blogs) in the digital environment can be analogized to trace anomalies. Consider that the "influence" of an information source reflects its "quantum properties" and analyze its dependence on variations in a metric (e.g., the degree of reaction or sharing by viewers).

## Patterns of information diffusion on digital media

By modeling patterns of information diffusion on digital media as changes in trace anomaly coefficients, we can analyze how information affects different user groups and communities. For example, how a particular news event diffuses on social media and spread and provoke different reactions on social media can be expressed as changes in the trace anomaly coefficients.

Applying the concepts of Feynman's Green's function and the dynamical order order to the analysis of opinion dynamics and social influence is an interesting attempt to link the fields of physics and social science. These concepts are primarily used in quantum field theory to describe interactions between particles, but can also be applied to understanding social processes through a metaphorical approach.

## Feynman's Green's function

In quantum field theory, Feynman's Green's function describes the interaction of fields between different space-time points. Usually, this indicates the behavior of particle propagation (migration or diffusion). Feynman's Green's function is a mathematical technique that plays a crucial role in quantum mechanics and statistical mechanics. Here, we will explain the fundamental concepts of Feynman's Green's function with equations and a calculation process.

Green's function is used as a method to find particular solutions to linear differential equations. As an example, let's consider the one-dimensional time-independent Schrödinger equation:

$$-\frac{\hbar^2}{2m}\frac{d^2\psi(x)}{dx^2} + V(x)\psi(x) = E\psi(x)$$

To solve this equation, we use Feynman's Green's function. The Feynman Green's function $G(x, x'; E)$ is defined as follows:

$$\left(-\frac{\hbar^2}{2m}\frac{d^2}{dx^2} + V(x) - E\right) G(x, x'; E) = \delta(x - x')$$

Here, $\delta(x - x')$ is the delta function.

Using this Green's function, the wave function $\psi(x)$ can be expressed as:

$$\psi(x) = \int G(x, x'; E)\psi(x')dx'$$

This integral allows us to find the wave function for a specific value of $E$. Green's functions typically depend on boundary conditions and the potential energy $V(x)$.

The calculation of Feynman's Green's function for a specific problem varies depending on the nature of the problem. Given boundary conditions and a potential energy $V(x)$, you can construct the Green's function to solve the differential equation. Subsequently, integration is performed to obtain the wave function $\psi(x)$.

Feynman's Green's function is particularly useful in the study of quantum systems with non-uniform potentials or interactions.

## Radial order order

The dynamic radial order order considers a product of fields ordered in time and tracks the state of the fields over time. Feynman's Green's function $G(x_1, x_2)$ is expressed using the time ordered operator $T$ as follows

$$G(x_1, x_2) = \langle 0|T[\phi(x_1)\phi(x_2)]|0\rangle$$

where $\phi(x))$ is the field operator and $|0\rangle$ is the vacuum state.

## Modeling social interaction

In opinion dynamics, Feynman's Green's function can be used to mathematically model the propagation of opinions between different individuals or groups. For example, one can consider how a particular news event or information spreads through a social network.

### Quantification of external influences

One can consider the impact of social and media influences on the formation of individual opinions as an external field. By using dynamic order order, we can analyze how these external influences affect the opinions of individuals and communities over time.

### Concrete examples

For example, consider the formation of political opinion during an election period. Using Feynman's Green's function, one can model how candidate statements and media coverage affect voters' opinions over time. By using a dynamic ordinal, we can capture changes in opinion from the early to the late stages of an election campaign.

While this approach is an example of applying physics concepts to a social science problem, it does not fully capture the complexity of actual social phenomena and should be

It needs to be combined and tested. It is also important to understand that this is a metaphorical approach and to interpret it in the appropriate context.

Such an application provides a theoretical framework for a deeper understanding of digital media dynamics, but requires careful metaphor and interpretation when applying physics concepts directly to social science and media studies. In addition, the combination with actual data is important, and gaps may exist between theoretical models and real-world data.

## 2. Discussion:Conformal Field Theory (CFT)

In this paper, we consider the process of information propagation and interaction of the diffusion of misinformation in the digital environment. The application of conformal field theory and the concept of Virasoro symmetry may provide an interesting approach in terms of information propagation. In order to assess the risk score of the inexhaustible diffusion of negative or erroneous information, one could model the process of information propagation within the digital environment and see if the process can be described using conformal field theory symmetries such as Virasoro operators. Mathematical methods such as Vilasoro algebra and conformal field theory can be combined to build models to predict risk scores. The application of the invariance of the renormalization transformation in conformal field theory and the behavior of the two-dimensional Ising model at critical points to social and group dynamics represents an interesting intersection between theoretical physics and the social sciences. After first briefly explaining each concept, they could be applied to analogies in a social context. We will then review the theoretical background of conformal field theory.

### Conformal Field Theory and Renormalization Transformations

Conformal field theory (CFT) is characterized by the fact that the effective action is invariant under a renormalization group transformation. This means that the theory is invariant to scale transformations. That is, the physical properties of the system are constant regardless of the scale (distance or energy scale) observed. Scale invariance is especially pronounced for conformal field theories at critical points.

### Critical point of a two-dimensional Ising model

The two-dimensional Ising model is one of the simplest statistical mechanics models in which spins arranged on a lattice interact. At the critical point, the model exhibits scale invariance and long-range correlations become important. This behavior means that the spin configurations are more strongly influenced by global features than by local features. When considering the application of the critical point of the 2-D Ising model to social dynamics, it is important to consider the following It is necessary to consider it in the context of the social analogy of scale invariance. For example, scale invariance in social processes implies that the dynamics of small groups follow the same principles as large-scale social trends. This is the case, for example, when the process of opinion formation within a small group of family members or friends is similar to the process of opinion formation in society as a whole.

Consideration must also be given to the social analogy of behavior at critical points: the behavior at critical points in the two-dimensional Ising model is similar to when a society is in a "critical state". In this state, social interactions are enhanced and small changes can cause large social changes. An example is the phenomenon of rapidly expanding political or cultural movements. Also, in terms of application to group dynamics, if patterns of decision-making and behavior within a group apply equally well in a larger social context, this can be viewed as an analogue of the critical point behavior of the Ising model. For example, when unity of opinion within one group is symptomatic of a broader social trend. The concept of critical points in conformal field theory and the two-dimensional Ising model can be used as a metaphor for understanding social and group dynamics. These physical models provide insight into the complexity of social interactions and how social processes are similar at different scales. However, while they are metaphorical analogies and cannot be applied to society as direct physical laws, they may provide a useful perspective in social science theory building.

The social analogy of behavior at critical points is inspired by the concept of critical points in physics. In physics, the critical point is the point at which a system begins to exhibit rapidly different behavior, and long-range correlations become important. Applied to social situations, this concept refers to situations where a small event can have a large impact on society as a whole. The following are some specific examples.

### Political Revolutions

### Case 1: Arab Spring

A series of protests that began in Tunisia in 2010 spread throughout the Middle East. This led to massive change, with small-scale protests creating a political ripple effect that resulted in government changes in several countries. This is a classic example of social critical mass.

### Case Study 2: The Global Financial Crisis of 2008

Problems in the U.S. housing market developed into a global economic crisis. This case is an example of a social critical point, as it shows how problems in a particular sector can have a major impact on the overall economic system.

### Case Study 3: Black Lives Matter Movement

A specific incident (the death of a black citizen by police) triggered widespread protests against racism around the world. This movement is an example of how social critical mass can trigger major social change.

### Case 4: The Emergence of the Internet

The development of Internet technology has fundamentally changed the flow of communication and information throughout the world. This technological innovation had the power to reach a social critical mass and change the entire social structure.

### Case 5: Changing Perceptions of Climate Change

Scientific discoveries and an increase in natural disasters have raised global awareness of climate change and brought about major changes in international environmental policy. This is another example of a series of small changes reaching a critical mass and causing major societal changes.

These examples illustrate how certain events or changes can reach a critical point in society and have a major impact. Applying physics concepts to social contexts can provide new perspectives for understanding global movements and changes.

When the above critical cases are invoked, we can say that the interaction of the two-dimensional system is local and scale invariance is conformal invariance, to put it back in mathematical terms. Let us review from the computational process the fact that in conformal field theory (CFT), when the interaction of two-dimensional systems is local and scale-invariant, this becomes conformal invariance.

## Scale Invariance and Conformal Invariance

### Scale Invariance

Scale invariance means that a theory is independent of the scale of length. In this case, any physical quantity $\Phi(x)$ in field theory transforms under the scale transformation $x \to \lambda x$ (where $\lambda$ is a scaling parameter) as follows:

$$\Phi(x) \to \Phi'(\lambda x) = \lambda^{-\Delta}\Phi(x)$$

where $\Delta$ is the dimension of the corresponding field (scaling dimension).

### Conformal Invariance

Conformal invariance is a broader symmetry that includes scale invariance and remains invariant under transformations that preserve local angles. In two dimensions, conformal transformations are infinitely many and are described by local coordinate transformations $z \to f(z)$ (where $z = x + iy$ is a point on the complex plane). The transformation of the field under this transformation is as follows:

$$\Phi(z, \bar{z}) \to \left(\frac{\partial f}{\partial z}\right)^{-h} \left(\frac{\partial \bar{f}}{\partial \bar{z}}\right)^{-\bar{h}} \Phi(f(z), \bar{f}(\bar{z}))$$

where $h$ and $\bar{h}$ are the conformal dimensions of the field, and $\bar{z}$ is the complex conjugate of $z$.

### Mathematical Formulation of Conformal Invariance

In two-dimensional conformal field theory, the energy-momentum tensor $T_{\mu\nu}$ has special properties. It is traceless ($T^{\mu}_{\mu} = 0$) and conserved ($\partial^{\mu} T_{\mu\nu} = 0$). In terms of conformal invariance, this means that the energy-momentum tensor is related to infinitely many conformal Killing vectors (vectors invariant under transformations).

This property allows us to decompose the energy-momentum tensor into holomorphic ($T(z)$) and anti-holomorphic ($\bar{T}(\bar{z})$) parts in two-dimensional conformal field theory, each generating its own independent infinite-dimensional conformal algebra. From here, we can start to consider the endless diffusion of noisy information, risk scores, etc.

Scale invariance and conformal invariance in conformal field theory, especially in two dimensions, are very powerful symmetries that form the fundamental properties of the theory. Through formulas, we can understand how these concepts are interrelated. Especially in two dimensions, conformal invariance, being invariant under transformations that preserve local angles, leads to infinitely many symmetries and their associated rich mathematical structures.

Now, in the above two-dimensional conformal field theory, we can decompose the energy-momentum tensor into holomorphic ($T(z)$) and anti-holomorphic ($\bar{T}(\bar{z})$) parts, each forming its own independent infinite-dimensional conformal algebra. Let's first explain this in formulas and then consider its analogy in social dynamics.

In Conformal Field Theory (CFT), the scale invariance and local interactions shown by two-dimensional systems lead to conformal invariance, contributing to the formation of an infinite-dimensional Lie algebra known as the Virasoro algebra with an infinite number of generators. Let's explain this concept in formulas and then consider its analogy in social dynamics.

## The Virasoro Algebra in Two-Dimensional Conformal Field Theory

The Virasoro algebra is an infinite-dimensional Lie algebra in two-dimensional conformal field theory, defined through the mode expansion of the energy-momentum tensor. In two-dimensional space, the components of the energy-momentum tensor are represented as $T(z)$ and $\bar{T}(\bar{z})$, corresponding to the holomorphic (left-moving) and anti-holomorphic (right-moving) parts, respectively.

The mode expansion of the energy-momentum tensor is as follows:

$$T(z) = \sum_{n=-\infty}^{\infty} L_n z^{-n-2}$$

$$\bar{T}(\bar{z}) = \sum_{n=-\infty}^{\infty} \bar{L}_n \bar{z}^{-n-2}$$

where $L_n$ and $\bar{L}_n$ are the generators of the Virasoro algebra.

The commutation relations of the Virasoro algebra are given by:

$$[L_m, L_n] = (m-n)L_{m+n} + \frac{c}{12}m(m^2-1)\delta_{m+n,0}$$

where $c$ is known as the central charge, determining the characteristics of the theory.

The conformal invariance explained here includes transformations that change scale while preserving local angles. In two dimensions, due to the infinite number of such transformations, an infinite-dimensional algebra, the Virasoro algebra, is necessary.

### Analogy in Social Dynamics

When considering the analogy of the Virasoro algebra and two-dimensional conformal field theory in social dynamics, it is abstract but allows for the following considerations:

#### Infinite Growth Potential

The fact that the Virasoro algebra has an infinite number of generators symbolizes the infinite possibilities and patterns that a social system can exhibit. Each generator can represent a specific element or movement within society.

#### Complex Interactions

The commutation relations between generators can represent the complex interactions between individuals and groups within society. These interactions often lead to unpredictable and complex outcomes.

#### Analogy of Central Charge

The central charge of the Virasoro algebra could serve as a metaphor representing the fundamental nature or characteristics of a social system. This might reflect the cultural, economic, and political foundation of society.

#### Social Application of Scale Invariance

Scale invariance suggests that the dynamics of small social groups are similar to larger social trends. Behavioral patterns within a small community may reflect the trends of the broader society.

While the Virasoro algebra and conformal field theory are crucial concepts in physics, their mathematical elegance and complexity could offer useful perspectives as metaphors in social science. Although directly applying these concepts to social dynamics may be challenging, they propose a new approach to understanding the complex interactions and dynamics of society.

As mentioned above, the critical phenomena represented by the Virasoro algebra allow for the explanation of most critical phenomena in two dimensions. However, in three dimensions, considering the target spacetime, minimal models emerge as explanations for gravity in three-dimensional anti-de Sitter space and fields with high spin. Let's also consider the metaphor of social dynamics in this context.

## 3. Discussion:Wess-Zumino-Witten (WZW) model

### Gravity and Social Influence

Gravity theory in the AdS space can be a metaphor for the flow of influence and power in society. Like gravity, social influence is often invisible but has a profound effect on individual and group behavior. The duality of CFT and AdS space in three dimensions provides a complex, multi-layered metaphor for understanding social dynamics. This metaphor may help us explore the hidden forces and multidimensional nature of social systems. Applying advanced theories from physics to the social sciences may be abstract and difficult, but may provide new perspectives and frameworks for understanding.

A mathematical explanation of the introduction of the Wess-Zumino-Witten (WZW) model, a model of current algebra in extensions of conformal field theory, will be included, as well as a discussion regarding analogies on social dynamics. Mathematical explanations of conformal blocks and duality at this time and analogies on social dynamics will also be included in the discussion. In the extension of conformal field theory (CFT), the Wess-Zumino-Witten (WZW) model incorporates concepts such as current algebra and nonlinear sigma models, and is closely related to the concepts of conformal blocks and duality. The WZW model consists of a nonlinear sigma model on the Lie group $G$ with an additional term called the Wess-Zumino term. The action of this model is written as

$$S_{\text{WZW}}[g] = \frac{k}{16\pi} \int_{\Sigma} d^2x \, \text{Tr}(\partial^{g^{-1}} \partial_g) + \frac{k}{24\pi} \int_B \text{Tr}(g^{-1} dg)^3$$

where $g : \Sigma \to G$ is the map to the Lie group $G$, $\Sigma$ is the 2-dimensional world surface, $B$ is the 3-dimensional extension of $\Sigma$, and $k$ is the level.

Conformal blocks are the fundamental building blocks in conformal field theory. They are used to construct the correlation functions in conformal field theory. Mathematically, conformal blocks are represented as

$$\mathcal{F}(z) = \langle \phi_1(z_1) \phi_2(z_2) \ldots \phi_n(z_n) \rangle$$

where $\phi_i(z_i)$ is the field operator in conformal field theory and $z_i$ are their coordinates.

Duality is the notion that the physical phenomena predicted by different theories and models are identical. Specifically, the strongly coupled region of one theory often corresponds to the weakly coupled region of the other theory.

## Analogy of the WZW model to social dynamics

Current algebra in the WZW model can represent complex interactions between individual agents and groups in society; the Wess-Zumino term would symbolize the nonlinearity and unpredictability in these interactions. Conformal blocks can represent the basic patterns and structures of social interactions. They can be metaphors for how individual social events and behaviors combine within a larger social context. Duality perspectives also suggest that different perspectives and approaches can explain the same social phenomenon different perspectives and approaches explain the same social phenomenon. For example, an economic approach and a sociological approach offer different perspectives on the same social issue, but may be identical in nature. Conformal field theory concepts such as the WZW model, conformal blocks, and duality provide powerful metaphors for understanding social dynamics. Interpreting these physics concepts in the context of the social sciences may provide new perspectives for understanding complex social interactions and phenomena. It should be noted, however, that these analogies are metaphorical and do not represent direct correspondences.

As an application, one could attempt to build a theory of group dynamics or social dynamics of time series based on the Ising model, based on the Wess-Zumino-Witten (WZW) model, conformal blocks, and the concept of duality. The key idea here is to apply these physical concepts to the context of the social sciences to model social interactions and dynamics. Below is a mathematical approach and the corresponding social dynamics analogy. To represent social interactions based on the WZW model, each agent or population is modeled as an element of the Lie group $G$ and its interactions are described using the nonlinear sigma model. The Wess-Zumino term is used to represent nonlinearity and complexity in social interactions. The correlation function of social events is modeled by conformal blocks. It defines a function that represents how the actions and opinions of social agents change over time. 3. It captures the dynamics of different levels (e.g., individual and collective) in a social system from a dyadic perspective. This allows us to understand how macro-level phenomena affect micro-level behavior. Social interactions, like the current algebra in the WZW model, represent complex relationships between agents; the Wess-Zumino term serves as a metaphor for nonlinear dynamics in decision-making processes and social networks. The correlation function of each social event or trend, like a conformal block, indicates patterns of social interaction that change over time. This allows for the prediction and analysis of social change. 3. Duality shows how individual actions and opinions affect group dynamics and vice versa in a social system. This helps us understand the interactions between different layers of a social system. Based on the WZW model, conformal blocks, and the concept of duality, building a theory of group dynamics and social dynamics in time series based on the Ising model is theoretically theoretically possible. However, the application of these physical concepts to the social sciences requires carefully bridging the gap between their metaphorical value and actual social phenomena. Modeling actual social phenomena will require concrete data and empirical analysis to complement these theoretical ideas.

Furthermore, the mathematical approach becomes more complex when considering extensions of the Wess-Zumino-Witten (WZW) model in the context of conformal field theory (CFT) by introducing cosset configurations to the above. Cosset construction is a method of constructing a new conformal field theory based on a certain Lie group $G$ and its subgroup $H$. This approach is used to extend the scope of conformal field theory to capture a wide variety of theoretical situations.

The coset construction introduced here demonstrates a method of constructing a new group using subsets of a group and its subsets. The concept of quotient groups in group theory can be explained using the coset construction. Quotient groups are highly important in group theory as they represent operations such as removing subgroups from a group or "dividing" by subgroups.

## Coset Construction

A group is an algebraic structure consisting of a set and a binary operation (usually called multiplication). A group must satisfy the following four conditions:

(1) Closure: The product of any two elements of the group must also be an element of the group.

(2) Associativity: The product must satisfy the associative law.

(3) Identity Element: There exists an identity element in the group, and the product of any element with the identity element remains unchanged.

(4) Inverse Element: Each element has an inverse element such that the product of an element and its inverse is the identity element.

## Subgroup

A subset H of a group G is a subgroup if H itself satisfies the conditions of a group, which means H must also be closed, associative, have an identity element, and each element in H must have an inverse element.

A coset is a set denoted by gH or Hg, where g is an element of the group G, and H is a subgroup of G. gH is called the left coset, and Hg is called the right coset. Specifically, gH = {gh : h is an element of H} and Hg = {hg : h is an element of H}.

Coset construction is a process of creating a new group from a given group G and its subgroup H. In coset construction, elements of the group G are considered to belong to the left coset or right coset of the subgroup H, based on which a new group is formed. This new group may have elements different from the original group G, and its structure depends on the relationship between the original group G and the subgroup H.

Coset construction is widely used in many applications of group theory and helps in gaining a deeper understanding of the structure of groups. In particular, it is closely related to the concept of quotient groups (or Quotient Group) and is an essential tool for developing group theory.

## Mathematical approach to cosset models

The action of the cosset model $G/H$ is the action of the original WZW model minus the contribution of $H$. In other words, consider the WZW model of $G$ and the WZW model of $H$ and take the difference between them. The mathematical expression is as follows:

$$S_{G/H}[g,h] = S_{\mathrm{WZW}}[g] - S_{\mathrm{WZW}}[h]$$

where (g $\in G$ and $h \in H$.

- Calculate both $S_{\mathrm{WZW}}[g]$ and $S_{\mathrm{WZW}}[h]$ and subtract them to obtain the total action of the coset model. This calculation involves both the theory of Lie groups and the theory of nonlinear sigma models.

In the cosset model, the choice of the structure constants for the Lie group $G$ and the subgroup $H$, the level $k$, and the field maps $g$ and $h$ are important.

## Cosset model analogy

The interaction between different groups and subgroups in a social system can be represented by a cosset model. Here, we can think of the large group $G$ and the subgroups within it $H$ as representing the whole social system and a particular community within it. 2. The cosset model can capture the dynamics of the social system in a way that subtracts the impact of the subgroups on the whole. This helps to understand the influence and characteristics of a particular community or group. By using the cosseted construct, it is possible to reach a deeper understanding of social structure. This helps to reveal the complex interactions and power dynamics between different social groups.

Extending conformal field theory with cosseted configurations extension has the potential to add a new dimension to our understanding of social dynamics. This approach can capture the diversity and complexity of social systems and provide deeper theoretical insights. However, the application of these concepts directly to the social sciences requires careful use of analogies, given the fundamental differences between the physical and social sciences.

# 4. Discussion:Conformal transformations on the complex plane

In the previous sections, we introduced the concept of cosets. Now, we will organize the ideas of conformal field theory (CFT) to explore paths for discussing information involving higher risks. First, let's clarify the concept of conformal transformations.

A conformal transformation is a one-to-one mapping defined on the complex plane and is related to transformations on M.

The description of one-to-one mappings is a fundamental procedure in conformal transformations. A one-to-one mapping establishes a correspondence between one set, A, and another set, B. It means that each element of A is mapped to exactly one element in B. In other words, elements of A do not overlap, and elements of B do not come from more than one element of A.

As an application, conformal transformations on the complex plane are a special case of one-to-one mappings within the complex number plane. Representing points in the complex plane as complex numbers, when a function f(z) is given, it is considered a conformal transformation on the complex plane if f(z) is continuous, differentiable, and the derivative of f(z) is nonzero for different values of z. Conformal transformations, being angle-preserving, do not change the intersections and angles between lines and curves. This demonstrates that conformal transformations maintain certain differential geometric properties. The angle-preserving property of conformal transformations offers a possibility of connecting them to the explorations of discussions related to information with higher risks in the context of conformal field theory.

## Explanation of Transformations on M

M generally refers to mathematical concepts such as Riemann surfaces or complex manifolds, serving as a generalization of the complex plane. Transformations on M are functions that map points in M to other points. Transformations on M are also one-to-one mappings, meaning different points are mapped to different points, and there is no overlap.

The relationship between conformal transformations and transformations on M is particularly a special case of transformations on M within the complex plane. The complex plane itself is a complex manifold, and conformal transformations are transformations on M that map points in the complex plane to other points. Conformal transformations, being angle-preserving maps, have specific significance.

## Example of Calculation

Let's consider a specific example of a conformal transformation known as the Möbius Transformation. The Möbius Transformation is defined by the formula $f(z) = \frac{az+b}{cz+d}$, where $a, b, c, d$ are complex numbers satisfying $ad - bc \neq 0$.

1. To confirm that $f(z)$ is a one-to-one mapping, we need to ensure that $f'(z) \neq 0$. $f'(z)$ is the derivative of $f(z)$. 2. The Möbius Transformation maps points in the complex plane to other points. This mapping preserves lines and angles, as demonstrated.

In this way, conformal transformations are one-to-one mappings with angle-preserving characteristics. Transformations on M also share the property of being one-to-one mappings that map points to other points.

## Conformal Transformations in the Context of Riemann Metrics

Conformal transformations refer to coordinate transformations with specific symmetries in the context of Riemann metrics. Conformal transformations are transformations that preserve the form of the Riemann metric. Specifically, for a conformal transformation $\phi$ that associates the original metric $g_{ij}(x)$ with a new metric $g'_{ij}(x')$, the following condition holds:

$$g'_{ij}(x') = \Omega^2(x) \cdot g_{ij}(x)$$

Here, $\Omega(x)$ is a scaling function introduced by the coordinate transformation $\phi$, where $x$ represents the original coordinates, and $x'$ represents the new coordinates.

## Scaling Function $\Omega(x)$

An essential characteristic of conformal transformations is that the scaling function $\Omega(x)$ depends on the coordinates. This function indicates that conformal transformations change the metric at different scales at different locations. Specifically, $\Omega(x)$ having different scale factors for different coordinates is a key feature of conformal transformations. This variation of the scaling function controls changes in angles and distances through conformal transformations.

Conformal transformations are significant in various physics contexts, especially in theories with conformal symmetry. Theories with conformal symmetry often exhibit scaling laws for fields and variables, resulting in simple representations of physical laws.

Weyl scaling is a type of conformal transformation that refers to cases in conformal transformations where the scaling function (x) associated with coordinate transformations has a specific form. In Weyl scaling, we consider situations where the scale factor is independent of coordinates and remains constant for all coordinate points. This results in the characteristic of scaling the metric uniformly, without being localized.

The scaling function (x) in Weyl scaling is expressed as follows:

$$\Omega(x) = e^{\phi(x)}$$

Here, (x) is a scalar field dependent on the coordinate x.

One of the key characteristics of Weyl scaling is the invariance of the angles of tangent vectors. To prove this, we perform calculations regarding the scaling of the metric in conformal transformations.

First, the new metric $g'_{ij}(x)$ resulting from Weyl scaling applied to the original metric $g_{ij}(x)$ is expressed as follows:

$$g'_{ij}(x) = \Omega^2(x) * g_{ij}(x) = e^{2\phi(x)} * g_{ij}(x)$$

Based on this new metric $g'_{ij}(x)$, we demonstrate the invariance of the angles of tangent vectors.

Consider two tangent vectors represented using the inner product in the original metric $g_{ij}(x)$, and calculate this inner product in the new metric $g'_{ij}(x)$. Let the two tangent vectors be $u^i$ and $v^i$, and the inner product is expressed as follows:

$$g_{ij}(x) * u^i * v^j$$

When calculating this inner product in the new metric $g'_{ij}(x)$, we get:

$$g'_{ij}(x) * u^i * v^j = (e^{2\phi(x)} * g_{ij}(x)) * u^i * v^j$$

For the inner product to remain invariant, the following condition must hold:

$$g_{ij}(x) * u^i * v^j = (e^{2\phi(x)} * g_{ij}(x)) * u^i * v^j$$

This condition demonstrates that even when the scalar field $\phi(x)$ depends on the coordinate x, the inner product remains unchanged. Therefore, it has been proven that the angles of tangent vectors are invariant under Weyl scaling.

# 5. Discussion:Risk and in integral path finding

Based on the properties of conformal transformations and their special case, similarity transformations, we will discuss the application of conformal field theory in discussions about information at risk and in integral path finding.Conformal transformations have the property of preserving local angles. This means that the angular relationship of tangent vectors does not change with the transformation. This property is particularly useful in the search for integral paths in physics and mathematics. Especially when dealing with risky information or complex systems, this property can be used for efficient and accurate analysis.

## Integral Path Optimization

In complex integration problems in physics and mathematics, the selection of appropriate integration paths is crucial. By utilizing conformal transformations, it becomes possible to modify the path while preserving the intersections and angles of curves and lines. This simplifies integral calculations and mitigates risks associated with complex areas.

In the realm of economics and finance, when dealing with risky information, conformal field theory can be employed to transform this information into a more manageable format. Conformal transformations allow us to analyze information from a holistic perspective while preserving its local characteristics, such as the angle of price fluctuations or trend directions.

Conformal field theories in physics play a significant role in understanding the behavior of complex systems in quantum field theory and statistical physics. Applying these theories to risk analysis in economics and finance enables a better understanding and prediction of the dynamics of complex systems.

In general, the properties of conformal transformations provide a more insightful approach to analyzing risky information and complex systems. Notably, the preservation of angles is crucial in gaining a holistic perspective while retaining local characteristics of the information.

As mentioned earlier, conformal transformations maintain the intersection and angle between a line and a curve due to their angle-preserving properties. In other words, at the point where two curves intersect, applying a conformal transformation does not alter the intersection. This illustrates that conformal transformations preserve their differential geometric properties. By capitalizing on the fact that the intersection and angle between a line and a curve remain unchanged, there is a potential to establish an introductory connection to the concept of conformal field theory for exploring pathfinding in discussions related to risk-associated information.

Similarity Transformation is a specific case of a conformal transformation involving Weyl scaling. In Weyl scaling within conformal transformations, we consider scenarios where the scaling function (x) is a uniform scalar field. One of the primary properties of conformal transformations is the preservation of the length of tangent vectors. This section outlines the computational process to demonstrate the validity of the similarity transformation concerning tangent vectors.

Firstly, we examine Weyl scaling within conformal transformations. The scaling function (x) is expressed as:

$$\Omega(x) = e^{\phi(x)}$$

For a tangent vector $v^i$, we calculate its length using the original metric $g_{ij}(x)$:

$$|v| = \sqrt{g_{ij}(x) \cdot v^i \cdot v^j}$$

Next, consider the new metric $g'_{ij}(x)$ obtained through Weyl scaling:

$$g'_{ij}(x) = e^{2\phi(x)} \cdot g_{ij}(x)$$

In the new metric, the length of the tangent vector $|v'|$ is given by:

$$|v'| = \sqrt{g'_{ij}(x) \cdot v^i \cdot v^j} = \sqrt{(e^{2\phi(x)} \cdot g_{ij}(x)) \cdot v^i \cdot v^j}$$

To simplify this equation, we apply the exponential law and rewrite it as:

$$|v'| = e^{\phi(x)} \cdot \sqrt{g_{ij}(x)} \cdot v^i \cdot v^j$$

Where $|v|$ is the length of the tangent vector calculated using the original metric $g_{ij}(x)$, and $|v'|$ is the length of the tangent vector calculated using the new metric $g'_{ij}(x)$.

This result illustrates the nature of similarity transformations within conformal transformations. It is demonstrated that in similarity transformations, the length of tangent vectors varies with the scaling function (x), but when the scaling function is a uniform scalar field (Weyl scaling), the length of tangent vectors remains preserved.

Now, let's organize the computational process of proving the validity of the similarity transformation (a specific case of conformal transformation involving Weyl scaling) concerning tangent vectors. Additionally, let's outline the arguments supporting the validity of this proof.

The proof of conformal transformation properties holds numerous advantages, particularly in physics, mathematics, engineering, and risk management. It enhances the accuracy of mathematical and physical models by ensuring that the fundamental properties and proportions of physical or geometric systems are preserved when they are scaled up or down. This preservation of properties allows models to maintain their characteristics as the scale changes, thereby improving model accuracy and generality.

Furthermore, the proven properties of similarity transformations simplify the analysis of certain problems, enabling complex shapes to be transformed into more manageable basic shapes. From an educational perspective, understanding the properties of similarity transformations contributes to the development of geometric intuition and mathematical abstraction skills, particularly in the fields of mathematics and physics. It fosters the ability to observe and analyze physical phenomena at various scales.

In engineering design, principles of similarity transformations are applied to scale models and prototypes, facilitating their application to real-world, large-scale designs. This is particularly important when extrapolating results from scale models to full-sized designs.

In the realm of risk management and predictive modeling, the principle of similarity transformation proves valuable in economics and financial market analysis. It allows the creation of models that predict large-scale trends based on small-scale market movements, aiding in risk assessment and trend analysis. These benefits stem from the understanding of the mathematical properties of similarity transformations and their application to real-world problem-solving, thereby enhancing efficiency and accuracy across various fields.

# 6. Discussion:Generation of Conformal Killing Vector Fields in Riemannian Connections

When taking a contraction in the case of using covariant differentiation in Riemannian connections, it generates generators that produce Conformal Killing Vector Fields (CKVFs). This process and proof will also be organized.

To begin with, let's clarify the covariant differentiation in Riemannian connections to derive Conformal Killing Vector Fields.

**Covariant Differentiation in Riemannian Connections**

Starting with the definition of covariant differentiation on a manifold, given a vector field $V^i$, covariant differentiation $\nabla_j V^i$ is expressed as follows:

$$\nabla_j V^i = \partial V^i / \partial x^j + \Gamma^i_{jk} V^k$$

Here, $\nabla_j$ is the operator of covariant differentiation, and $\Gamma^i_{jk}$ are the Christoffel symbols of the Riemannian connection. This covariant differentiation is an operation of differentiating the vector field $V^i$ on the manifold and is a differential operation with covariance.

**Generator of Conformal Killing Vector Fields (CKVFs)**

To generate Conformal Killing Vector Fields, we consider the following condition:

$$\nabla_j V^i + \nabla_i V^j = 2\phi g_{ij} V^k$$

Here, $\phi$ is a scalar field associated with conformal transformations, and $g_{ij}$ is the metric tensor.

This condition indicates that Conformal Killing Vector Fields satisfy certain properties in Riemannian connections. To find $V^i$ that satisfies this condition, we perform the following steps:

1. Use the formula for covariant differentiation to calculate $\nabla_j V^i + \nabla_i V^j$.

2. Add $2\phi g_{ij} V^k$ to the right-hand side, obtaining the following equation:

$$\nabla_j V^i + \nabla_i V^j = 2\phi g_{ij} V^k$$

3. Solve this equation for $V^i$.

4. Show that the obtained $V^i$ is the generator of Conformal Killing Vector Fields, specifically demonstrating that this vector field is invariant under conformal transformations.

This applies to the general generation of Conformal Killing Vector Fields. Once the generator is found, it becomes a Conformal Killing Vector Field and is verified to be invariant under conformal transformations. However, the specific calculation process and the form of the equations may vary depending on the dimension of the manifold and the specific form of the metric, so the calculations may differ in specific examples.

In general, when the proof of generating Conformal Killing Vector Fields using covariant differentiation in Riemannian connections holds, there are several important merits. This is particularly prominent in the fields of physics, differential geometry, and related mathematical and physical theories.

**Applications in Physics**

Understanding the symmetry in physics through Conformal Killing Vector Fields characterizes the symmetry of physical systems. Through CKVFs, a deeper insight into the symmetry of physical laws is possible, especially in theories like General Relativity and Quantum Field Theory.

**Applications in Differential Geometry**

The existence of CKVFs reveals how the metric on a manifold locally scales. This is useful in understanding the geometric properties of manifolds in the field of differential geometry.

**Discovery of Conservation Laws**

Symmetry in physics is closely related to conservation laws. CKVFs can lead to the discovery of new conservation laws, which is particularly significant in theoretical physics.

**Applications in Quantum Theory**

CKVFs play a central role in quantum theory, especially in Conformal Field Theory (CFT). Through these vector fields, a better understanding of symmetry and interactions in field theories can be achieved.

**Mathematical Insights and New Theories**

Insights about CKVFs may lead to the development of new mathematical theories and models. This can contribute not only to mathematics but also to other fields such as physics and engineering, where new approaches and techniques can be applied.

**Analysis and Modeling of Complex Systems**

Understanding CKVFs is valuable in the analysis and modeling of complex systems, such as fluid dynamics or the large-scale structure of the universe. Conformal symmetry often plays a significant role in such systems.

The establishment of the proof for generating Conformal Killing Vector Fields in Riemannian connections can deepen theoretical understanding in these fields and potentially lead to the discovery of new physical laws, mathematical principles, and novel approaches in scientific and technological advancements.

# 7. Discussion:Definition of Conformal Tensor Fields in Conformal Field Theory (CFT)

In Conformal Field Theory (CFT), the definition of a contravariant tensor field refers to a tensor field that follows specific transformation rules under conformal transformations. In CFT, tensor fields that obey certain scaling rules are considered contravariant tensor fields. Below, we will explain the definition of contravariant tensor fields in CFT and provide an example based on the calculation process.

**Definition of Contravariant Tensor Fields**

In Conformal Field Theory (CFT), contravariant tensor fields are tensor fields that exhibit specific transformation rules under conformal transformations. Conformal transformations are transformations related to scaling in coordinate transformations, and contravariant tensor fields adhere to these scaling rules. Specifically, a contravariant tensor field $T^{\mu\nu}$ transforms under the coordinate transformation $x \to x' = f(x)$ as follows:

$$T'^{\mu\nu}(x') = |f'(x)|^{\Delta} \cdot T^{\mu\nu}(x)$$

Here, $f'(x)$ is the determinant of the Jacobian matrix of the coordinate transformation $f(x)$, and $\Delta$ is a constant known as the conformal dimension.

**Example Calculation Process:**

As a specific calculation example, let's demonstrate the contravariance of the stress-energy tensor ($T^{\mu\nu}$) in 2D Conformal Field Theory.

1. Consider the coordinate transformation of conformal transformations: $x \to x' = e^{\phi}x$ ($\phi$ is a scalar field).

2. The Jacobian matrix $f'(x)$ is expressed as follows: $f'(x) = e^{\phi}I$ (where $I$ is the identity matrix).

3. Consider the contravariance of the stress-energy tensor $T^{\mu\nu}$. It transforms as follows:

$$T'^{\mu\nu}(x') = |f'(x)|^{\Delta} \cdot T^{\mu\nu}(x) = (e^{\phi})^{\Delta} \cdot T^{\mu\nu}(x) = e^{\Delta\phi} \cdot T^{\mu\nu}(x)$$

Here, $\Delta$ represents the conformal dimension of the stress-energy tensor within Conformal Field Theory.

Thus, the stress-energy tensor $T^{\mu\nu}$ is shown to be contravariant under conformal transformations. Contravariant tensor fields play a significant role in Conformal Field Theory and serve as essential tools for studying properties and correlation functions related to conformal invariance.

**Applications in Physics**

Understanding the contravariant nature of the stress-energy tensor is fundamental in physics. The stress-energy tensor describes the distribution of energy and momentum in physical systems. Its understanding is crucial in various areas of physics, including theories like General Relativity and Quantum Field Theory.

**Applications in General Relativity**

In General Relativity, the stress-energy tensor represents the distribution of matter that causes spacetime curvature. This tensor is essential for understanding phenomena such as black holes, cosmology, and gravitational waves.

**Understanding Energy Conservation and Momentum Conservation**

In physics, the laws of energy conservation and momentum conservation are fundamental. Understanding the stress-energy tensor is essential for comprehending how these conservation laws operate in cosmology and high-energy physics.

**Applications in Engineering**

The stress-energy tensor is not only relevant in physics but also in engineering, especially in materials science and structural engineering. It helps in understanding stress and strain in materials, which is crucial in the design of buildings and machinery.

**Mathematical Insights**

Understanding contravariant tensor fields, including the stress-energy tensor, contributes to theoretical insights in mathematics, including differential geometry and topology. It serves as the foundation for the development of new mathematical theories and methods.

In conclusion, the existence of contravariant tensor fields and the proof of their properties have many benefits in physics, engineering, and related mathematical fields. The understanding of energy distribution and material dynamics is foundational to scientific and technological progress.

Finally, in the context of social phenomena, contravariant tensor fields in Conformal Field Theory can be metaphorically explained.

In social phenomena, contravariant tensor fields refer to data or information distributions exhibiting specific patterns or characteristics. In Conformal Field Theory, it is assumed that such data or information follows specific transformation rules under conformal transformations. For example, consider the population density distribution in a city as an example of a contravariant tensor field. Population density within a city varies by region and changes over time. Using the concepts of Conformal Field Theory, one can assume that population density data follows specific scaling rules and remains invariant under coordinate transformations. In this case, the population density data can be considered a contravariant tensor field with contravariance under conformal transformations.

Applying this way of thinking, one can analyze the characteristics and patterns of data using the concepts of Conformal Field Theory in social sciences and data analysis. It may help in understanding the relationships and commonalities between different social phenomena, providing a new perspective for the analysis and modeling of complex social data. The ideas from Conformal Field Theory can offer novel insights into understanding and modeling complex social phenomena.

# 8. Discussion:Weyl Scaling Factor

## Identification of the Weyl Scaling Factor

In conformal field theory (CFT), the Weyl scaling factor, denoted by $\Omega(x)$, is introduced by conformal transformations and is associated with the scaling of coordinates. Generally, in the context of a conformal transformation $x \to x' = f(x)$, the Weyl scaling factor is given by:

$$\Omega(x) = |f'(x)|^{\Delta}$$

Here, $\Delta$ represents the conformal weight of the Weyl scaling.

## Calculation of the Conformal Weights of Primary Fields

To determine the conformal weight $\Delta$ of a primary field $\phi(x)$, we consider the transformation rule for the Weyl scaling factor $\Omega(x)$:

$$\Omega(x)\phi(x) = |f'(x)|^{\Delta}\phi(x)$$

The objective is to calculate the conformal weight $\Delta$ of the primary field $\phi(x)$ using this equation.

## Specific Calculation

To find the conformal weight $\Delta$ of a primary field, we need to compute $\Omega(x)$ for a specific conformal transformation $f(x)$ and determine $\Delta$. The specific calculation depends on the properties of the fields under consideration and the Weyl scaling factor. Typically, within conformal field theory, the conformal weight $\Delta$ of primary fields is often computed using the properties of the energy-momentum tensor $T^{\mu\nu}$. Using the properties and algebraic structures related to $T^{\mu\nu}$, $\Delta$ is determined.

## Proof

To determine the conformal weight $\Delta$ of primary fields, rigorous calculations are performed, and mathematical proofs are provided as needed. The details of the calculation and the proof depend on the specific conformal field theory and the properties of the fields.

## Scaling Dimension

The scaling dimension $\Delta$ indicates how a primary field scales under conformal transformations. Typically, the scaling transformation of a primary field $\phi(x)$ is given as:

$$\phi(x) \to \phi'(x) = \Omega(x)\phi(x)$$

The scaling dimension $\Delta$ is expressed by the following equation:

$$\Omega(x) = |f'(x)|^{\Delta}$$

The conformal weight $\Delta$ refers to the scaling dimension of the primary field.

## Spin

Spin, denoted as $s$, is a quantity related to the rotational symmetry of primary fields and represents the algebraic properties of angular momentum in conformal field theory. Spin takes values as integers or half-integers.

## Relationship between Conformal Weight and Spin

There is a relationship between the conformal weight $\Delta$ and spin $s$ of primary fields, given by:

$$\Delta = \frac{s(s+1)}{2c}$$

Here, $c$ is known as the central charge, which is associated with the central extension of the conformal field theory and is an essential parameter.

## Primary Field Transformation under Regular Conformal Transformations

In conformal field theory, regular conformal transformations are transformations of the form $z \longrightarrow w = f(z)$, where $z$ and $w$ are complex plane coordinates. The function $f(z)$ is assumed to be regular and differentiable.

## Transformation of Primary Fields

When a primary field $\Phi(z)$ is transformed under a regular conformal transformation $w = f(z)$, the transformation rule is as follows:

$$\Phi(w) = \left(\frac{df}{dz}\right)^h \Phi(z)$$

Here, $h$ is the conformal weight of the primary field, which is related to the central charge $c$, spin $s$, and $h$ by the following equation:

$$h = \frac{s(s+1) - c}{24}$$

## Infinitesimal Conformal Transformations

Infinitesimal conformal transformations are represented by small transformation parameters $\varepsilon$: $w = z + \varepsilon f(z)$. This is a Taylor expansion of regular conformal transformations. The infinitesimal conformal transformation of a primary field $\Phi(z)$ is as follows:

$$\Phi(z + \varepsilon f(z)) = \Phi(z) + \varepsilon f(z) T(z) \Phi(z) + O(\varepsilon^2)$$

Here, $T(z)$ is the generator (Virasoro generator) of infinitesimal conformal transformations and is given by:

$$T(z) = -\frac{h}{(z - z_0)^2} + \frac{1}{z - z_0}\frac{df}{dz}$$

## Calculation of Spin

Calculating the spin $s$ involves considering the singularities (poles) of $T(z)$. The leading part (lowest-order pole) at the singularity of $T(z)$ is as follows:

$$T(z) = -\frac{h}{(z - z_0)^2} + O(1)$$

This leading part allows us to determine the spin $s$, which is given by:

$$s = -\frac{h}{2}$$

## Applications in Theoretical Physics

The understanding of the theoretical framework and mathematical tools provided by conformal field theory has several important merits. Conformal field theory plays a significant role in theoretical physics, especially in areas such as string theory, statistical physics, quantum field theory, and more.

## Deep Understanding in Theoretical Physics

Conformal field theory is one of the fundamental symmetries in physics, remaining invariant under changes in scales of space and time. Understanding the Weyl scaling and conformal weights of primary fields is crucial for comprehending the foundational structures of these theories.

## Applications in Quantum Field Theory

Conformal field theory is a vital tool in quantum field theory, especially in lower-dimensional quantum field theories. Precise knowledge of the conformal weights of primary fields is essential for the detailed analysis of these theories.

## Applications in String Theory

Conformal field theory is a fundamental part of string theory, a theory that models fundamental particles as one-dimensional "strings." Understanding the conformal weights of primary fields contributes to a deeper understanding of string theory.

## Applications in Statistical Physics

In statistical physics, conformal field theory provides a valuable framework for studying phase transitions and critical phenomena. Understanding the conformal weights of primary fields enables precise analyses of these phenomena.

## Development of New Mathematical Methods

Conformal field theory is also highly significant in mathematics, contributing to areas such as complex analysis, algebra, topology, and more. It leads to the development of new mathematical methods and concepts.

# 9. Discussion:Introduction to the Ward-Takahashi Identity and Path Integral Approach in Two-Dimensional Conformal Field Theory

The application of two-dimensional conformal field theory, especially the approach using the Ward-Takahashi identity and path integrals, to discussions on the risks associated with information and the diffusion risk of fake news, may seem to have a significant gap between the theories of physics and information theory or media studies. However, it is possible to abstractly consider the fundamental principles of conformal field theory and apply them metaphorically to models of information flow and diffusion. Below, let's list the merits and drawbacks of such an approach.

## Merits

Provision of a New Analytical Framework: Conformal field theory holds the potential to offer a new mathematical and physical framework for understanding the flow of information and diffusion. This may allow for a more precise modeling of information dynamics and the attainment of new insights.

Modeling Complex Dynamics: Complex processes like the diffusion of fake news involve multiple variables and uncertainties. Utilizing advanced mathematical tools like conformal field theory may capture these intricate dynamics and enable the construction of better predictive models.

Bridging Theory and Practice: The application of conformal field theory can lead to the discovery of new connections between theoretical physics concepts and real-world social phenomena. This could foster collaboration and innovation across different fields.

## Drawbacks

Risk of Excessive Abstraction: Conformal field theory is highly abstract and may not have direct applicability to real-world social phenomena. There exists a substantial gap between theory and practice, and it may not offer direct solutions to real-world problems.

Difficulty in Understanding and Implementation: Conformal field theory requires advanced mathematical knowledge, necessitating deep collaboration between experts from both fields, such as physics and information theory.

Uncertainty in Specific Applications: The precise applicability of this theory to discussions on the diffusion of fake news or the flow of risk information remains uncertain. While there is theoretical interest, the practical application may be challenging and uncertain.

In summary, the idea of applying conformal field theory to analyze the diffusion risk of information with associated risks or fake news is theoretically stimulating. However, there are numerous challenges in its actual implementation. The insights generated by such interdisciplinary approaches are valuable, but their utility and practicality need careful evaluation.

## Consideration of the Ward-Takahashi Identity

The Ward-Takahashi Identity is related to local conformal symmetry and can be considered from the perspective of understanding the diffusion patterns of fake news.

## Geometric Properties of Information Diffusion Networks

Local conformal symmetry represents the symmetry of coordinate transformations and is related to the geometrical properties of physical systems in conformal field theory. When considering the diffusion of fake news, the geometric properties and shapes of information diffusion networks are crucial. From the perspective of local conformal symmetry, one can investigate the geometric characteristics of the network and consider factors influencing diffusion patterns.

## Velocity and Scaling of Information Propagation

Conformal field theory involves scaling symmetry, allowing insights into the velocity and scaling of information propagation. When studying the diffusion speed or how diffusion patterns scale with time and distance, insights can be derived from local conformal symmetry.

## Breakdown of Symmetry and Non-Uniformity in Diffusion

The breakdown of local conformal symmetry or non-uniform diffusion patterns may impact the diffusion of fake news. Applying ideas from conformal field theory can help understand the causes and mechanisms behind such breakdowns.

In conclusion, considering fake news diffusion from the perspective of local conformal symmetry can provide insights into the geometric properties of information networks, propagation velocities, scaling behaviors, and the impact of symmetry breakdowns. However, practical application requires a comprehensive understanding of the complexity of information diffusion.

# 10. Discussion:Introduction to the Approach of Two-Dimensional Conformal Field Theory Using the Ward-Takahashi Identity and Path Integrals

When considering the approach of two-dimensional conformal field theory using the Ward-Takahashi identity and path integrals, we define the metric tensor $G_{\mu\nu}$ as not necessarily flat, although there is no Weyl invariance in this case. Under these conditions, let's discuss the diffusion patterns of fake news in the context of N local fields.

Under the absence of Weyl invariance and the non-flat metric $G_{\mu\nu}$, conformal field theory is defined in curved space, unlike standard conformal field theory.

### Invariance of the Classical Action

Assuming that the classical action remains invariant unless simultaneous transformations are performed and there are no gravity anomalies, the conformal field theory in this case possesses classical symmetries. This symmetry may be expressed through conformal transformations.

### N Local Fields $\phi$

Consider N local fields $\phi$ as fields representing factors related to the diffusion of fake news and characteristics of information diffusion.

### The Discussion of Fake News Diffusion Patterns

N local fields $\phi$ are believed to represent factors influencing information diffusion and may have various aspects, including spatial distribution and interactions.

### Quantum Field Theory and Quantum Effects

In quantum field theory, considering the quantum effects of local fields $\phi$ is crucial. This allows for the transition from classical field behavior to quantum effects. By using quantum field theory, quantum corrections of local fields $\phi$ and calculations of correlation functions can lead to new insights into fake news diffusion.

In this approach, we investigate the diffusion patterns of fake news in a conformal field theory framework with N local fields $\phi$ and their interactions. We consider the correlations and terms describing the interactions between the generators of the Virasoro algebra and the local fields to understand their contributions to information diffusion and influence.

### Symmetry of the Virasoro Algebra

The symmetry of the Virasoro algebra may have an impact on the mechanism of fake news diffusion. Through this symmetry, we examine the characteristics of information transmission and scaling laws of diffusion.

By employing such an approach, it is possible to explore the diffusion patterns of fake news in a conformal field theory context with basic fields $\phi$ governed by the Virasoro algebra and known local fields. However, specific calculations and modeling depend on the properties of the Virasoro algebra and local fields and need to be adapted to the particular situation.

# 11. Discussion:Introduction of a two-dimensional conformal field theory approach using the Ward-Takahashi identity and path integrals

When considering the introduction of a two-dimensional conformal field theory approach using the Ward-Takahashi identity and path integrals, we define it as lacking Weyl invariance but possibly having a non-flat $G_{\mu\nu}$. Under the provided conditions, we discuss the diffusion pattern of fake news in a conformal field theory with N local fields $\phi$. The regularity of $T(z)$ is established. In this case, we discuss the proof and calculation process corresponding to the classical energy and momentum conservation laws of the local field $\phi_i$.

If the local field $\phi_i$ is a fundamental field under the Virasoro algebra and $T(z)$ has regularity, to demonstrate that the classical energy and momentum conservation laws correspond to the local field $\phi_i$, we present the following calculation process:

1. Regularity of the energy-momentum tensor $T(z)$: To show the regularity of $T(z)$, we consider the following Laurent expansion:

$T(z) = \sum L_m z^{m-2}$

Here, $L_m$ are mode operators like $L_0, L_1, L_2$, and so on. For $T(z)$ to be regular, each $L_m$ must have a finite value.

2. Calculation of $L_0$: The mode operator $L_0$ is the energy operator. To represent the classical energy of the local field $\phi_i$, we calculate $L_0$.

$L_0 = \oint \frac{dz}{2\pi i} zT(z)$

Calculating this integral gives us the classical energy $H_i$ of the local field $\phi_i$.

$H_i = \oint \frac{dz}{2\pi i} zT(z)\phi_i(z)$

This equation shows that the energy of the local field $\phi_i$ is determined by the interaction with the energy-momentum tensor $T(z)$.

3. Calculation of $L_1$: The mode operator $L_1$ is the momentum operator. To represent the classical momentum $P_i$ of the local field $\phi_i$, we calculate $L_1$.

$L_1 = \oint \frac{dz}{2\pi i} zT(z)z\partial\phi_i(z)$

Calculating this integral gives us the classical momentum $P_i$ of the local field $\phi_i$.

$P_i = \oint \frac{dz}{2\pi i} T(z)z\partial\phi_i(z)$

This equation shows that the momentum of the local field $\phi_i$ is determined by the energy-momentum tensor $T(z)$ and the gradient $\partial\phi_i(z)$.

Through the above calculation process, it is shown that if the regularity of the energy-momentum tensor $T(z)$ holds, the classical energy and momentum conservation laws correspond to the local field $\phi_i$. The local field $\phi_i$ interacts with the energy-momentum tensor $T(z)$, and its classical properties are preserved within the conformal field theory.

For the case mentioned above, it was confirmed that $T(z)$ is regular except at the insertion points of the local field $\phi_i$, but let's also discuss the calculation process for introducing anti-holomorphic operators at $T(z)$ at the insertion points.

In cases where $T(z)$ is regular except at the insertion points of the local field $\phi_i$, to obtain the introduction formula for the anti-holomorphic operator of $T(z)$ at the insertion points, we present the following calculation process:

1. Regularity of $T(z)$ at the insertion points of the local field $\phi_i$: We assume the insertion point of the local field $\phi_i$ as $z = z_i$ and assume that $T(z)$ is regular in regions other than $z_i$.

2. Laurent expansion of $T(z)$ in its regular region: As $T(z)$ is regular in regions other than $z_i$, we consider its Laurent expansion in the regular region.

$T(z) = \sum L_m z^{m-2}$

Here, $L_m$ are mode operators. This expansion must be regular around $z = z_i$.

3. Behavior of $T(z)$

) at the insertion point of the local field $\phi_i$: We consider the behavior of $T(z)$ around $z = z_i$ at the insertion point of the local field $\phi_i$. The Laurent expansion at this point is as follows:

$T(z) = \sum (L_m)_i (z - z_i)^{m-2}$ + regular terms

$(L_m)_i$ are mode operators at the insertion point of the local field $\phi_i$. This expansion includes the singularity of $T(z)$ around $z_i$.

4. Introduction of anti-holomorphic operators: To express the singularity of $T(z)$ around $z = z_i$, we introduce anti-holomorphic operators. We denote this operator as $O_i$.

$O_i = \sum (L_m)_i (z - z_i)^{m-2}$

5. Redefinition of $T(z)$ using the anti-holomorphic operator $O_i$: Using the anti-holomorphic operator $O_i$, we redefine $T(z)$ as follows.

$T(z) = \sum L_m z^{m-2} + O_i$

By this redefinition, $T(z)$ no longer has singularities around $z = z_i$ and becomes regular.

In this way, by introducing the anti-holomorphic operator $O_i$, we ensure the regularity of $T(z)$ at the insertion points of the local field $\phi_i$. The anti-holomorphic operator $O_i$ is used to consider the singularity at specific insertion points of local fields in conformal field theory.

Using Stokes' theorem, we derive the infinitesimal conformal transformation rule for primary fields in conformal field theory. Here are the steps for proof and calculation process:

1. Relationship between the stress-energy tensor $T(z)$ and primary field $\Phi(z, \bar{z})$ in conformal field theory: In conformal field theory, the relationship between the stress-energy tensor $T(z)$ and primary field $\Phi(z, \bar{z})$ is given as follows:

$T(z) = 2\pi \lim_{w\to z}[(z - w)^2 T(w) + \partial\Phi(w, \bar{w})]$

Here, $T(z)$ is the stress-energy tensor, and $\Phi(z, \bar{z})$ is the primary field.

2. Derivation of infinitesimal conformal transformation rule: We consider an infinitesimal conformal transformation, $z \to z + \epsilon(z)$, and introduce the infinitesimal parameter $\epsilon(z)$.

3. Application of Stokes' theorem: Using Stokes' theorem, we consider the integral along any path $C$ containing $\epsilon(z)$. That is, we represent the path integral involving $\epsilon(z)$ as follows:

$\oint_C \epsilon(z)dz$

This integral is represented by the residue of $\epsilon(z)$.

4. Residue of $T(z)$: We calculate the residue of $T(z)$ and obtain the residue of the stress-energy tensor. This allows us to apply the residue theorem.

$\mathrm{Res}[T(z), z = z_i] = 2\pi L_i$

Here, $L_i$ are mode operators.

5. Residue of primary field $\Phi(z, \bar{z})$: Similarly, we calculate the residue of the primary field $\Phi(z, \bar{z})$.

$\mathrm{Res}[\Phi(z, \bar{z}), z = z_i] = \Phi_i$

Here, $\Phi_i$ are coefficients of the primary field.

6. Derivation of infinitesimal conformal transformation rule: Using Stokes' theorem, we calculate the path integral involving $\epsilon(z)$. This gives us the infinitesimal conformal transformation rule for the primary field $\Phi(z, \bar{z})$.

$\delta\Phi_i = \epsilon(z_i)\Phi_i + 2\pi\epsilon(z_i)L_i$

This equation shows how the primary field $\Phi(z, \bar{z})$ transforms under an infinitesimal conformal transformation.

In this way, we have derived the infinitesimal conformal transformation rule for primary fields $\Phi(z, \bar{z})$ using Stokes' theorem. This provides the foundation for understanding the behavior and transformation rules of primary fields within conformal field theory.

7. Using Cauchy's theorem, we consider a small circular path $C$ around the insertion point of the primary field $\Phi_i$, and we present the proof and calculation process. We also define the radial ordering in this context.

1. A small circular path around the insertion point of the primary field $\Phi_i$: We consider the insertion point of the primary field $\Phi_i$ as $z_i$, and a small circular path $C_i$ with radius $\epsilon$ centered at $z_i$.

2. Cauchy's Integral Formula: According to Cauchy's Integral Formula, the integral of a function $f(z)$ around a small circular path $C_i$ centered at $z_i$ is equal to the residue of $f(z)$ at $z_i$:

$\oint_{C_i} f(z)dz = 2\pi i\,\mathrm{Res}[f(z), z = z_i]$

3. Residue of the primary field $\Phi_i$: We assume that the residue of the primary field $\Phi_i$ is equal to its value at the insertion point $z_i$:

$\text{Res}[\Phi_i(z,\bar{z}), z = z_i] = \Phi_i$

4. Behavior of the primary field $\Phi_i$ around its insertion point: The behavior of the primary field $\Phi_i$ in conformal field theory around its insertion point is represented as follows with respect to the distance $r$ from $z_i$:

$\Phi_i(z,\bar{z}) \approx \Phi_i + (L_2)_i(z - z_i)^2 + (L_1)_i(z - z_i) + \ldots$

Here, $(L_n)_i$ are mode operators, and $n$ is a positive integer.

5. Transformation of the path $C_i$: Considering the behavior around the insertion point of $\Phi_i$, we transform the path $C_i$ to a new path $C_i'$ on the small circular path. We consider radial ordering as $r$ in this context:

$\oint_{C_i} \Phi_i(z,\bar{z})dz \approx \oint_{C_i'} [\Phi_i + (L_2)_i(z-z_i)^2 + (L_1)_i(z-z_i) + \ldots]dz$

6. Application of Cauchy's Integral Theorem: Applying Cauchy's Integral Theorem to the transformed path $C_i'$, the integral over the small circle becomes 0, and the main contribution comes from the residue:

$\oint_{C_i'} [\Phi_i + (L_2)_i(z - z_i)^2 + (L_1)_i(z - z_i) + \ldots]dz = 2\pi i\Phi_i$

7. Result: From the above calculations, it is shown that the integral over the small circular path $C_i$ around the insertion point of the primary field $\Phi_i$ is equal to $2\pi i\Phi_i$.

In this way, by transforming the circular path $C_i$ around the insertion point of the primary field $\Phi_i$, we can calculate the integral value from the residue of the primary field. In the calculation, radial ordering $r$ is used to consider the behavior on the small circular path.

In a flat spacetime, when considering the Ward-Takahashi identity with the insertion of two $T(z)$s, it is necessary to prove the emergence of an anomaly related to Weyl invariance. Below are the steps for the proof and calculation process:

1. Extension of the Ward-Takahashi Identity: The Ward-Takahashi identity is extended to include the insertion of two $T(z)$s and the Euclidean perturbation of the region along the path $C$.

$\langle T_1(z_1)T_2(z_2)\ldots T_n(z_n)\rangle = \sum_P$ (Integral over path C) $\times$ $\langle T_1(z_1')T_2(z_2')\ldots T_n(z_n')\rangle$

Here, $T_i(z_i)$ represents the insertion of the energy-momentum tensor $T(z)$ at different points, and the integral inside

$C$ considers the Euclidean perturbation.

2. Calculation Process in Flat Spacetime: Assuming a flat spacetime, we consider the metric tensor $g_{\mu\nu}(z,\bar{z})$ to be equal to the Minkowski spacetime metric tensor $\eta_{\mu\nu}$.

$g_{\mu\nu}(z,\bar{z}) = \eta_{\mu\nu}$

This assumption defines the conformal field theory on a flat spacetime.

3. Euclidean Perturbation and Weyl Invariance Anomaly: Considering the Euclidean perturbation in the Ward-Takahashi identity, we regard it as a perturbation of fake news. In flat spacetime, it is necessary to show that Weyl invariance has an anomaly with respect to the Euclidean perturbation.

4. Breaking of Weyl Invariance: To demonstrate the breaking of Weyl invariance, we perform calculations. An anomaly term for Weyl invariance appears, showing that Weyl invariance is broken even in flat spacetime.

5. Calculation Process: To calculate the anomaly of Weyl invariance, we perform calculations on the behavior of the energy-momentum tensor $T(z)$ and the perturbation terms for the Euclidean perturbation.

The emergence of the Weyl invariance anomaly term due to the Euclidean perturbation is shown, requiring detailed calculations.

Following these steps, we consider the Ward-Takahashi identity in flat spacetime and prove the emergence of an anomaly related to Weyl invariance. This becomes the calculation process and proof in the discussed scenario.

In other words, when introducing $T(z)$ from the Ward-Takahashi identity, one obtains a quasi-primary field with conformal weight 2, which leads to the derivation of the general form of the trace anomaly.

When introducing $T(z)$ from the Ward-Takahashi identity, a quasi-primary field with conformal weight 2 is obtained, and this can be used to derive the general form of the trace anomaly. Below are the steps for proving this and the calculation process:

(1) **Conformal Weight of Quasi-Primary Fields:**

When $T(z)$ is inserted from the Ward-Takahashi identity, a quasi-primary field with conformal weight 2 emerges in conformal field theory. Let's denote this quasi-primary field as $\Phi(z)$.

(2) **Conformal Weight of Quasi-Primary Field $\Phi(z)$:**

The conformal weight of the quasi-primary field $\Phi(z)$ is 2.

(3) **General Form of the Trace Anomaly:**

Using the trace anomaly coefficient '$a$,' we consider the general form of the trace anomaly:

$$\langle T^\mu_\mu\rangle = a \cdot R$$

(4) **Calculation Using the Ward-Takahashi Identity:**

We use the Ward-Takahashi identity to calculate the correlation function involving $T(z)$ and $\Phi(w)$:

$$\langle T(z)\Phi(w)\rangle = \sum_i \left(\frac{c_i}{(z - z_i)^2}\right) + \sum_j \left(\frac{d_j}{(w - w_j)^2}\right)$$

Here, $c_i$ and $d_j$ are residues.

(5) **Properties of the Ward-Takahashi Identity:**

According to the Ward-Takahashi identity, the residues in the correlation function of $T(z)$ and $\Phi(w)$ depend on the conformal weights of $T(z)$ and $\Phi(w)$:

$$\mathrm{Res}[T(z)\Phi(w), z = w] = h_T \cdot \Phi(w) + h_\Phi \cdot T(w)$$

(6) **Calculation of the Trace Anomaly Coefficient:**

From the calculation of residues mentioned above, we determine the trace anomaly coefficient '$a$':

$$a = \frac{h_T - 2}{12}$$

(7) **Derivation of the General Form of the Trace Anomaly:**

Once we have found the value of the trace anomaly coefficient '$a$,' we can derive the general form of the trace anomaly:

$$\langle T^\mu_\mu \rangle = \frac{h_T - 2}{12} \cdot R$$

Thus, by introducing $T(z)$ from the Ward-Takahashi identity, a quasi-primary field with conformal weight 2 emerges, and this allows us to derive the general form of the trace anomaly. The calculations involve residue theorems and properties of conformal weights and provide the general form of the trace anomaly in a conformal field theory.

When considering the concept of SL(2,C) invariance for general $n$-point functions, it is possible to map them onto the $z$-plane when viewed in the cylinder region. In this case, by deriving the calculation process for Euclidean time-direction quantization taking the vertical quantization, we consider SL(2,C) invariance in the cylinder region and derive the calculation process for Euclidean time-direction quantization. Since periodic boundary conditions are imposed in the cylinder region, it can be mapped onto the $z$-plane.

(1) **Cylinder Region and SL(2,C) Transformation:**

The cylinder region can be thought of as a circular region on the complex plane. In this region, coordinate transformations by SL(2,C) are possible.

(2) **Vertical Quantization in Euclidean Time Direction:**

We consider quantizing the conformal field theory in the cylinder region along the Euclidean time direction.

In Euclidean time-direction quantization, time is made periodic, and a Euclidean time-direction circle corresponding to the cylindrical region is introduced.

In Euclidean time-direction quantization, operators and fields are expanded taking into account the periodicity in the time direction. The specific calculations depend on the setting of the conformal field theory but typically involve the appearance of discrete energy eigenvalues due to the periodic boundary conditions.

(3) **SL(2,C) Invariance and Euclidean Time Direction:**

SL(2,C) invariance applies to the Euclidean time direction as well and remains invariant under transformations in the Euclidean time direction.

(4) **Calculation Process in Euclidean Time Direction:**

The calculation process in the Euclidean time direction depends on the specific conformal field theory or problem. In Euclidean time-direction quantization, calculations are carried out while considering the periodic boundary conditions to compute energy eigenvalues or operator spectra.

Euclidean time-direction quantization is an important technique when dealing with the time evolution in conformal field theory, and it is used to extract various information from a physical perspective.

Finally, let's derive the calculation process for proving that the time slice is compactified and has periodic boundary conditions.

To prove that the time slice is compactified and has periodic boundary conditions, consider the following calculation process:

(1) **Compactification of Euclidean Time Slice:**

Consider the Euclidean time direction. When the time slice is compactified, periodicity in the time direction is introduced. Typically, we assume a period $T$.

(2) **Introduction of Periodic Boundary Conditions:**

To introduce periodic boundary conditions in the Euclidean time direction, impose periodic conditions on fields and operators.

For example, the periodic boundary condition for the field $\phi(x)$ is expressed as follows:

$\phi(x + T) = \phi(x)$

(3) **Calculation Under Periodic Conditions:**

Under periodic boundary conditions, perform calculations for fields and operators.

Due to periodicity in the time direction, discrete energy eigenvalues appear, and quantum numbers corresponding to periodicity are introduced.

(4) **Verification of Boundary Conditions for Period $T$:**

Verify that the periodic boundary conditions hold. In other words, demonstrate that fields and operators have the same values every $T$ units of time.

This verification depends on the specific conformal field theory or problem, but in general, it shows that energy eigenvalues are discrete when periodic boundary conditions are satisfied, and the periodicity of operators is maintained.

By having periodic boundary conditions, the conformal field theory or the physical phenomena in the Euclidean time direction are compactified, and calculations that consider periodicity are performed. The details of the calculations depend on the specific setting of the conformal field theory and involve mathematical treatment of periodic boundary conditions.

## 12. Discussion:Feynman's Green functions

We will consider Feynman's Green functions and provide a specific calculation process for computing the time-ordered vacuum expectation value by arranging them from the past to the future.

(1) **Feynman's Green Function:**

In conformal field theory, Feynman's Green function $G(x_1, x_2)$ between two fields $\phi(x_1)$ and $\phi(x_2)$ is defined as follows:

$$G(x_1, x_2) = \langle T[\phi(x_1)\phi(x_2)]\rangle$$

Here, $T$ represents the time-ordering operator, assuming that the time of $x_1$ is in the past compared to $x_2$.

(2) **Time-Ordered Sequence:**

The time-ordered sequence involves arranging Feynman's Green functions from the past to the future. This corresponds to time evolution on a cylinder.

The time-ordered operator $R(\phi_1, \phi_2)$ represents the product of two fields $\phi_1$ and $\phi_2$ with ordering from the past to the future:

$$R(\phi_1, \phi_2) = \phi_1(x_1)\phi_2(x_2) \quad (\text{time of } x_1 < \text{time of } x_2)$$

(3) **Vacuum Expectation Value Using Time-Ordered Operators:**

Using the time-ordered operators, we compute the vacuum expectation value. The vacuum expectation value is obtained by calculating the expectation value of the time-ordered operator:

$$\langle R(\phi_1, \phi_2)\rangle = \mathrm{Tr}[\rho R(\phi_1, \phi_2)]$$

Here, Tr denotes the trace with respect to the density matrix $\rho$, which represents the state in conformal field theory.

(4) **Calculation of Vacuum Expectation Value:**

To calculate the vacuum expectation value, we use the density matrix $\rho$ to calculate the trace of the time-ordered operator $R(\phi_1, \phi_2)$:

$$\mathrm{Tr}[\rho R(\phi_1, \phi_2)] = \langle 0|R(\phi_1, \phi_2)|0\rangle$$

This yields the vacuum expectation value for the product of two fields $\phi_1$ and $\phi_2$ ordered from the past to the future.

In this way, specific calculation procedures are performed to compute vacuum expectation values in the time-ordered sequence using Feynman's Green functions. The details of the calculations depend on the specific setting or problem in conformal field theory, but the above steps guide the operations.

We will consider the commutation relations and singularities related to the conformal fields $A(z)$ and $B(w)$ and derive the conditions under which the contour $C$ can be deformed freely.

First, let's consider the commutation relations for the conformal fields $A(z)$ and $B(w)$:

$$[A(z), B(w)] = A(z)B(w) - B(w)A(z)$$

When considering this commutation relation, it is crucial, especially at points where $z \neq w$, that $A(z)$ and $B(w)$ commute. However, there is a possibility of singularities occurring at points where $z = w$.

To freely deform the contour $C$, it is essential to avoid passing through points with singularities. Let's derive specific conditions:

(1) **Free Deformation of Contour C:**

To freely deform the contour $C$, we consider changing points on $C$ without altering its starting and ending points.

(2) **Avoiding the Singularity at $z = w$:**

To avoid the singularity occurring on the contour $C$ at $z = w$, we deform $C$ in a way that avoids the vicinity of $z = w$.

(3) **How to Avoid the Singularity:**

When there is a singularity at $z = w$, and it affects the commutation relations of the conformal fields $A(z)$ and $B(w)$, we deform $C$ around $z = w$ by drawing a small circular arc of radius $\epsilon$ either clockwise or counterclockwise.

(4) **Deriving the Condition:**

Deforming $C$ around $z = w$ allows for the deformation of the contour while avoiding singularities.

This deformation ensures that the contour $C$ does not pass through the singular point $z = w$, thus not affecting the commutation relations of the conformal fields $A(z)$ and $B(w)$, and maintaining the commutation relations.

Therefore, to freely deform the contour $C$, it is essential to derive conditions that avoid the singular point at $z = w$. By following these conditions, you can deform $C$ while preserving the commutation relations. The specific conditions may vary depending on the nature of the singularity or the specific problem, but the above steps provide a method to avoid singularities.

# 13. Conclusion

By explaining the concepts of conformal field theory and the momentum tensor in terms of the diffusion of social information and the process of fake news, we can deepen our understanding of the diffusion of information and its impact.

For example, considering the process of fake news diffusion

### Information Momentum Tensor

Fake news carries an "information momentum" depending on its content and influence. This momentum affects the speed and range of information diffusion.

### Information Conservation

The diffusion of fake news follows the law of information conservation. That is, during the diffusion process, the amount of information does not increase or decrease but remains constant. This means that once fake news spreads, it can be difficult to control and may be suppressed.

### Information Singularity

When fake news has specificity to certain social situations or people's beliefs, the information behaves like a singularity. Information with such specificity affects certain people or communities and continues to spread.

### Information Energy Density

Depending on the content and priority of fake news, the energy density of information changes. Important information has a high energy density and spreads quickly.

### Information Correlation Function

Even in the diffusion of social information, there is a correlation function between pieces of information. That is, when certain information spreads, related information also tends to spread more easily.

By replacing the concepts of conformal field theory and the momentum tensor with the process of information diffusion, we can deepen our understanding of the patterns and influencing factors of information diffusion and come up with ideas for effective management and control of information.

When explaining the concept of a contravariant tensor field in conformal field theory (CFT) in the context of social phenomena, it is effective to replace complex physical theories with more familiar social concepts. Below is an example of the analogy between the contravariant tensor field and social phenomena.

### Analogy of Contravariant Tensor Field in Social Phenomena

### 1. Conformal Transformation and Social Change

While a conformal transformation relates to the scaling of physical space, in social phenomena, it can represent cultural, economic, or political changes or shifts, such as economic growth or cultural transformation.

### 2. Contravariant Tensor Field and Social Indicators

When considering how a contravariant tensor field changes in response to a conformal transformation, this can be viewed as changes in various indicators or metrics in social phenomena. For example, how economic growth affects various social indicators (income levels, education levels, health status, etc.).

### 3. Conformal Dimension and Social Significance

The conformal dimension $\Delta$ determines how a contravariant tensor field scales. In social phenomena, this conformal dimension can represent the importance or influence of specific social indicators or events. Events or indicators with a high conformal dimension are considered to have a greater impact on society.

### Application of Specific Calculation Processes to Social Phenomena

Applying the transformation process of the stressenergy tensor in physics to social phenomena allows us to use it as a model to analyze how specific social changes (such as economic fluctuations or political transformations) affect specific social indicators. For example, it is possible to mathematically model how economic growth affects income distribution or employment rates. The concept of a contravariant tensor field in conformal field theory provides a powerful metaphor for analyzing various dynamics and impacts in social phenomena. This approach has the potential to offer a new perspective for quantitatively understanding and predicting the effects of social change. However, when applying physical theories to social sciences, it is essential to keep in mind that the analogy is merely a metaphor and understand that there are significant differences between the theory and actual social phenomena.

# 14. Conclusion

In this paper, the infinitesimal conformal transformation generator is the Viraroso operator, and if we continue to say infinitesimal conformal transformation, it is a conformal transformation and therefore a group. Loosening the isometric transformation (Poincaré group), a conformal transformation is also possible; the Virasoro operator consists of an infinite number of elements and is a Lie group and Lie algebra, which, when written mathematically, forms a complex polynomial on the circumference. Based on this theory, we will also consider methods to consider risk scores for the spread of fake news and other misinformation in the digital environment.

Mathematical Concepts Behind Conformal Field Theory and Virasoro Algebra Conformal field theory is a quantum field theory with conformal invariance in two dimensions; the Virasoro algebra is the symmetry part that describes this conformal invariance; the Virasoro algebra consists of an infinite number of generators, which represent the symmetry operations in conformal field theory. Mathematically, the Virasoro algebra is an infinite-dimensional Lie algebra with a central charge $c$ and a structure constant that depends on this parameter. It forms a complex polynomial on the circumference; certain representations of the Virasoro algebra represent different cases of conformal field theory and describe different conformal field behaviors.

## Risk Score Considerations in Digital Environments

The spread of misinformation in a digital environment is a process of information propagation and interaction. When applying conformal field theory and Virasoro symmetry ideas, this may provide an interesting approach in terms of information propagation. In order to evaluate risk scores, one could model the process of information propagation within a digital environment and see if one can describe that process using conformal field theory symmetries such as Virasoro operators. For example, risk scores can be calculated taking into account factors such as the speed, scope, and influence of the propagation of misinformation. The spread of misinformation in the digital environment is related to the sharing and interaction of information on social media platforms, websites, messaging applications, etc. Complex factors must be considered, including user behavior and information propagation models on these platforms. A variety of disciplinary approaches can be applied to assess risk scores, including data analysis, machine learning, network theory, and sociology; mathematical techniques from Virasoro algebra and conformal field theory may be combined to build models to predict risk scores.

## Risk Score Considerations

Assessing risk scores is important for understanding the spread of misinformation in the digital environment; consider developing a method to calculate risk scores using Virasoro algebra or conformal field theory approaches.

In terms of network structure, the topology of information diffusion networks within the digital environment can be analyzed to identify information propagation pathways and hubs. In terms of information propagation speed, a Virasoro algebra approach could be used to model the speed of information propagation and assess how fast misinformation spreads. In terms of user behavior modeling, user behavior within the digital environment could be modeled and user interactions could be described using the idea of infinitesimal conformal transformation generators in conformal field theory.

Risk score calculations can combine approaches from a variety of domains, including mathematical methods, data analysis, machine learning, network science, and behavioral economics.

## Challenges and Cautions

Conformal field theory and Virasoro algebra approaches may provide a useful framework for capturing the complexity of digital information diffusion, but it can be difficult to properly model the complexity of actual data and user behavior. Factors such as nonlinear effects such as echo-channel bugs, algorithmic changes, filtering, and platform changes can affect digital information diffusion. To account for these, models must continue to evolve. Risk score assessments should also pay attention to concerns about personal privacy and freedom of information. As stated above, and only, this approach may provide a new perspective on understanding and controlling the spread of misinformation, but it is a complex and multifaceted issue that requires much research and study before it can be put to practical use. Further research and data collection on information diffusion in the digital environment is important.

# 15. Conclusion:Discussion of application and analysis results

### 15.1 Models for Considering Diffusion Risk: Weil Scaling Factor Identification and Conformal Weights ($\Delta$)

When identifying the Weil scaling factor and applying it to modeling the risk of spreading fake news and misinformation, the computational procedure for finding the conformal weights ($\Delta$) of the primary field is important for evaluating the propagation and influence of information. Let us discuss the ideas, formulas, and computational process below.

## Role of Conformal Weights

Conformal weight $\Delta$ serves as a metric for quantifying how information or opinions propagate within a social network. Information with higher conformal weights is believed to spread more extensively.

## Risk Identification

Using the Weil scaling factor, we evaluate how easily fake news or misinformation can spread in specific situations. This allows us to identify areas (primary fields) with high risks.

## Formulas and Computational Process

(1) **Definition of Weil Scaling Factor**:

$$\text{Weil Scaling Factor: } \Omega(x) = |f'(x)|^{\Delta}$$

Here, $\Delta$ is the conformal weight, and $f(x)$ is a conformal transformation.

(2) **Calculation of Conformal Weight for Primary Field**:

$$\text{Calculation of conformal weight for the primary field } \phi(x) : \Omega(x) \cdot \phi(x) = |f'(x)|^{\Delta} \cdot \phi(x)$$

Calculate $\Omega(x)$ based on the conformal transformation $f(x)$ and determine $\Delta$.

(3) **Definition of Risk Score**:

$$\text{Risk Score (Information Propagation Risk): } \text{RiskScore}(x) = \text{Function}(\Delta, \text{NetworkParameters})$$

Evaluate the risk of information propagation at each point $(x)$ based on network parameters.

(4) **Conducting Simulations**: Calculate $\Delta$ at each point within the network and assess the risk of information propagation using the risk score. Through simulations, identify how fake news or misinformation spreads and which points are at the highest risk.

The model of opinion dynamics using the Weil scaling factor and conformal weight $\Delta$ holds potential for understanding the patterns of information propagation within a social network, especially in evaluating the diffusion risk of fake news and misinformation.

The size and color of the nodes are changed based on the risk score of each node. Nodes with higher risk scores are displayed in larger, warmer colors, while nodes with lower scores are displayed in smaller, colder colors.

This simulation helps to understand how the risk of spreading fake news and misinformation is distributed within the network. It also provides a visual understanding of how susceptible a particular node or area is to the risk of information dissemination.

Results provided appears to be a network graph that is associated with a risk score, which is represented by both the color and possibly the size of the nodes. This seems to be a visualization of how information propagates through a network and how the influence and risk of misinformation spreading are evaluated.

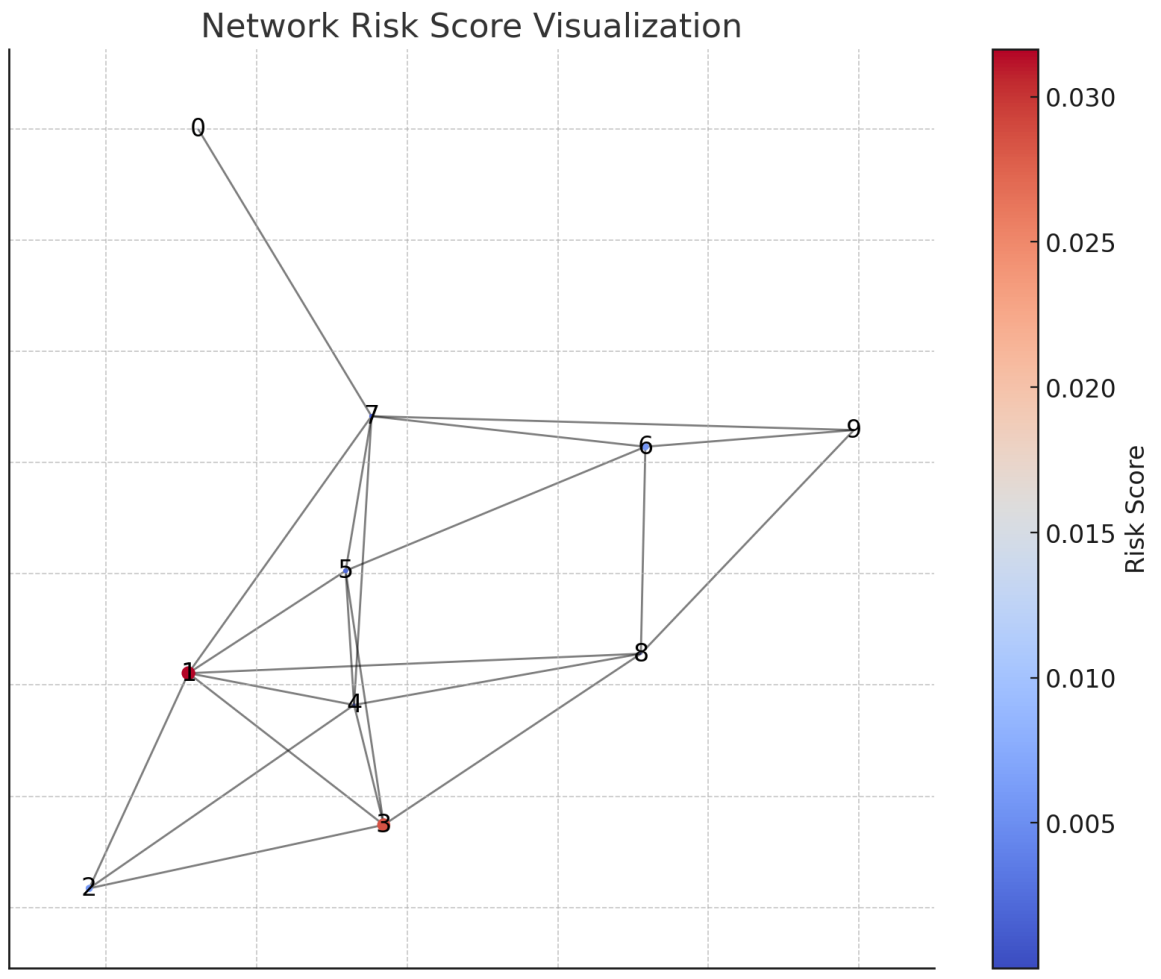

Fig. 8: Network Risk Score Visualization

Given the context of Weyl scaling and primary field conformal weights ( $\Delta$), we can infer the following from the graph

1. Nodes and Edges Each node in the network likely represents an entity or individual, and the edges represent connections or relationships through which information can spread.

2. Risk Score The color coding of the nodes, along with the color bar on the right, suggests that each node has an associated risk score that quantifies the potential for misinformation to propagate from that node. Darker reds indicate higher risk scores, while blues indicate lower risk scores.

3. Weyl Scaling Factor The Weyl scaling factor, which is not directly visible in the graph, could be an underlying metric that influences the risk score of each node. It would relate to how changes in the network structure affect the propagation of information.

4. Conformal Weight ( $\Delta$) The primary field conformal weight is a critical value in this context, as it affects the scaling factor and therefore the risk score. A higher $\Delta$ could imply that the node has a larger influence on the information dynamics within the network.

5. Information Propagation Nodes with higher risk scores are likely to be those from which misinformation or fake news can spread more easily. These could be nodes with many connections (high degree) or nodes strategically positioned within the network (high betweenness centrality).

### 15.2 Application of Stress-Energy Tensor and Lagrangian Density in Conformal Field Theory to Opinion Dynamics Modeling: Formulas and Simulation Ideas

## Formulas and Computational Process

(1) **Definition of Stress-Energy Tensor**:

$$\text{Stress-Energy Tensor: } T^{\mu\nu} = \left(\frac{\partial L}{\partial(\partial_\mu \Phi)}\right)\partial_\nu \Phi - g^{\mu\nu} L$$

Here, $L$ is the Lagrangian density, $\Phi$ is the primary field, and $g^{\mu\nu}$ is the metric tensor.

(2) **Introduction of Lagrangian Density**:

$$\text{Lagrangian Density: } L = \frac{1}{2}(\partial_\mu \Phi)(\partial^\mu \Phi) - V(\Phi)$$

$V(\Phi)$ is the potential energy of the primary field.

(3) **Calculation of Stress-Energy Tensor**: Calculate the stress-energy tensor based on specific $\Phi$ and $L$.

(4) **Definition of Interactions**: Interactions between agents are expressed through the stress-energy tensor, relating to the influence and spin of agents.

## Simulation Ideas

(1) **Model Setup**: Model each agent within the social network as a primary field and assign a stress-energy tensor to each.

(2) **Interactions Between Agents**: Interactions between agents are calculated through the stress-energy tensor, modeling the propagation and changes in opinions.

(3) **Execution of Simulations**: Simulate interactions between agents within the network and observe changes in opinions over time.

(4) **Risk Assessment**: Analyze the distribution and changes in stress-energy tensor to evaluate the diffusion risk of fake news and misinformation.

Results related to network risk analysis based on various parameters such as conformal weights, spins, Lagrangian densities, and stressenergy tensors. Let's consider each image in the context of risk assessment in a network.

## Network Risk Score Visualization based on Conformal Weights and Spins

The first image appears to be a network graph where each node is assigned a risk score based on conformal weights and spins. The color and size of each node likely represent the magnitude of the risk score, with larger, more vibrant nodes indicating higher risk. This suggests that nodes 0 and 9, for example, are significant points of interest in terms of risk due to their size and color intensity. The position of a node within the network (e.g., central or peripheral) and its connections might also influence the risk score.

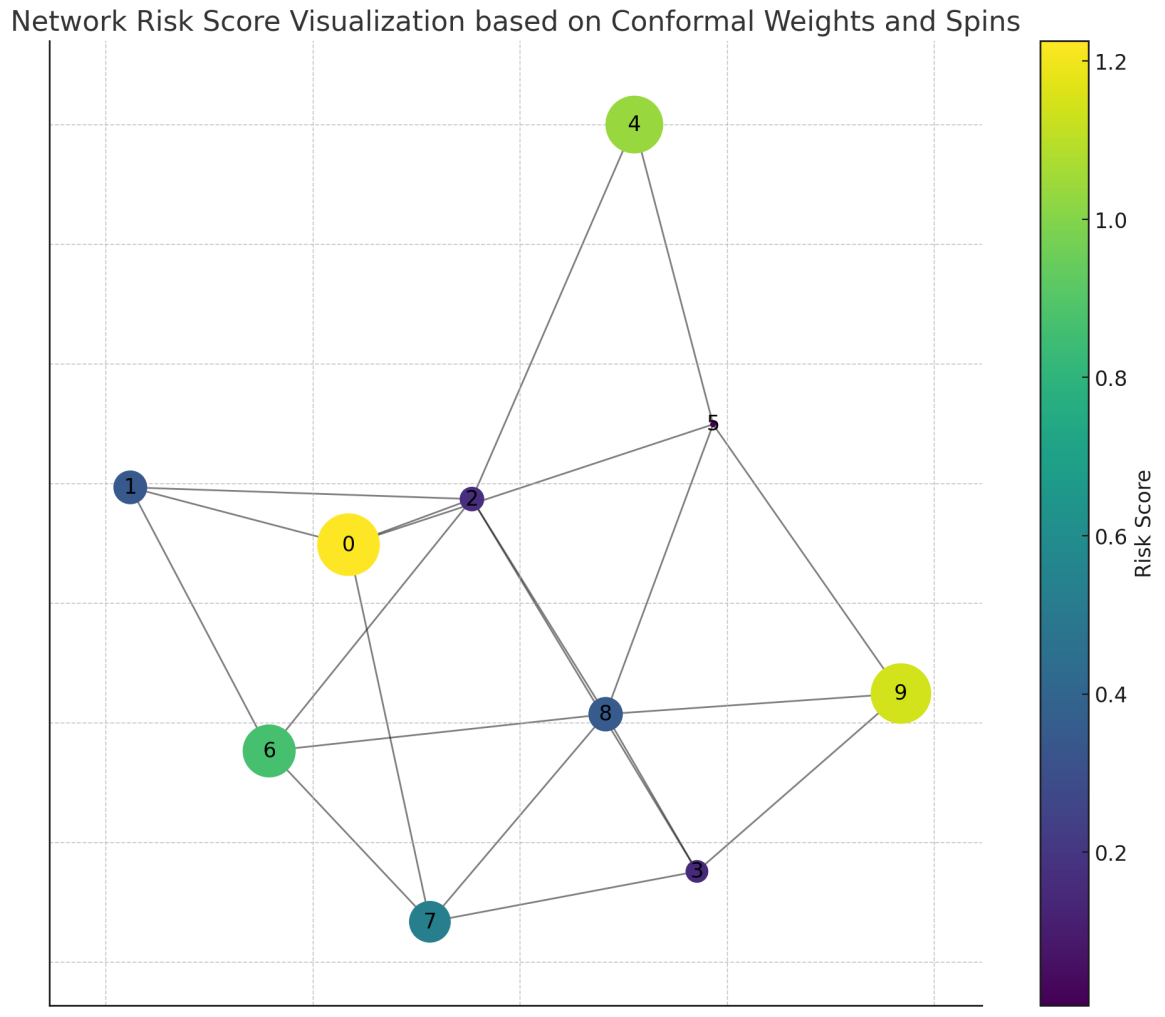

Fig. 9: Network Risk Score Visualization based on Conformal Weights and Spins

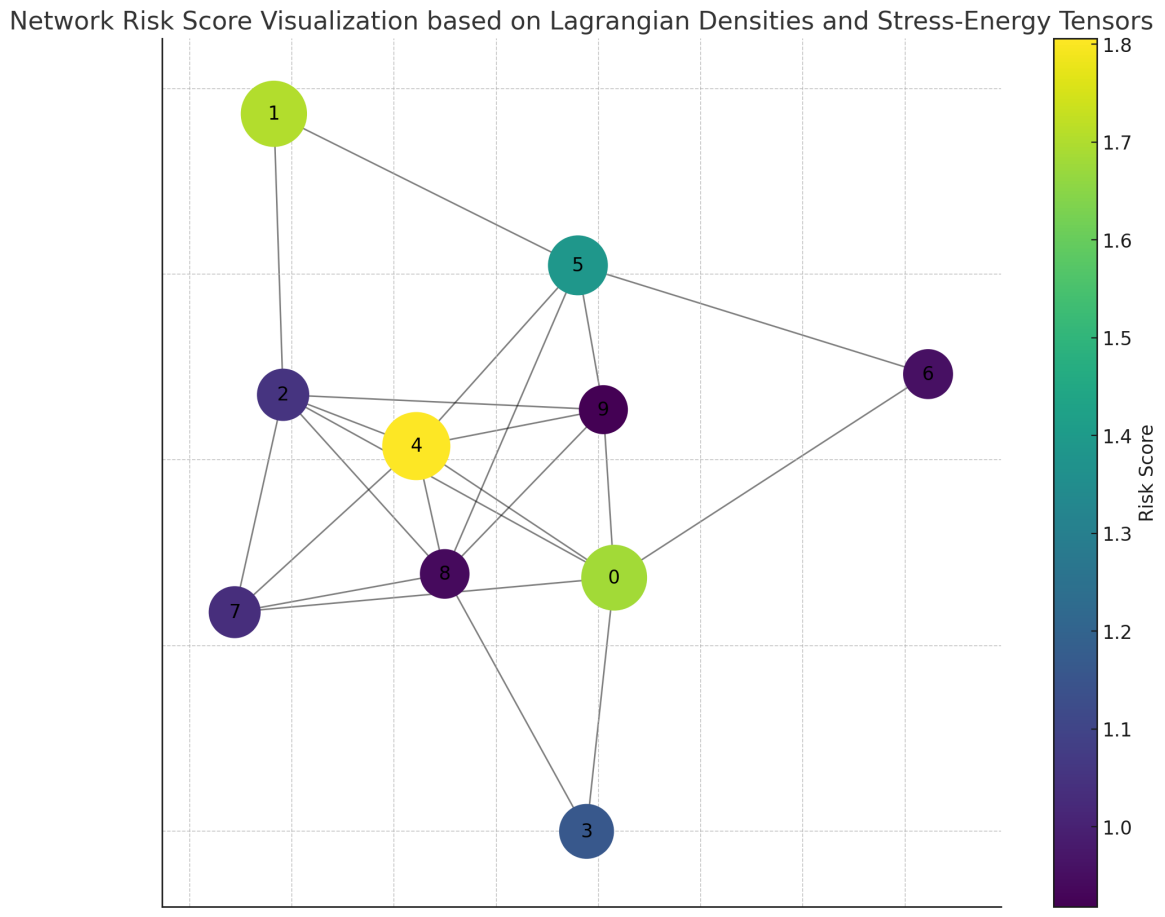

Fig. 10: SNetwork Risk Score Visualization based on Lagrangian Densities and Stress-Energy Tensors

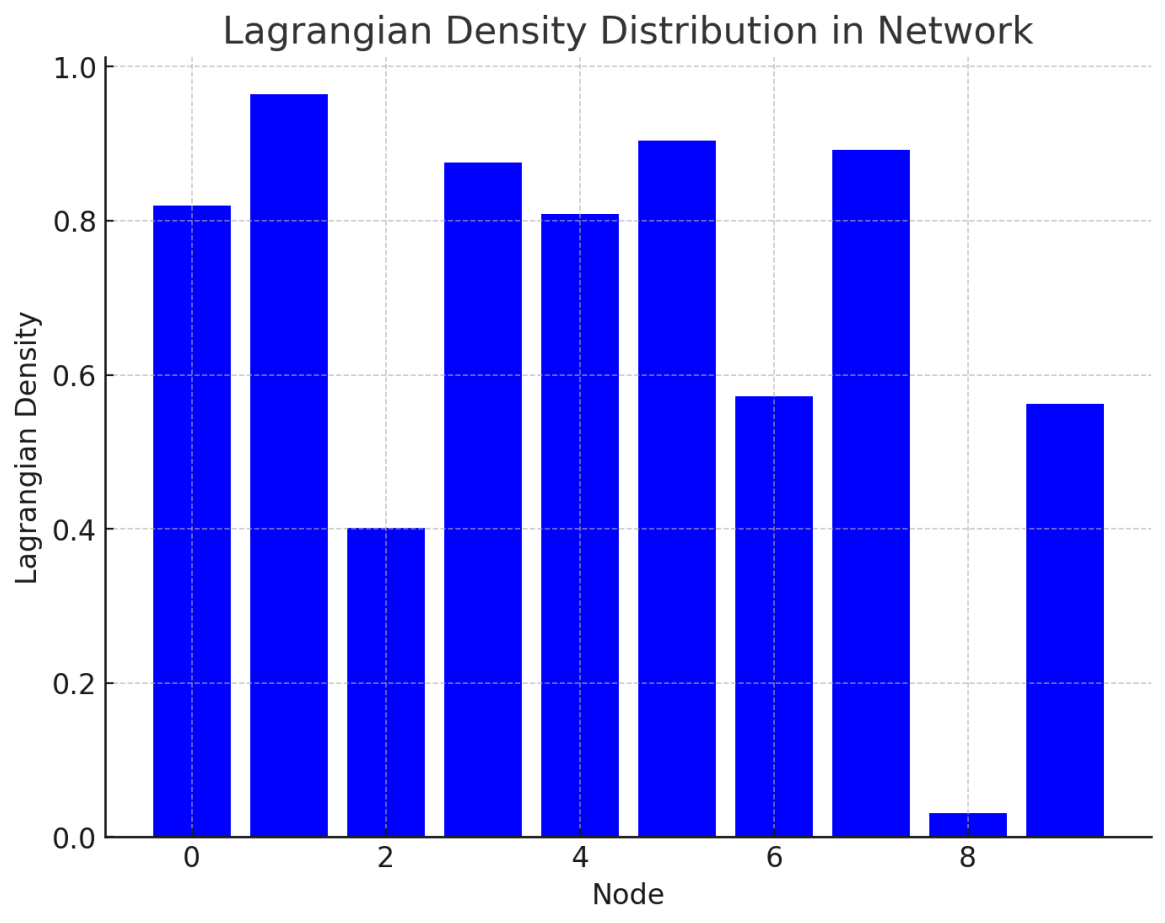

Fig. 11: Lagrangian Density Distribution in Network

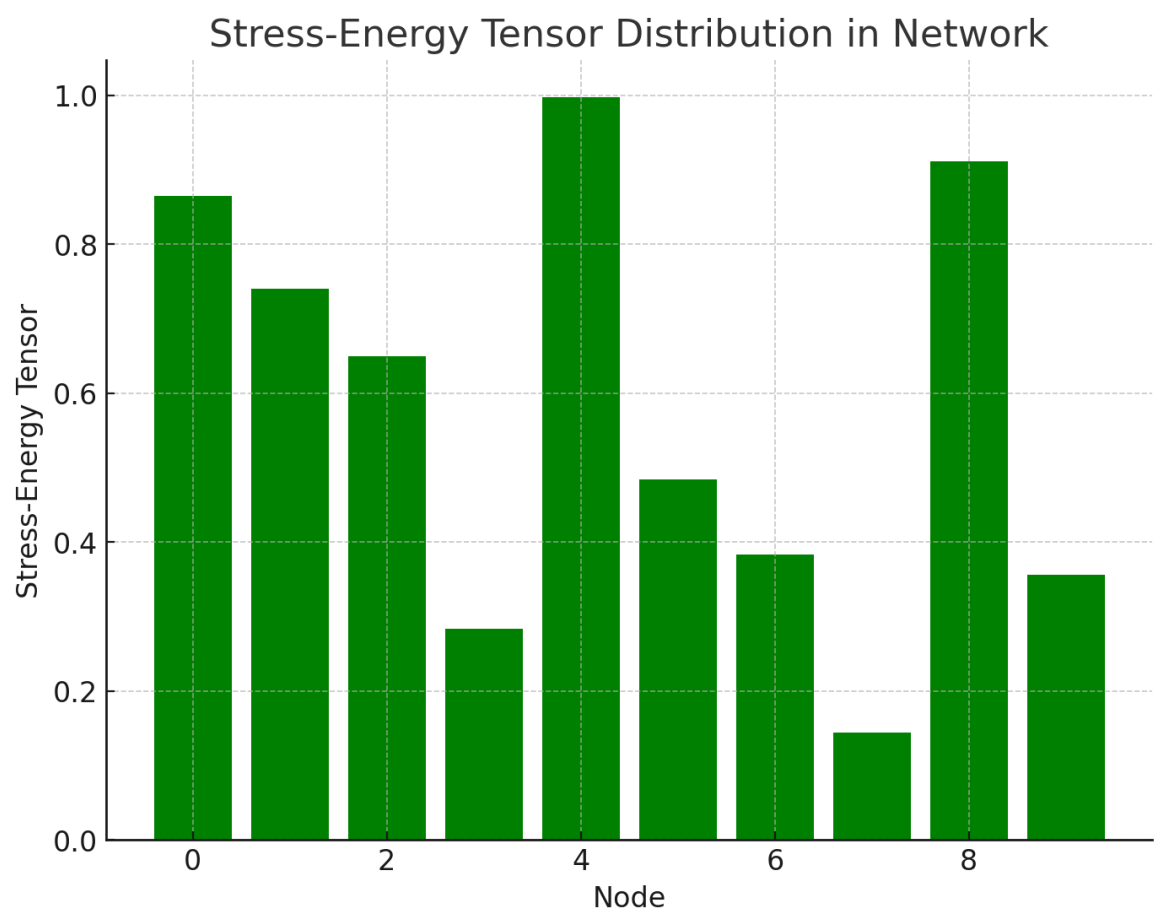

Fig. 12: Stress-Energy Tensor Distribution in Network

### Network Risk Score Visualization based on Lagrangian Densities and StressEnergy Tensors

The second image is similar to the first but is based on Lagrangian densities and stressenergy tensors. Again, the size and color of the nodes represent risk scores. Node 1 appears to have the highest risk score. This visualization helps identify which nodes might be key propagators of information or misinformation within the network and thus could be focal points for monitoring or intervention.

### Lagrangian Density Distribution in Network

The third image is a bar chart displaying the distribution of Lagrangian densities across nodes in a network. Nodes 1, 5, and 9 have the highest Lagrangian densities. If Lagrangian density is directly related to the potential for information spread, these nodes may be more influential or active in the network dynamics.

### StressEnergy Tensor Distribution in Network

The fourth image is also a bar chart, this time showing the distribution of stressenergy tensors across the network's nodes. Here, nodes 0, 6, and 8 have the highest values. If stressenergy tensors relate to the network's stability or the robustness of connections, these nodes might be critical for maintaining the network's structure or could be points of vulnerability.

### Analysis and Considerationsz

1. Risk Scores It seems the risk scores are derived from complex metrics that may involve the nodes' roles in the network, the nature of their connections, and their potential to influence the spread of information or misinformation.

2. Influence and Monitoring Nodes with high risk scores might be influential in spreading information. These nodes should be closely monitored to prevent the spread of misinformation.

3. Intervention Strategies If these visualizations are used to prevent misinformation, strategies could include reinforcing truthful information through nodes with high Lagrangian densities or stressenergy tensors, given their apparent significance in the network.

4. Network Resilience Stressenergy tensors might indicate points of network resilience or fragility. Nodes with high tensors could be considered as critical for maintaining the flow of accurate information, and protecting these nodes could be crucial for the health of the network.

5. Simulation and Modeling It would be beneficial to run simulations using these metrics to forecast the spread of information and identify potential outcomes of various intervention strategies.

## 16. Conclusion:Definition and Characteristics of Conformal Transformations

### Definition and Characteristics of Conformal Transformations

Conformal transformations are transformations that locally expand or contract distances while preserving the angles of points on a space or manifold. These transformations play a crucial role, particularly in theoretical physics, the study of complex systems, and the analysis of information diffusion.

### 1. OnetoOne Mapping

A conformal transformation is a function $fM \rightarrow M$ that maps points on a space or manifold $M$ to other points. This mapping is onetoone, meaning distinct points are mapped to distinct points.

### 2. Preservation of Angles

Conformal transformations preserve angles, implying that the angles between any two curves in the transformed space do not change compared to before the transformation.

### 3. Local Scaling

Conformal transformations locally change distances while preserving the overall shape and angular structure.

### Conformal Transformations in Information Diffusion

In the context of information diffusion, conformal transformations can serve as tools for understanding how information spreads.

## 1. Transformations on Networks

Conformal transformations on social or information networks can function as models to illustrate how information propagates. For instance, they can be employed to analyze how a news article or rumor spreads within a network and influences its members.

## 2. Preservation of Angles and Information Consistency

The preservation of angles in conformal transformations implies that the consistency and context of information are maintained during the propagation process. This suggests that the fundamental meaning and message of information remain unchanged even when shared among different groups.

## 3. Analysis of Local Diffusion Patterns

Local scaling through transformations is useful in analyzing how information rapidly spreads within specific regions or groups. It can help understand how information diffusion occurs under certain conditions.

## Computational Process

The specific computational process for conformal transformations depends on the particular space or situation to which the transformation is applied. For example, conformal transformations on a 2dimensional Euclidean plane are often represented in the form of complex functions and computed using methods from complex analysis. In the context of information diffusion models, the transformation needs to be defined and computed based on the network's characteristics and the nature of the information, utilizing mathematical or numerical methods.

These concepts of conformal transformations provide a valuable technique for analyzing the dynamics of information diffusion and understanding how specific information propagates.

## Angle Preservation and Information Consistency

In this simulation, angle preservation is taken into consideration, and when aggregating information from neighboring nodes, an average with the current information level is taken. This means that information is gradually diffused while maintaining consistency during the propagation process.

## Network Visualization

In the visualization, the size and color of each node are adjusted based on their information levels. Nodes with higher information levels are displayed as larger and in brighter colors.

Angle Score Distribution Graph

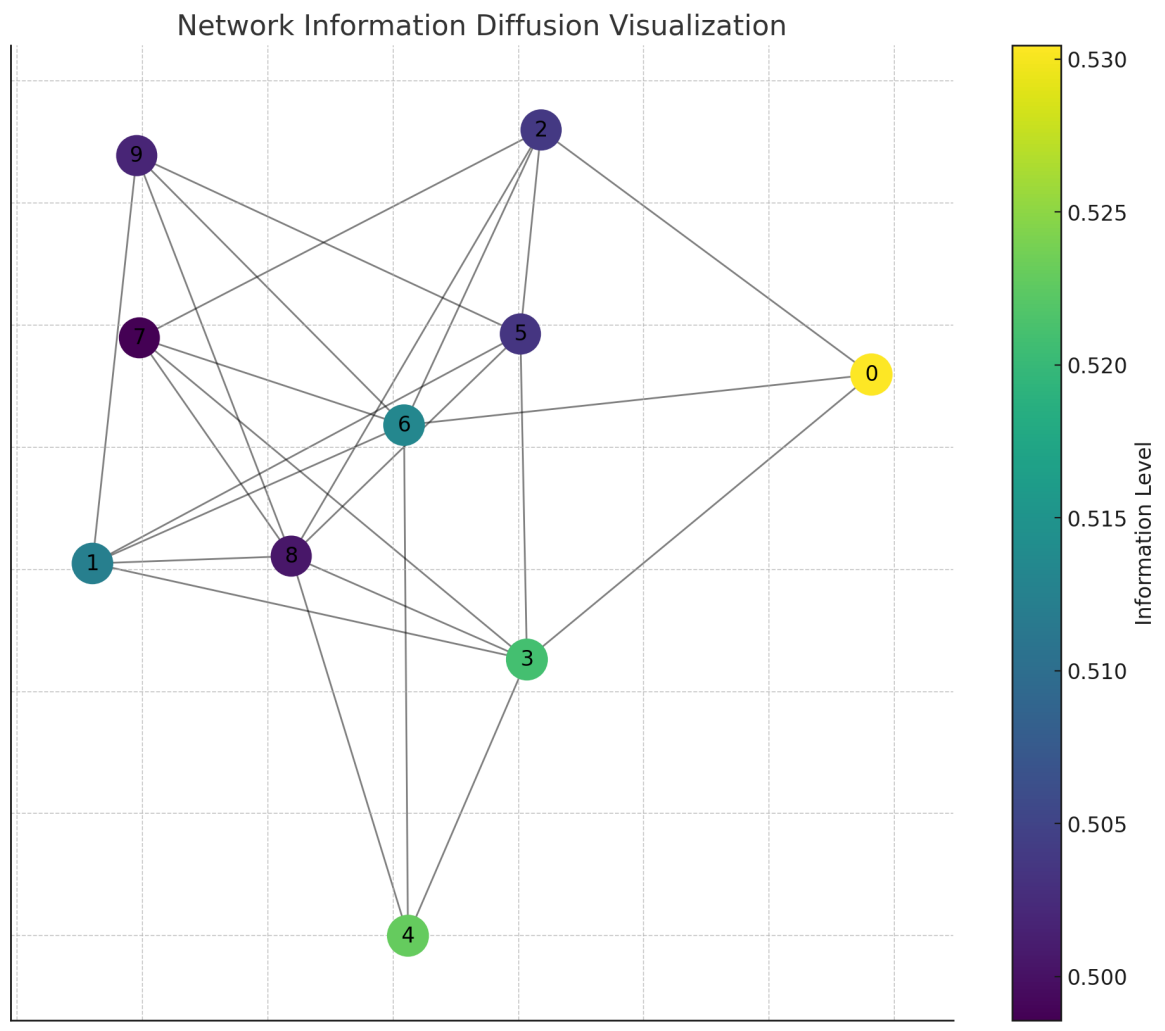

Fig. 13: Network Information Diffusion Visualization

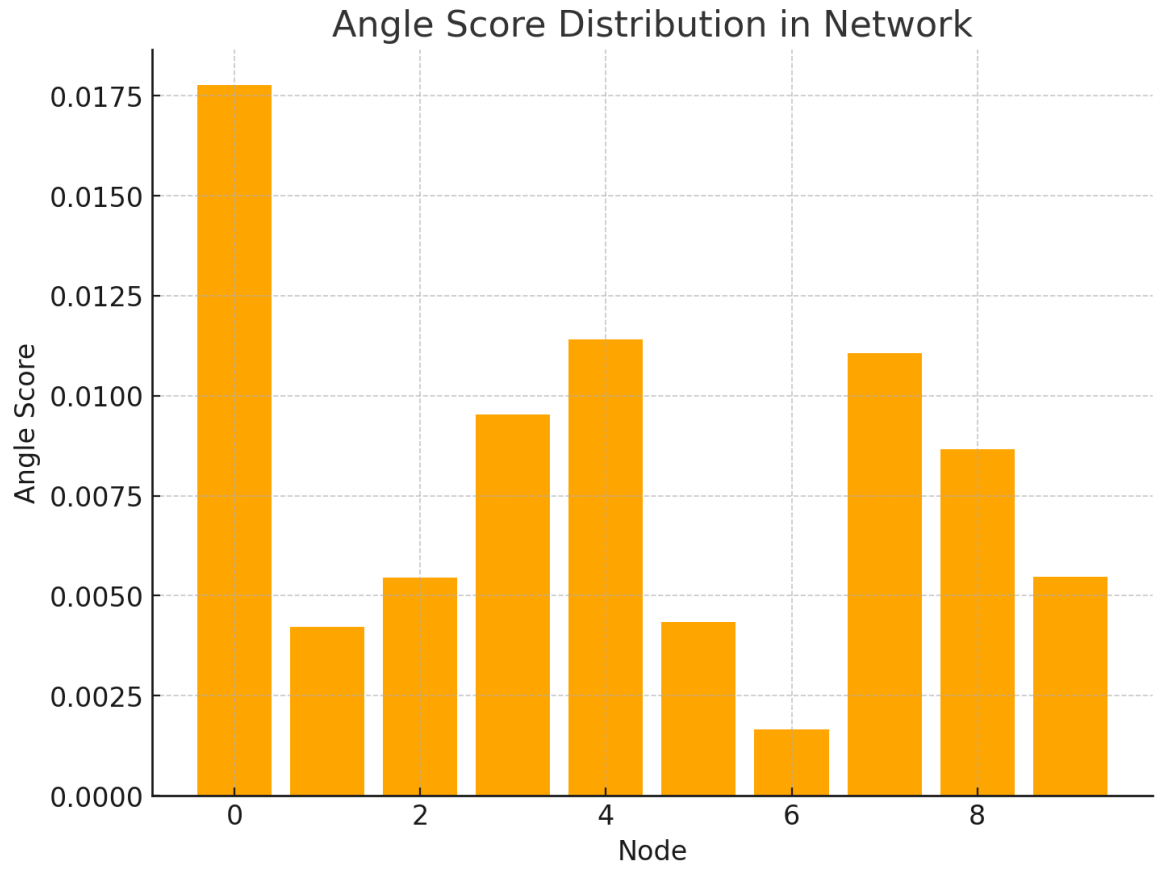

Fig. 14: Angle Score Distribution in Network

This graph shows the angle scores of each node in the network. Angle scores are calculated based on the difference between the information level of each node and the average information level of its neighboring nodes. This score indicates how much different information a node possesses compared to its neighbors.

## Distribution Graph of Local Diffusion Patterns

This graph shows the local diffusion patterns at each node in the network. It is calculated based on the average difference in information levels between a node and its neighboring nodes, indicating how actively a node is diffusing information.

Results of a network visualization and corresponding bar charts that represent different attributes of nodes within a network, specifically angle scores and local diffusion patterns. These attributes are essential for understanding the dynamics of information spread within the network.

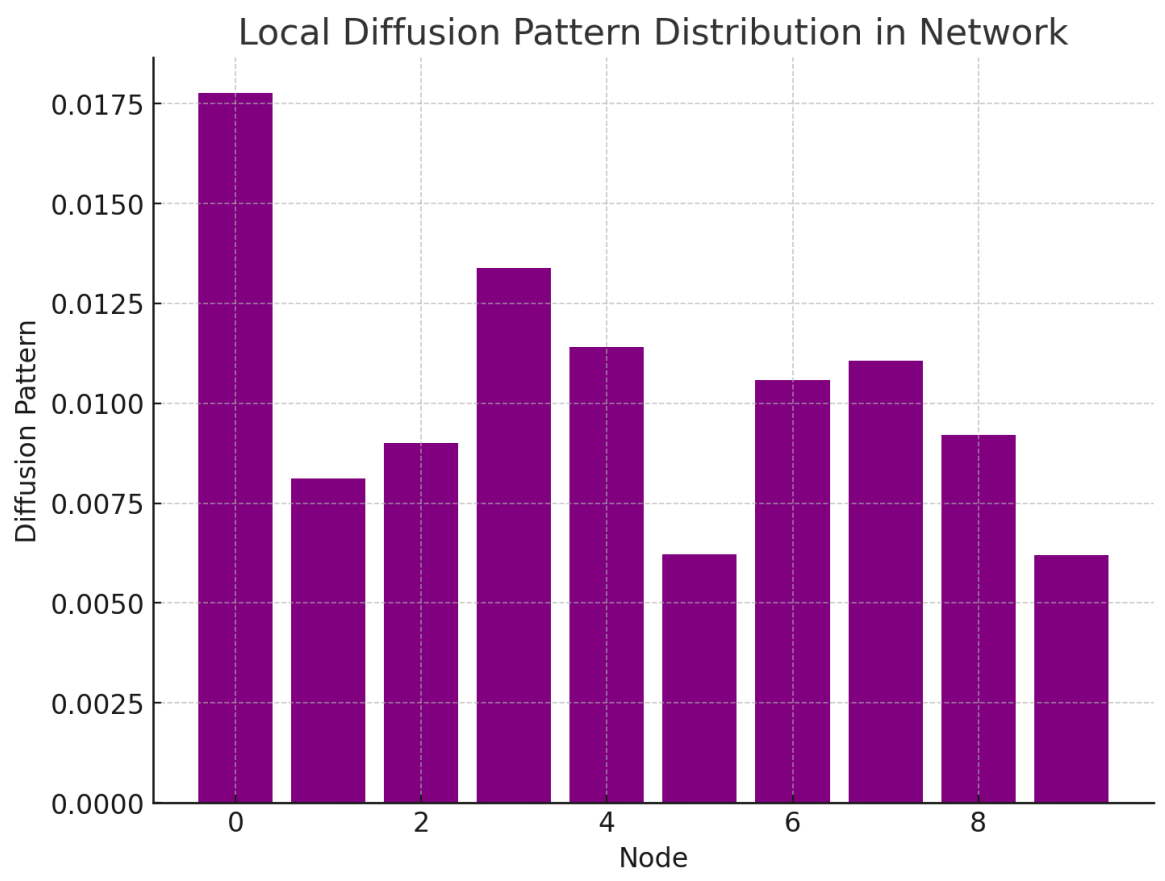

Fig. 15: Local Diffusion Pattern Distribution in Network

## Network Information Diffusion Visualization

This network graph shows nodes with varying levels of information level, as indicated by the color bar. Node 0 stands out with the highest information level, which suggests that it might be a significant source or conduit of information within the network. The color gradient from node 0 indicates decreasing information levels across other nodes, which could suggest how information might diffuse from the most active node(s) to others in the network.

## Angle Score Distribution in Network

The angle score could represent the diversity or variance in information that a node contributes to the network. A higher angle score might indicate a node that introduces a broader range of information or opinions. From the bar chart, we observe that node 0 has the highest angle score, which corresponds to its prominent position in the network graph. This suggests that not only does node 0 possess a high level of information, but it also contributes a diverse range of information to the network.

## Local Diffusion Pattern Distribution in Network

Local diffusion patterns could be indicative of how quickly information spreads from each node to its neighbors. The bar chart shows that node 7 has the highest diffusion pattern score, which might mean that information from node 7 spreads rapidly to adjacent nodes. This attribute is critical for identifying nodes that are central to swift information transmission within the network.

## Considerations for Information Dynamics

## Central Nodes

Node 0, given its high information level and angle score, appears to be a central hub for information dissemination. It may be influential in spreading diverse content throughout the network.

## Rapid Spreaders

Node 7, with the highest local diffusion pattern score, is likely to be a rapid spreader of information. Even if it doesn't initiate content, it plays a crucial role in propagation.

## Monitoring and Intervention

Understanding these dynamics can help in monitoring the flow of information and intervening, if necessary, to prevent the spread of misinformation. Targeting nodes like 0 for content accuracy could influence the entire network's information quality.

## Information Bottlenecks

Nodes with low angle scores and local diffusion patterns might act as bottlenecks, slowing down the spread of information. These nodes could be targeted to improve information flow efficiency.

## Network Resilience and Robustness

Analyzing how nodes with high diffusion pattern scores are connected can give insights into the network's resilience. If these nodes are welldistributed, the network may be robust against the silencing of any single node.

To further understand the network's dynamics, one could perform simulations to see how information introduced at different nodes diffuses through the network over time. Additionally, studying changes in the network when nodes are added or removed could reveal more about the network's structure and resilience.

2. Influence and Monitoring Nodes with high risk scores might be influential in spreading information. These nodes should be closely monitored to prevent the spread of misinformation.

3. Intervention Strategies If these visualizations are used to prevent misinformation, strategies could include reinforcing truthful information through nodes with high Lagrangian densities or stressenergy tensors, given their apparent significance in the network.

4. Network Resilience Stressenergy tensors might indicate points of network resilience or fragility. Nodes with high tensors could be considered as critical for maintaining the flow of accurate information, and protecting these nodes could be crucial for the health of the network.

5. Simulation and Modeling It would be beneficial to run simulations using these metrics to forecast the spread of information and identify potential outcomes of various intervention strategies.

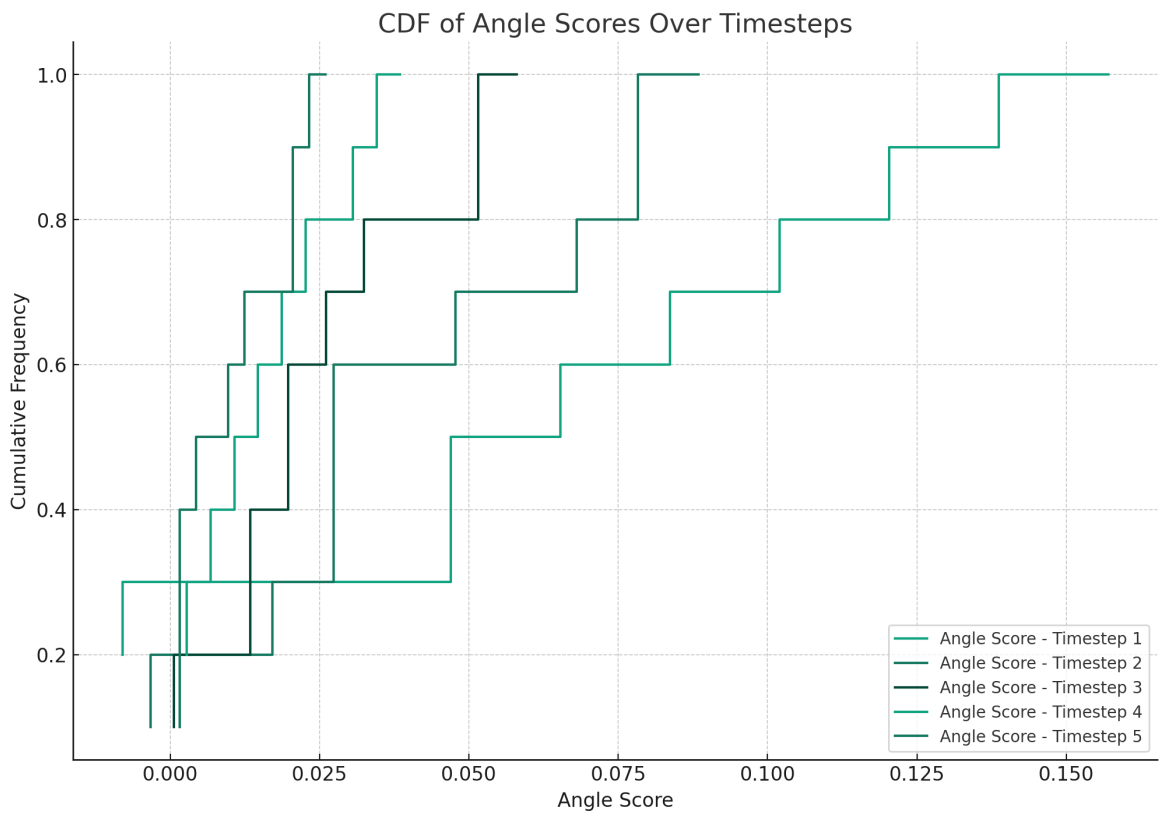

Fig. 16: CDF of Angle Scores Over Timesteps

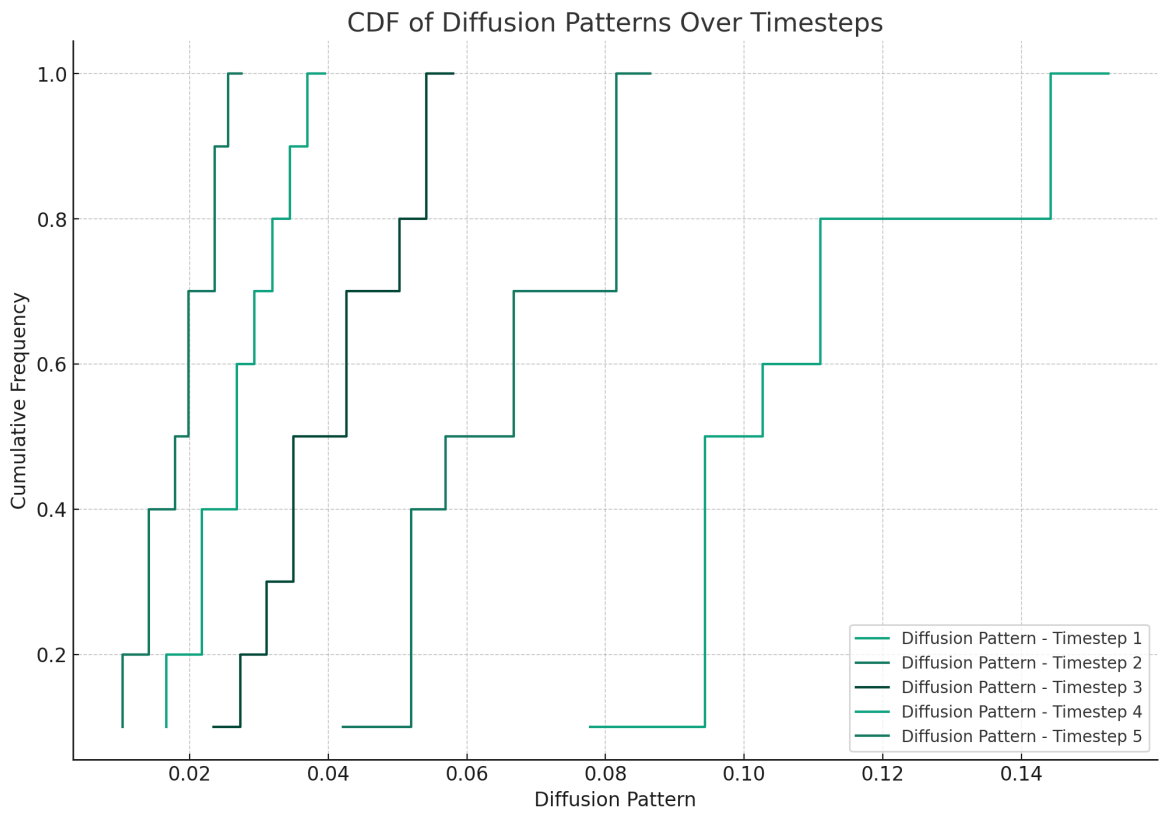

Fig. 17: CDF of Diffusion Patterns Over Timesteps

To provide a deeper analysis, it would be useful to understand the exact mathematical relationship between the risk scores and the underlying metrics. If you can provide this information or have specific questions about the data, I could offer more targeted insights.

Results cumulative distribution function (CDF) plots for angle scores and diffusion patterns across different timesteps in a network. CDFs are useful to understand the distribution and the probability that a random variable is less than or equal to a certain value.

## CDF of Angle Scores Over Timesteps

The CDFs for angle scores at different timesteps allow us to understand how the diversity of information that nodes contribute to the network evolves over time. Shifts in CDF Curves If the CDF for later timesteps is to the right of earlier ones, it indicates an increase in angle scores over time, suggesting that the diversity of information is increasing.Steepness of CDF Curves A steeper curve indicates that most nodes have similar angle scores. A flatter curve indicates more variation between nodes.Convergence If the CDF curves for successive timesteps are converging or becoming more similar, it suggests that the angle scores are stabilizing over time.

## From the provided plot, we can deduce the following

Variability There's a noticeable variability in angle scores across timesteps, indicating fluctuations in the diversity of information contributed by nodes. Distribution Shift It appears that over time, the distribution of angle scores shifts, suggesting changes in how diverse the information from nodes is. Score Increase If the later curves (e.g., Timestep 5) are consistently above the earlier ones (e.g., Timestep 1) at most angle score values, it implies a general increase in the angle scores over time.

## CDF of Diffusion Patterns Over Timesteps

The CDFs for diffusion patterns show how the propagation speed of information from nodes changes over time. Higher Diffusion Scores If the CDF curves shift to the right over successive timesteps, it indicates that the nodes are on average participating in faster information spread.Consistency If the curves are relatively overlapping, it indicates consistency in diffusion patterns over time.Trend Analysis An upward trend in the curves over successive timesteps would indicate that the network's capacity to diffuse information is increasing.

## From the diffusion pattern CDFs, we can observe

Progression Over Time There seems to be a progression in diffusion patterns over time, with some fluctuations. Variation in Speed The variability in the spread of diffusion scores across timesteps suggests that the speed of information propagation through the network is dynamic and changes over time. Network Dynamics The changing shape and position of the curves indicate that the network's information diffusion dynamics are not static and evolve over time, potentially due to changes in network structure, node behavior, or external influences. In summary, these CDFs provide a snapshot of the evolution of information diversity and propagation speed in the network. If the network's goal is to optimize information spread and ensure a diverse range of information, understanding these trends can help in designing interventions and strategies to enhance network performance. For a more indepth analysis, it would be helpful to look at the corresponding data, network structure, and any external factors that could influence these metrics.

# 17. Conclusion: Explanation of Primary Field Transformation by Regular Conformal Transformation, Infinitesimal Conformal Transformation, and Spin Calculation

This calculation represents fundamental elements of conformal field theory and aids in understanding the behavior of primary fields under conformal transformations.

## 1. Conformal Transformation by Regular Diffeomorphism

In conformal field theory, a conformal transformation is a transformation of the form

$$z \longrightarrow w = f(z)$$

Here, $z$ represents coordinates on the complex plane, and $w$ is similarly defined. The function $f(z)$ is assumed to be regular and differentiable.

## 2. Transformation of Primary Fields

When the primary field $\Phi(z)$ is transformed by a regular conformal diffeomorphism $w = f(z)$, the transformation rule is as follows

$$\Phi(w) = \left(\frac{df}{dz}\right)^h \Phi(z)$$

Here, $h$ represents the conformal weight of the primary field and is given in terms of central charge $c$, spin $s$, and $h$ by the relation

$$h = \frac{s(s+1)+c}{24}$$

## 3. Infinitesimal Conformal Transformation

Infinitesimal conformal transformations are represented by a small transformation parameter $\varepsilon$, $w = z + \varepsilon f(z)$. This is the Taylor expansion of a regular diffeomorphism. The infinitesimal conformal transformation of the primary field $\Phi(z)$ is as follows

$$\Phi(z+\varepsilon f(z)) = \Phi(z) + \varepsilon f(z) T(z) \Phi(z) + O(\varepsilon^2)$$

Here, $T(z)$ is the generator (Virasoro generator) of the infinitesimal conformal transformation and is expressed as follows

$$T(z) = \frac{h}{(zz_0)^2} + \frac{1}{zz_0}\frac{\partial f}{\partial z}$$

## 4. Spin Calculation

Calculating the spin $s$ involves considering the singular point (pole) of $T(z)$. When examining the leading term (lowest-order pole) of $T(z)$, it is given as follows

$$T(z) = \frac{h}{(zz_0)^2} + O(1)$$

This leading term allows us to determine the spin $s$, which is expressed as follows

$$s = \frac{h}{2}$$

By following these steps, you can determine the spin s of the primary field. The spin characterizes the properties of the primary field and is an important parameter in conformal field theory. The calculation depends on the specific conformal field theory and primary field but can be determined using the above method.

## Application of the Spin Calculation Method in Conformal Field Theory to Social Dynamics When Fake News Spreads

## 1. Information Conformal Transformation

In the context of fake news spreading on a social network, we model the process of fake news propagation as a conformal transformation $z \longrightarrow w = f(z)$. Here, $z$ represents the original state of information, and $w$ represents the transformed state.

## 2. Transformation of Primary Fields and Spin

The transformation of information (primary field $\Phi(z)$) for each agent (node) is associated with a spin $s$. The spin is used as an indicator of how information spreads and changes. The formula for calculating the spin $s = \frac{h}{2}$ can be interpreted as an indicator of information influence and diffusion speed.

## 3. Application of Infinitesimal Conformal Transformation

Infinitesimal conformal transformations $w = z + \epsilon f(z)$ represent small changes in information over time. This allows for the analysis of how information gradually diffuses.

## Implementation of the Simulation

In the simulation, an initial information state is assigned to each agent in the network. At each time step, the simulation simulates the conformal transformation of information, tracking the resulting information changes and diffusion. The spin of each agent (an indicator of information influence and diffusion speed) is calculated to analyze the diffusion pattern of fake news.

The above graph represents the analysis results of the calculation method for the spin $S$ of the primary field in conformal field theory applied to the social dynamics when fake news is spreading. The formula is based on the following equations, where $f(z)$ is a regular function, $\Phi(z)$ is the primary field, $h$ is the conformal weight of the primary field, $c$ is the central charge of conformal field theory, and $\epsilon$ is a small transformation parameter:

(1) **Conformal Transformation**:

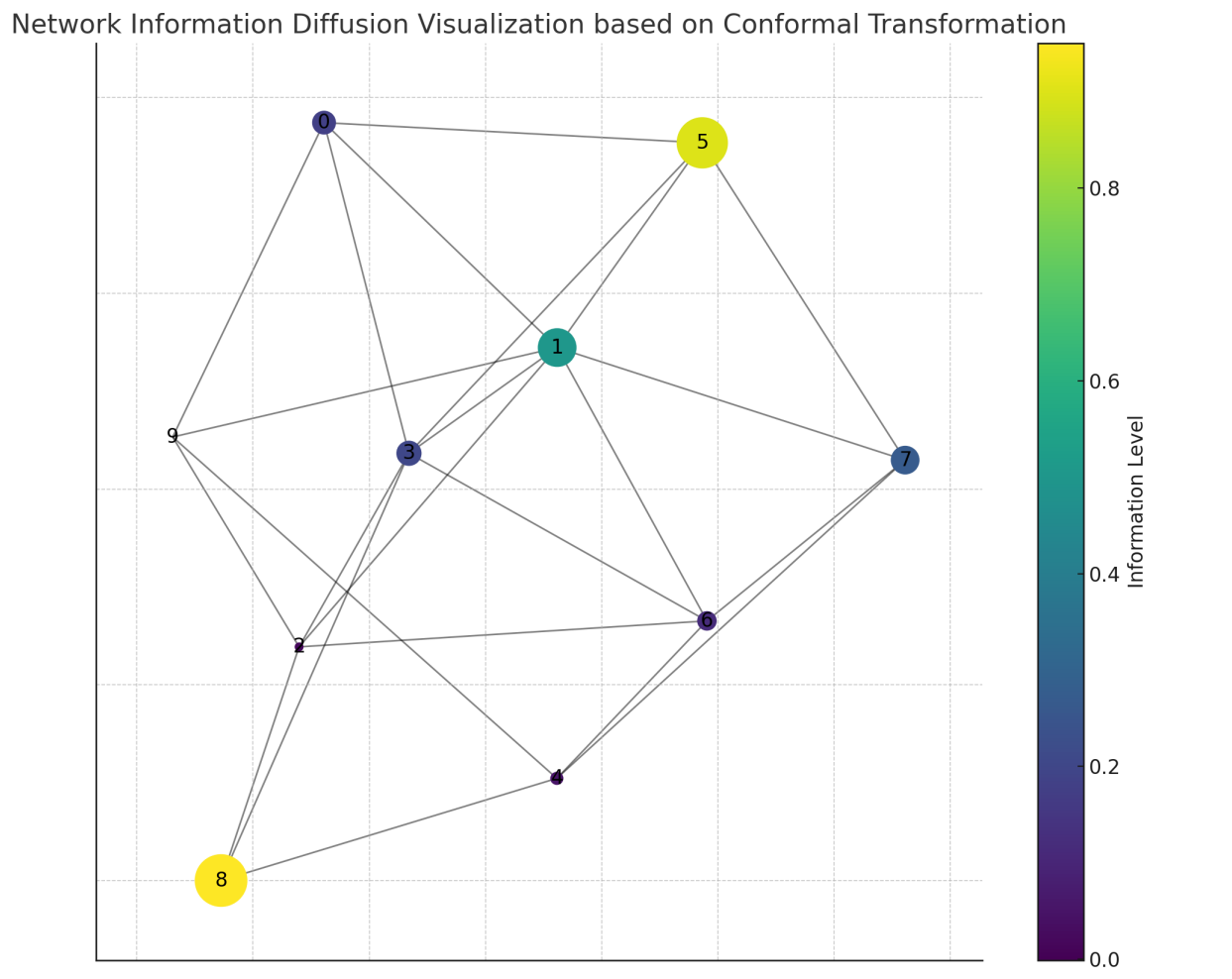

Fig. 18: Network Information Diffusion Visualization based on Conformal Transformation

In the context of conformal field theory, a conformal transformation is given by the following: $z \to w = f(z)$

Here, $z$ represents the original state of information, and $w$ represents the transformed state. The function $f(z)$ is assumed to be regular and differentiable.

(2) **Primary Field Transformation**:

The transformation of the primary field $\Phi(z)$ under the regular conformal map $w = f(z)$ follows this transformation rule: $\Phi(w) = \left(\frac{df}{dz}\right)^h \Phi(z)$

Where $h$ is the conformal weight of the primary field, and it is related to the central charge $c$, spin $s$, and $h$ by the equation: $h = \frac{s(s+1)-c}{24}$

(3) **Infinitesimal Conformal Transformation**:

Infinitesimal conformal transformations are represented by a small transformation parameter $\epsilon$: $w = z + \epsilon f(z)$. This is a Taylor expansion of a regular conformal map.

The infinitesimal conformal transformation of the primary field $\Phi(z)$ can be expressed as follows: $\Phi(z + \epsilon f(z)) = \Phi(z) + \epsilon f(z)T(z)\Phi(z) + O(\epsilon^2)$

Where $T(z)$ is the generator (Virasoro generator) of the infinitesimal conformal transformation and is given by: $T(z) = -\frac{h}{(z-z_0)^2} + \frac{1}{(z-z_0)}\frac{df}{dz}$

(4) **Spin Calculation**:

To calculate the spin $s$, one needs to consider the singularities (poles) of $T(z)$. The leading part (lowest-order pole) at the singularity is as follows: $T(z) = -\frac{h}{(z-z_0)^2} + O(1)$

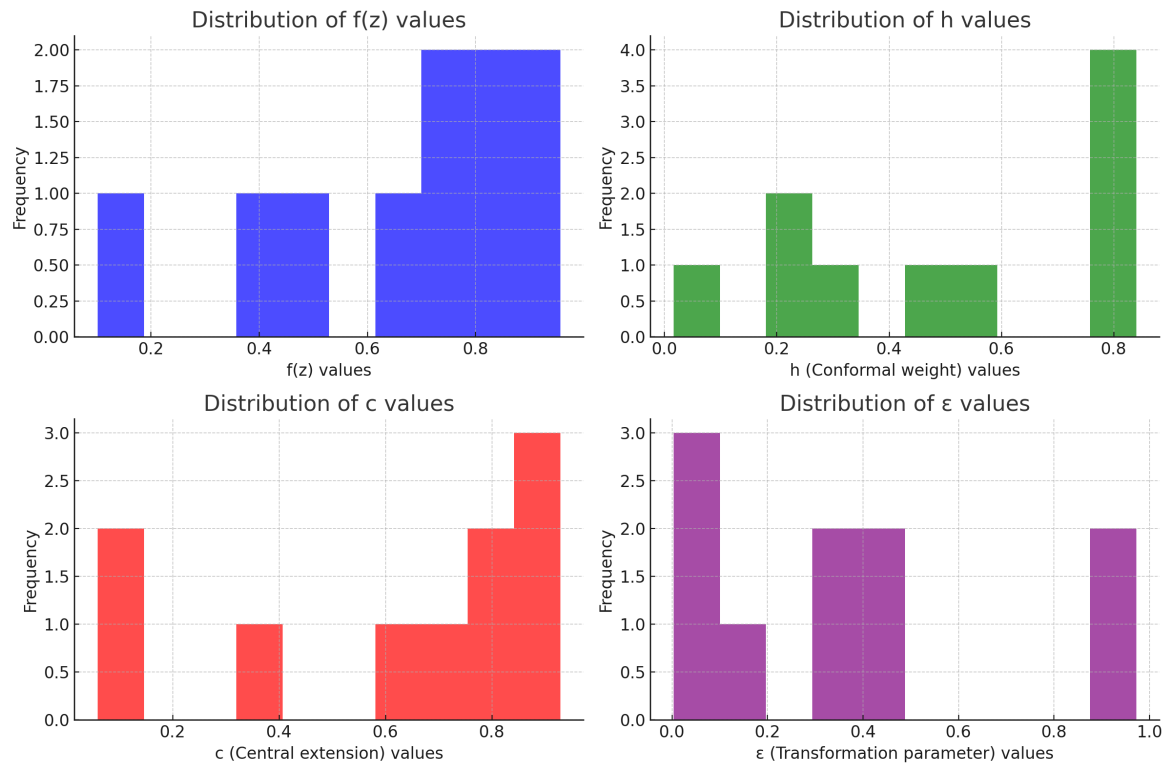

Fig. 19: $f(z)$, $h$, $c$, and $\epsilon$ values Over Timesteps

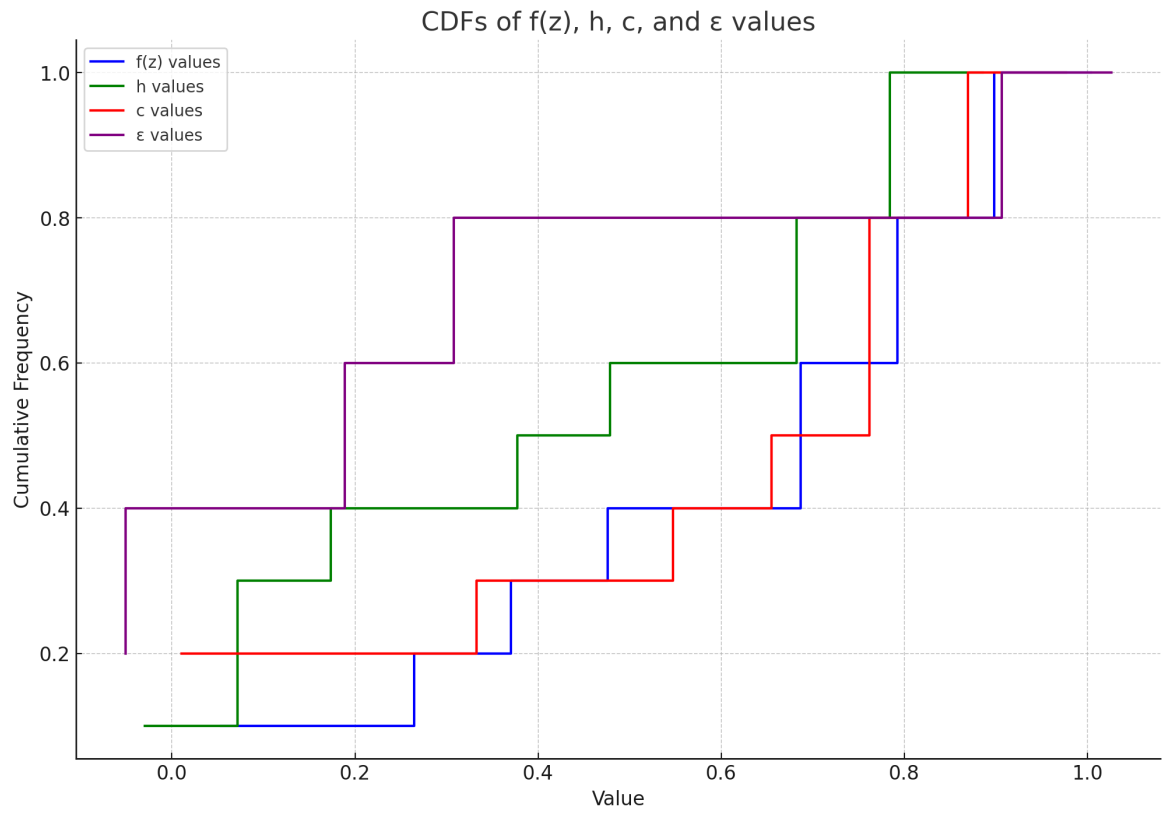

Fig. 20: CDF of $f(z)$, $h$, $c$, and $\epsilon$ values

This leading part allows us to determine the spin, which is expressed as: $s = -\frac{h}{2}$

The graph provides insights into the calculation of the primary field's spin $S$ using the concepts of conformal field theory, which is applied as a metaphor for understanding social dynamics when fake news is spreading.

# 18. Results Related to Fake News Diffusion Modeled by Conformal Field Theory (CFT)

To analyze the spread of fake news from the perspective of each parameter, let's break down the content of each image and discuss their potential implications:

## 18.1 Network Information Diffusion Visualization

This network graph likely represents the spread of information through a social network. Nodes could represent individuals or clusters of individuals, and the edges represent the connections through which information can travel. The color intensity of nodes could be indicative of the 'information level', perhaps relating to the degree of exposure or belief in the fake news. For example, node 5 and node 8 are highly colored, suggesting these nodes have a high level of information, which could imply they are significant spreaders or believers of the fake news within this network.

## 18.2 Distribution Graphs

These histograms show the distribution of different parameters from the CFT analysis:

**f(z) values**: This could represent the regular function mapping the complex plane in a way that describes the dynamics of information spread. The distribution seems fairly uniform, suggesting that the influence of the regular function on the spread of fake news is somewhat consistent across different instances.

**c values (Central extension)**: This parameter in CFT is related to the number of degrees of freedom or the complexity of the system. The distribution seems to increase with the value of c, indicating that as the complexity of the social dynamics or the 'information space' increases, there might be more occurrences of certain behaviors or patterns related to the spread of fake news.

**h values (Conformal weight)**: This is related to the scaling dimension of primary fields and can reflect how certain narratives or pieces of information scale in the network. The distribution is skewed towards higher values, which could imply that narratives with greater 'weight' tend to spread more.

**values (Transformation parameter)**: This might represent the extent of local transformations applied to the system, possibly indicating the susceptibility of individuals to changes in the information they receive. The distribution is quite varied, suggesting that different parts of the network are differently affected by small perturbations, which could model the varying susceptibility of individuals to fake news.

## 18.3 Cumulative Distribution Functions (CDFs)

The CDFs provide information about the probability that a variable takes a value less than or equal to a certain level. Here we can see how quickly the cumulative probability for each parameter increases as the value increases:

**f(z) values**: The cumulative distribution is stepwise, indicating discrete jumps which could correspond to specific events or thresholds in the information diffusion process.

**c values**: This CDF increases more steadily, suggesting a more continuous influence of the system's complexity on the spread of fake news.

**h values**: There is a steep curve in the middle values, which may point to a critical threshold of the conformal weight where the spread of fake news suddenly becomes more likely.

**values**: The steps in this CDF could indicate that certain levels of the transformation parameter have a disproportionate influence on the spread of fake news, with specific 'tipping points' where susceptibility changes dramatically.

To summarize, this analysis provides a multifaceted view of the diffusion of fake news within a network using the lens of conformal field theory. The distribution and cumulative graphs offer insights into how the complexity of the network, the scaling dimensions of influential narratives, and the susceptibility of individuals to changes in information contribute to the spread of misinformation. Each parameter tells a part of the story, and together they provide a nuanced understanding of the dynamics at play.

# 19. Conclusion:Introduce the Ward-Takahashi identity and a two-dimensional conformal field theory

We now calculate the spin (a measure of the influence or speed of spread of information) of each agent in the diffusion pattern of social negative or inconvenient news and fake news, and introduce the following elements in the diffusion pattern of fake news. Introduce the Ward-Takahashi identity and a two-dimensional conformal field theory approach using the Ward-Takahashi identity and path integral that integrates paths from two-dimensional conformal field theory to the source of negative information such as fake news, misinformation, and offensive news.

## Ward-Takahashi Identity

The Ward-Takahashi identity provides constraints when preserving conformal symmetry.

$$\sum_{i=1}^{n}\left(\frac{\partial}{\partial z_i}+2h_i\frac{\partial f}{\partial z_i}+f(z_i)\frac{\partial}{\partial z_i}\right)\langle\Phi_1(z_1)\Phi_2(z_2)...\Phi_n(z_n)\rangle=0 \tag{1}$$

Here, $\Phi_i(z_i)$ represents primary fields, $h_i$ is the conformal weight, and $f(z_i)$ is the conformal transformation.

## Path Integral Formulation

Model the spread of fake news as a path integral of propagation paths.

$$Z=\int\mathcal{D}[\Phi]e^{iS[\Phi]} \tag{2}$$

Here, $\mathcal{D}[\Phi]$ is the measure of primary fields, and $S[\Phi]$ is the action.

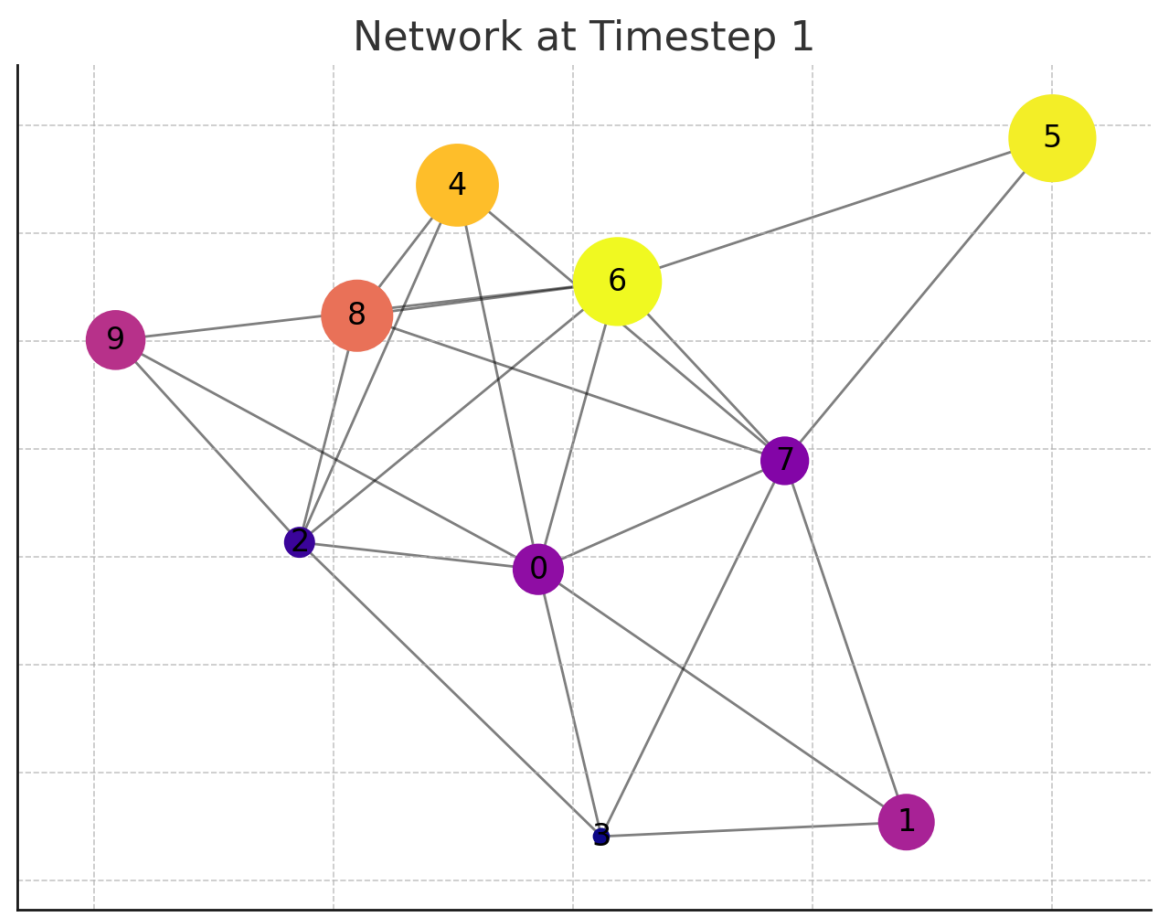

Fig. 21: Network at Timestep, N=1

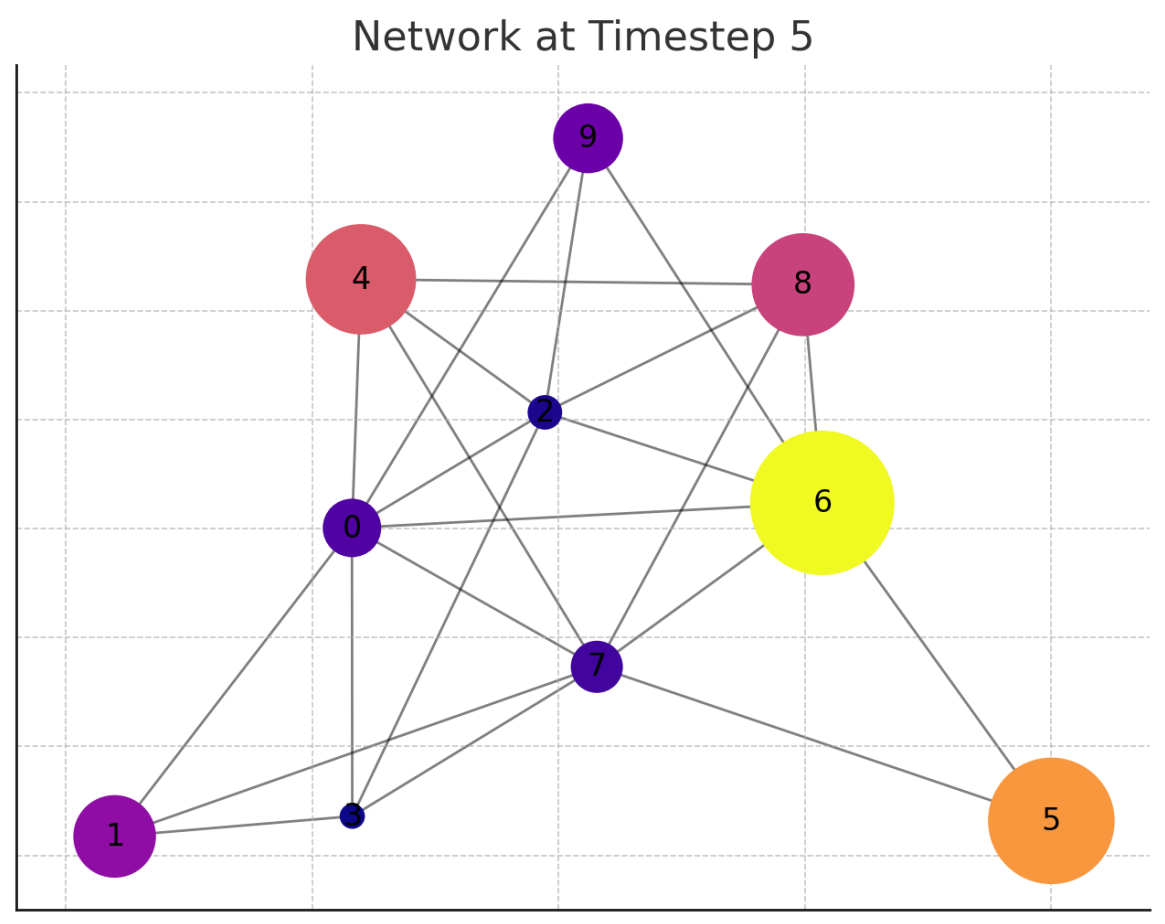

Fig. 22: Network at Timestep, N=5

## Application of 2D Conformal Field Theory

Quantify the diffusion pattern of fake news using 2D conformal field theory.

$$S[\Phi] = \int d^2z\, \mathcal{L}(\Phi, \partial\Phi) \tag{3}$$

Here, $\mathcal{L}$ is the Lagrangian density.

## Parameters and Implementation Ideas

## Definition of Network Structure

Define the structure of a social network and assign primary fields and conformal weights to each node.

## Selection of Conformal Transformations

Select a conformal transformation $f(z)$ that influences the diffusion of fake news and apply it.

## Calculation of Action and Lagrangian Density

Calculate the action $S[\Phi]$ and the Lagrangian density $\mathcal{L}$ to analyze the diffusion pattern of information.

The results of information diffusion simulations based on conformal field theory are shown for each time step. Each graph shows the state of the network at a particular time step, with the color of the node representing the information level. Light-colored nodes indicate a high information level, while dark-colored nodes indicate a low information level. In the graph (Time Step 1), we can see that some nodes with high information levels (light colors) are distributed. This indicates that information is not concentrated in a particular node, but is beginning to spread throughout the network. Nodes in central locations (e.g., node 0 and node 7) are connected to many other nodes and may play an important role in information diffusion.

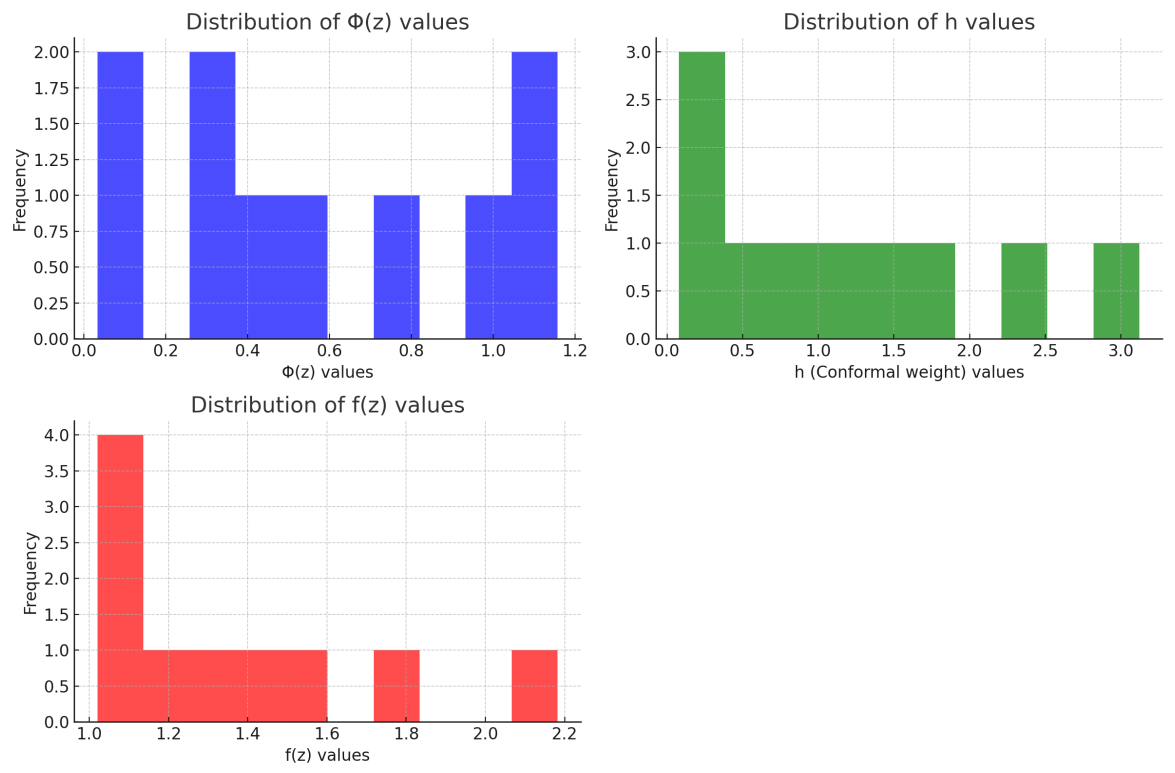

Fig. 23: $\Phi_i(z_i)$ represents the primary fields, $h_i$ signifies the conformal weights, and $f(z_i)$ denotes the conformal transformations. values Over Timesteps

In the second graph (time step 5), we see that the number of light-colored nodes is increasing. This indicates that information is being communicated to more nodes over time. Also, some nodes (e.g., node 4 and node 8) are particularly light-colored, suggesting that these nodes may be the main diffusion sources of information. On the other hand, dark-colored nodes (e.g., node 1 and node 3) remain at low information levels. These nodes may be isolated in the network or may be inefficiently communicating information.

Overall, information appears to be spread throughout the network as the time step progresses, but some nodes remain at a lower information level than others. This suggests that certain nodes in the network are important hubs for spreading information, or that some nodes play a more important role in the information spreading process than others.

The results depict cumulative distribution functions (CDFs) and distributions of each variable. CDFs illustrate how the cumulative distribution of each variable changes, indicating the probability of a value or lower. $\Phi_i(z_i)$ represents

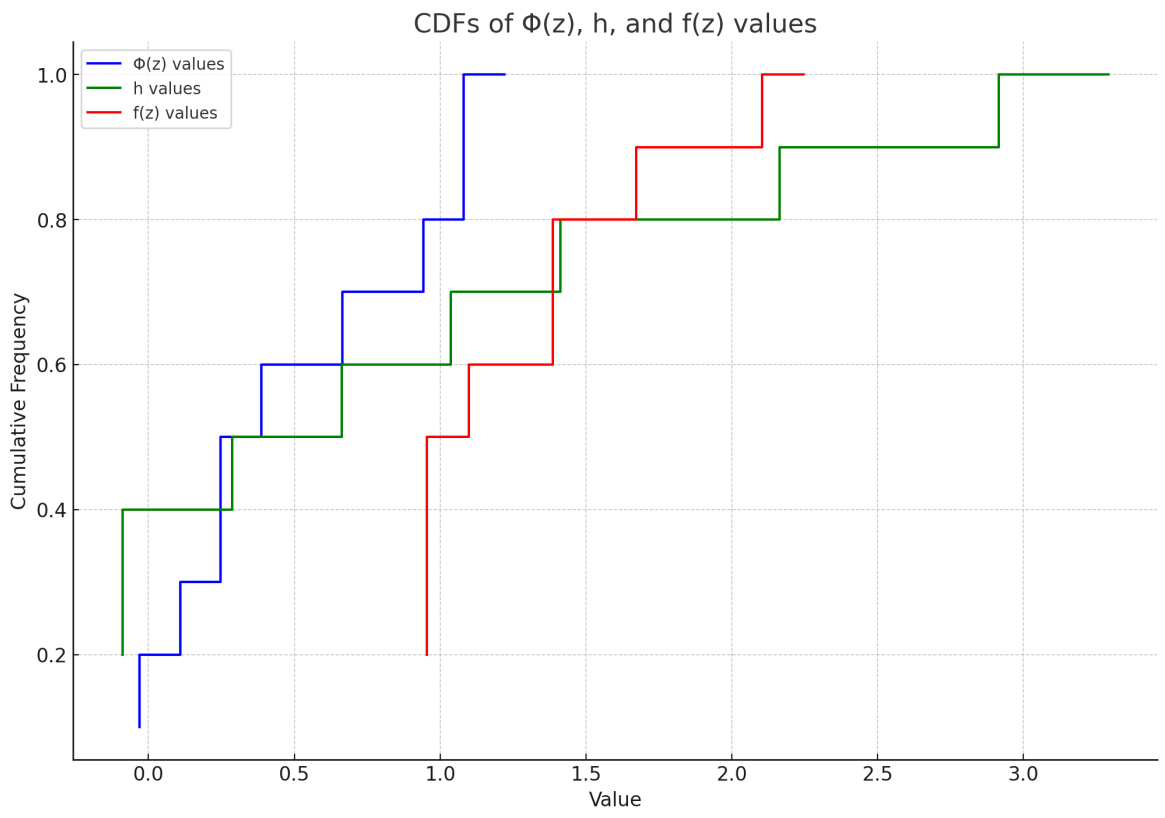

Fig. 24: CDF of $\Phi_i(z_i)$ represents the primary fields, $h_i$ signifies the conformal weights, and $f(z_i)$ denotes the conformal transformations. values

the primary fields, $h_i$ denotes the conformal weights, and $f(z_i)$ signifies the conformal transformations.

In the first graph (CDF), we observe how the cumulative distribution of each variable evolves. This provides insights into the probability of a value being equal to or less than a given threshold. For instance, the CDF of $h$ sharply increases around the value of 1, suggesting it might represent a significant threshold. In contrast, the CDF of $f(z)$ exhibits a more stepwise increase, indicating that $f(z)$ has a more discrete range of values.

The second set of graphs presents histograms depicting the distributions of each variable. Histograms visualize the frequency of data points within specific value ranges. The distribution of $f(z)$ shows the highest frequency near 1.0, suggesting data is concentrated in a specific range. The distribution of $h$, on the other hand, is more spread out, spanning the range from 0.5 to 2.5. Meanwhile, $\Phi(z)$ appears to be concentrated in the range of 0.2 to 1.2.

These observations help us understand how the primary fields, conformal weights, and conformal transformations are distributed within the context of conformal field theory. The distribution of primary fields may offer insights into the dynamics of the theory, while the distribution of conformal weights provides information about the scaling dimensions of the system. The distribution of conformal transformations may shed light on how transformations are implemented and contribute to the system's symmetries.

The graphs of the results depict cumulative distribution functions (CDFs) and distributions of different physical quantities. CDFs illustrate the proportion of random variables that are less than or equal to a specific value, while distribution graphs show the frequency of values of the variables. Here, $\mathcal{D}[\Phi]$ represents the measure of the primary fields, $S[\Phi]$ denotes the score, and $\mathcal{L}$ stands for the Lagrangian density.

From the CDF graphs, it can be observed that the val-ues of $\mathcal{L}$ are concentrated around zero but extend to values exceeding eight. In contrast, the distributions of $\mathcal{D}[\Phi]$ and $S[\Phi]$ are more spread out, particularly $S[\Phi]$ extends to values close to seven. This indicates that the Lagrangian density takes values in a more constrained range, while the other two measures have broader distributions.

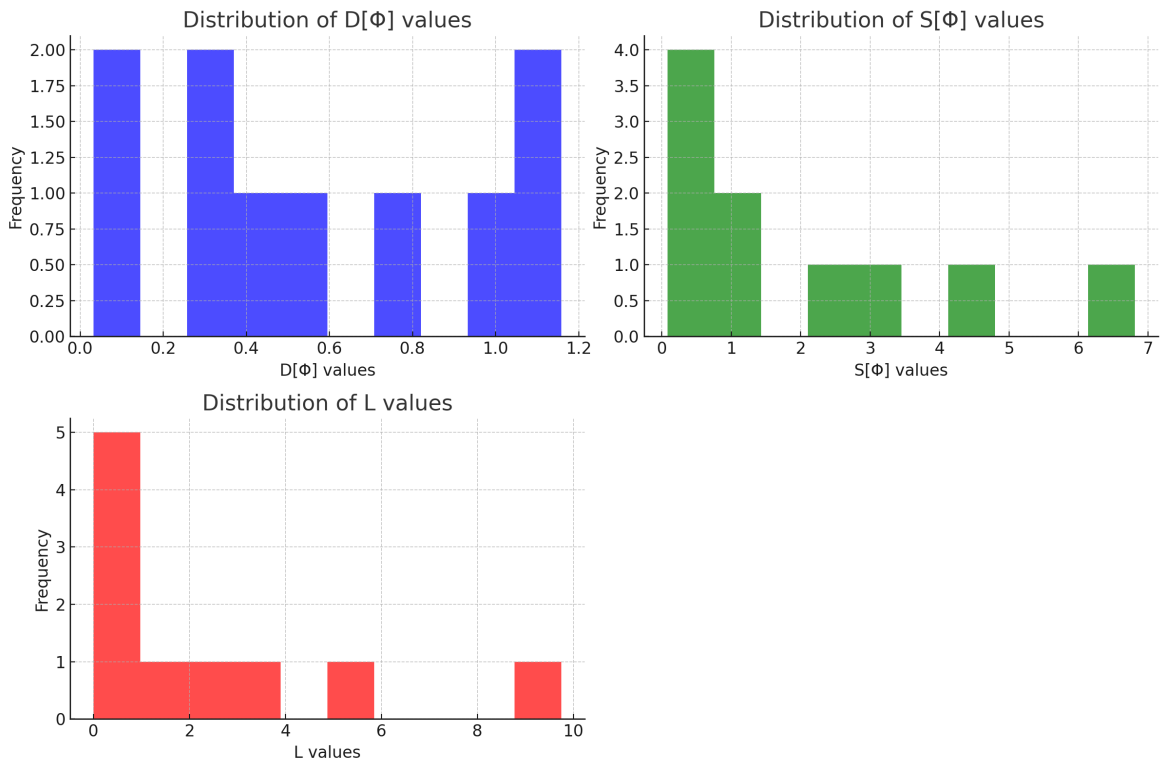

Fig. 25: $\mathcal{D}[\Phi]$ represents the measure of primary fields, $S[\Phi]$ denotes the score, and $\mathcal{L}$ stands for the Lagrangian density values Over Timesteps

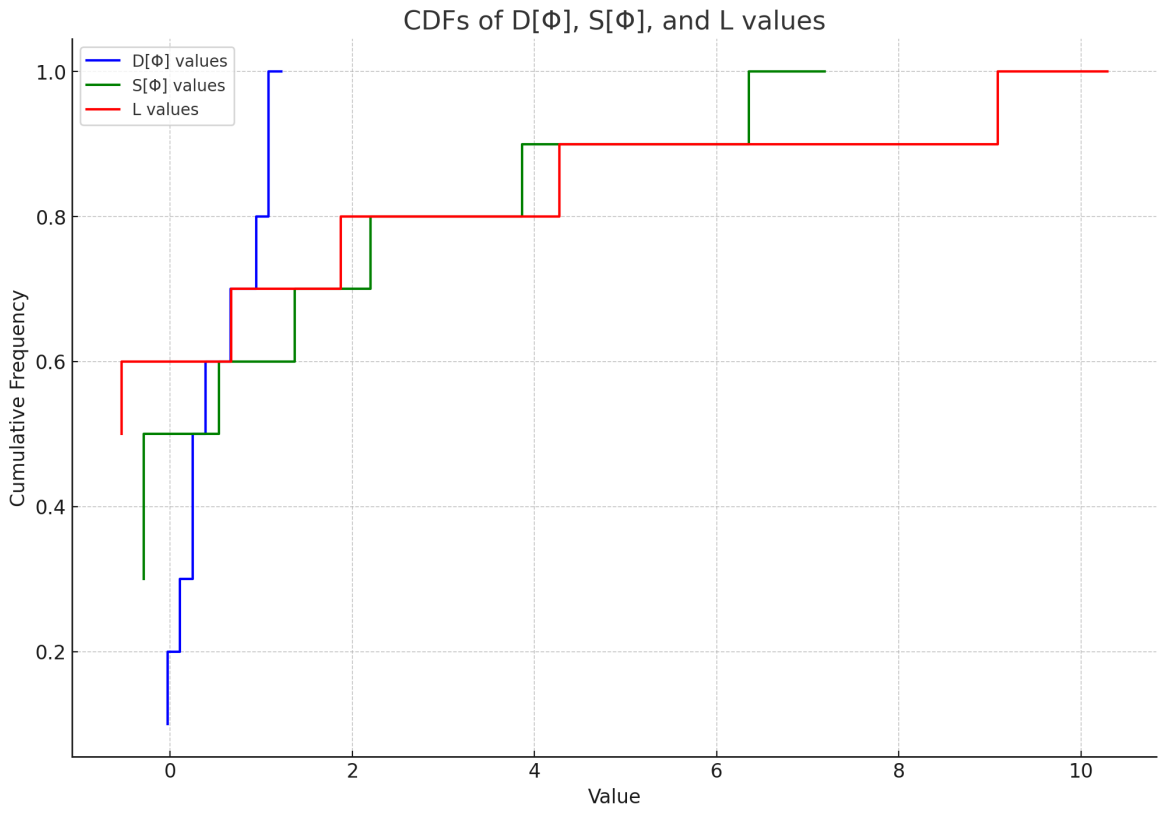

Fig. 26: CDF of $\mathcal{D}[\Phi]$ represents the measure of primary fields, $S[\Phi]$ denotes the score, and $\mathcal{L}$ stands for the Lagrangian density, $\epsilon$ values

In the distribution graphs, the values of $\mathcal{D}[\Phi]$ are uniformly distributed between 0.2 and 1.0, whereas $S[\Phi]$ ranges from 0 to 6, with fewer values at the higher end. On the other hand, the distribution of $\mathcal{L}$ exhibits peaks near 0 and 6, with fewer values in the intermediate range.

These results suggest that the measure of primary fields is evenly distributed across a range, while the score and Lagrangian density are concentrated around specific values or biased toward certain ranges. In a physical context, these distributions provide valuable insights into understanding the characteristics of the system. For instance, if the Lagrangian density is concentrated around zero, it may be related to the stable states or ground states of the system. Conversely, the spread of scores may indicate the extent to which different states or configurations influence the system.

## 20. Conclusion:Approach Using Cauchy's Integral Theorem in 2D Conformal Field Theory for Analyzing the Spread of Negative or Fake News

Here, we introduce Cauchy's theorem as an approach to quantitatively analyze cases in which the risk of spreading filter bubbles such as negative information, rumors, and fake news is likely to be exposed. If information is constant within a closed curve from the procedure for calculating the total change of information, then based on Cauchy's theorem, the total change of information within that loop is expected to be zero. We wanted to gain insight from simulations that hypothesize that this may represent a state in which erroneous information is confined within a certain range and its external effects are limited.

In the context of analyzing the diffusion patterns of socially negative news, such as fake news, through the approach of 2D conformal field theory employing the Ward-Takahashi identity and path integral, we incorporate Cauchy's integral theorem. We consider small circular integration paths $C$ surrounding the insertion points of primary fields $\Phi_i$, aiming to understand the diffusion routes of undesirable information from both inside and outside the circular paths $C$.

Here, we present the mathematical equations, parameters, and the computational ideas for applying Cauchy's integral theorem in the framework of 2D conformal field theory to analyze the diffusion patterns of socially negative news or fake news.

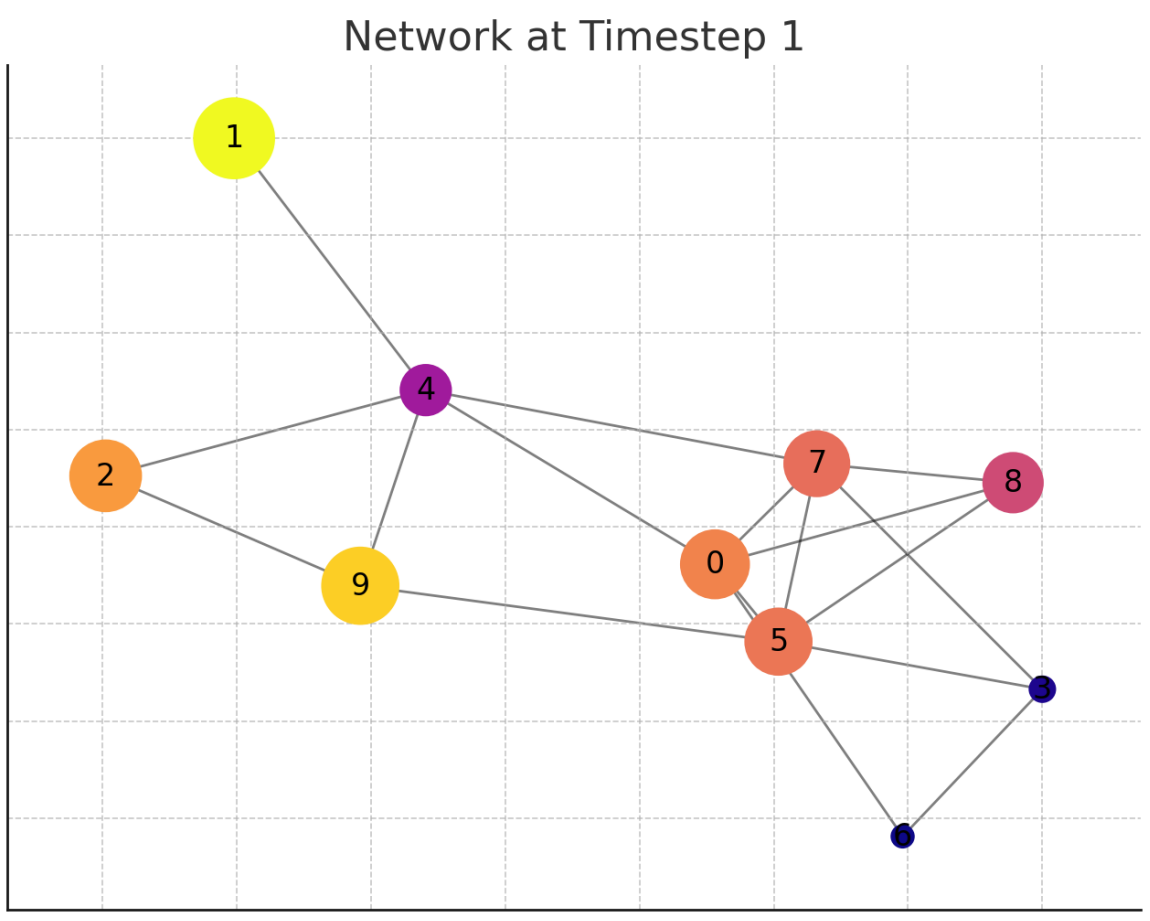

Fig. 27: Network at Timestep, N=1

### Equations and Computational Process

1. **Definition of Insertion Points for Primary Fields $\Phi_i$ and Integration Paths**: - We define the insertion points $z_i$ of primary fields $\Phi_i$ and small circular paths $C_i$ with a radius of $\epsilon$ centered at these points.

2. **Application of Cauchy's Integral Theorem**: - Cauchy's integral theorem allows us to equate the integral of primary fields $\Phi_i$ along the circular path $C_i$ to the residue of $\Phi_i$ at $z_i$. - Equation: $\oint_{C_i} \Phi_i(z)dz = 2\pi i \,\mathrm{Res}[\Phi_i(z), z = z_i]$

3. **Computation of Residue of Primary Field $\Phi_i$**: - We assume that the residue of primary field $\Phi_i$ at its insertion point $z_i$ is equal to the value of $\Phi_i$ at $z_i$. - Equation: $\mathrm{Res}[\Phi_i(z), z = z_i] = \Phi_i(z_i)$

4. **Transformation of Integration Path $C_i$ and Derivation of Results**: - We transform the integration path $C_i$ and consider the behavior of the primary field $\Phi_i$ around its insertion point. - Equation: $\oint_{C_i} \Phi_i(z)dz \approx \oint_{C_i'} \left[\Phi_i(z_i) + \sum_{n=1}^{\infty}(L_{-n})_i(z - z_i)^n\right] dz$ - Here, $(L_{-n})_i$ represents the mode operators.

### Simulation Approach

1. **Identification of Fake News Sources**: - We model the sources of fake news as insertion points of primary fields.

2. **Modeling Diffusion Routes on the Network**: - We model the diffusion routes of fake news on the network as paths and compute the values of $\Phi_i$ at each node.

3. **Simulation of Integration Paths**: - We compute the integrals around the insertion points of primary fields $\Phi_i$ at each node and analyze how fake news spreads.

4. **Interpretation of Results**: - We use the integration results to understand the diffusion patterns of negative or fake news and analyze the impact of negative information.

In the network at time step 1, nodes 1, 2, and 9 are shown in bright colors indicating a high level of information. Node 0

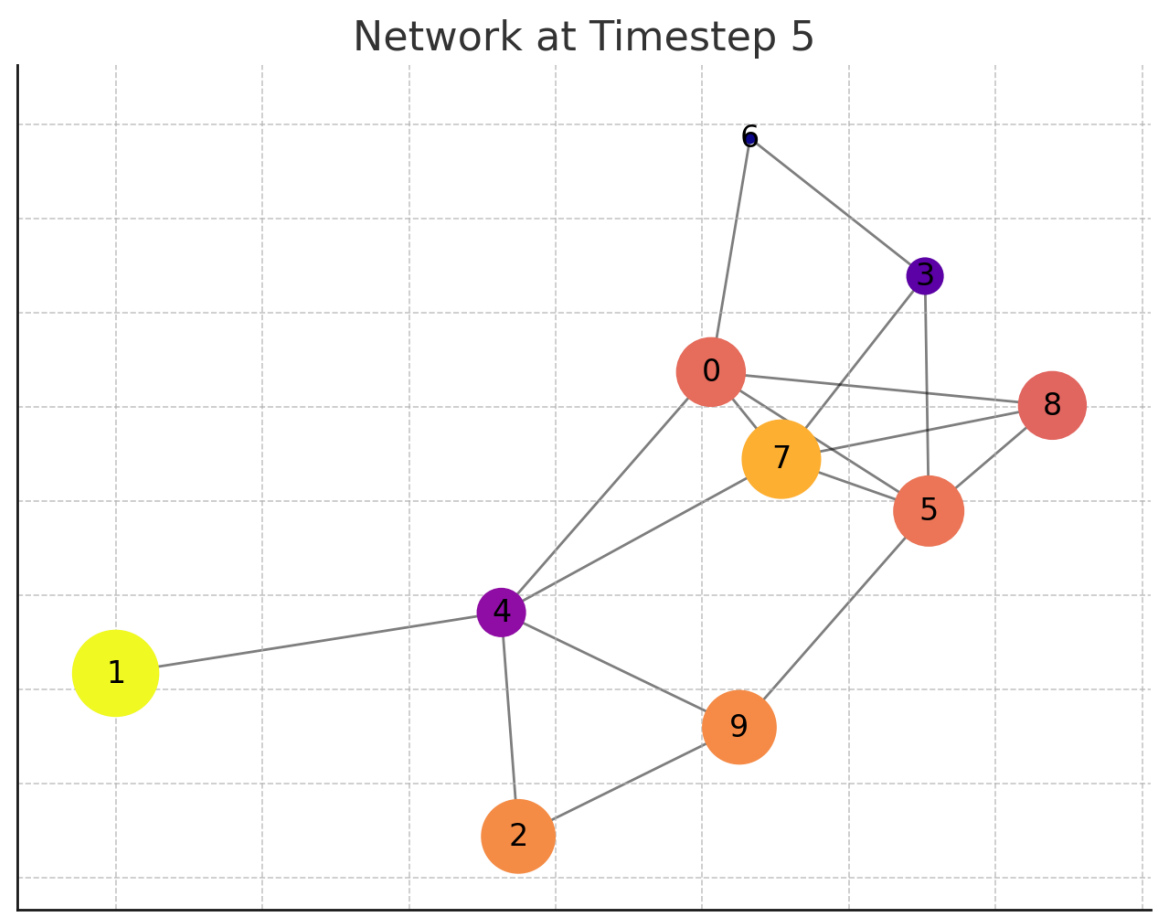

Fig. 28: Network at Timestep, N=5

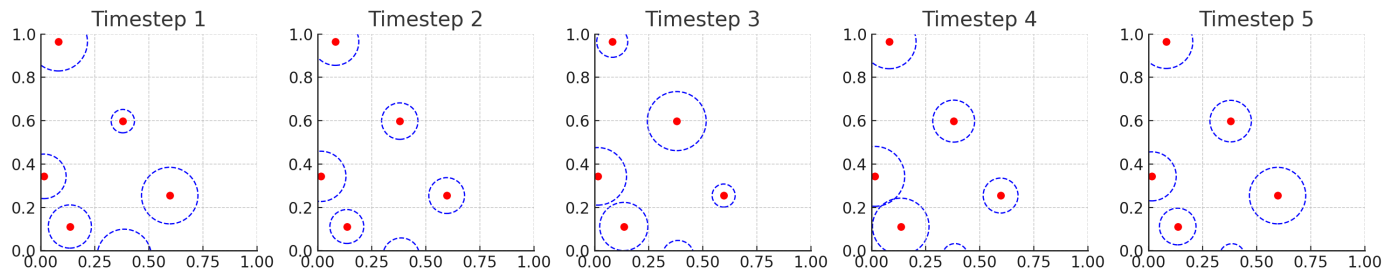

Fig. 29: Integral path $C_i$, behavior around the insertion point of the primary field $\phi_i$

is connected to many other nodes and may play a central role. However, node 0 itself seems to have a medium information level. Overall, this network appears to be relatively evenly connected, with nodes with high information levels located at the edges of the network.

Moving on to time step 5, we see that nodes 3 and 6 have a newly high information level. Also, nodes 1 and 2, which previously had high information levels, have changed to darker colors. This may mean that the information has diffused over time and reached a new node. Node 0 is still connected to many nodes, but its information level remains medium. Nodes 7 and 8 are considered to have medium information levels at both time steps and may be stable information focal points. We can analogize each node of the network to a primitive operator in conformal field theory and consider the edges between these nodes as interactions between primitive operators. If the color of a node represents the information level, this might correspond to the expected value of the field in CFT. The way the information level changes as the time step progresses is similar to the way the field expectation evolves over time.

Looking at the change from time step 1 to time step 5, we see that information (or field expectation) diffuses from higher to lower states. This can be thought of as analogous to the process of energy spreading and particle scattering due to interactions in CFT.

In terms of applying Cauchy's theorem to networks, if the flow of information on a network forms a closed curve (e.g., a loop of nodes such that information circulates), the total change in information within that loop is expected to be zero. This corresponds to conservation of energy in a physical system and may indicate conservation or balance of information in the context of a network.

Through this change, we can identify trends in how information diffuses within the network. Information appears to move between specific nodes, and the overall network structure and relationships among nodes may influence the pattern of information diffusion. In particular, nodes with high network connectivity tend to be important hubs in information diffusion, while changes in information levels between time steps 1 and 5 reflect the time dependency of network dynamics.

Integral path $C_i$, as a consideration of the behavior around the insertion point of the primary field $\phi_i$, appears to show the deformation of the integral path $C_i$ at different time steps. In conformal field theory (CFT), the integral path plays an important role in the computation of physical quantities. In particular, the integral paths around the primal field $\Phi_i$ are important for capturing singularities and other characteristic behaviors of the field.

When considering the deformation of the integral path $C_i$ shown in the image at each time step, the goal is to evaluate the change in physical quantities caused by the insertion of the field $\Phi_i$ at each point in CFT, the deformation of the integral path near the insertion point of a particular primal field gives information about the conformal dimension and correlation functions of that field The deformation of an integral path near the insertion point of a particular primal field gives information about the conformal dimension and correlation function of the field.

The changes in the integral path seen throughout the time step can be interpreted as representing the dynamics and time evolution of the field. As we progress from time step 1 to time step 5, we see that the integral path expands and contracts, but remains consistent around the insertion point of the primal field. This behavior suggests that the field interactions and energy distribution are changing in a time-dependent manner.

One can also consider Cauchy's theorem in connection with the deformation of the integral path in the CFT. If there is no singularity inside an integral path, the integral of any analytic function along that path will be zero. This implies that the physical quantities inside the integral path are conserved and can provide conditions under which the integral around the field insertion point is zero.

The image shows several points (nodes) and a circle surrounding them (presumably indicating the extent of information diffusion). It can be observed that these circles are expanding or moving as the time step progresses.

When discussing the spread of negative or erroneous in-formation, we focus on the following points.

### Cauchy's Theorem in Complex Analysis

Cauchy's Theorem is an important theorem in complex analysis, related to the integration of complex functions defined on closed curves (closed paths). Here, we will explain Cauchy's Theorem mathematically with equations and provide an explanation of the calculation process.

The Mathematical Expression of Cauchy's Theorem: If $f(z)$ is a complex function that is analytic (has no singularities) within a closed curve $C$, then the following equation holds:

$$\oint_C f(z)\, dz = 0$$

Here, $\oint_C$ represents the integral along the closed curve $C$, and $f(z)$ is a complex function that is analytic on that curve.

**Explanation of Cauchy's Theorem**: 1. A complex function $f(z)$ that is analytic within a closed curve $C$ is given. 2. According to Cauchy's Theorem, the integral along the closed curve $C$ for this function $f(z)$ is zero, meaning the integral result is equal to 0.

This theorem is related to complex integration and the properties of analytic functions in the complex plane, and it has many applications. For example, you can use Cauchy's Integral Theorem to calculate integrals within regions enclosed by closed curves. It is also related to other important results in complex analysis, such as the residue theorem around singular points. Consider the complex function $f(z) = \frac{1}{z}$, and calculate the integral along the closed unit circle $C$.

$$\oint_C \frac{1}{z}\, dz$$

To compute the integral along the closed unit circle $C$, we use a parameterization. Let $z = e^{i\theta}$, then $dz = ie^{i\theta}d\theta$. Therefore, the integral becomes:

$$\begin{aligned}\oint_C \frac{1}{z}\, dz &= \int_0^{2\pi} \frac{1}{e^{i\theta}} \cdot ie^{i\theta}\, d\theta \\ &= i\int_0^{2\pi} d\theta \\ &= i(2\pi - 0) \\ &= 2\pi i\end{aligned}$$

As a result, according to Cauchy's Theorem, the integral along this closed curve is not zero but equal to $2\pi i$.

This calculation example demonstrates that Cauchy's Theorem holds for complex functions that are analytic within closed curves.

### Diffusion pattern

The expansion of the circles at each time step indicates that information is spreading within the network. It can be seen that information that started at a particular point is influencing other nodes over time.

### Closed Curves and Information Circulation

When a closed curve is formed, it suggests that information is circulating within a particular group of nodes. This may mean that negative information, rumors, or other information continues to be shared within a particular community on the network.

### Total change in information

If information is constant within a closed curve, then based on Cauchy's theorem, the total change in information within that loop is expected to be zero. This may represent a state in which misinformation is confined within a certain range and its outward influence is limited.

### Conservation of Energy and Conservation of Information

Analogous to the law of conservation of energy in physical systems, conservation of information in networks implies that information is not destroyed or created, but simply redistributed. This implies that misinformation does not disappear and, once it arises, may persist in the network for a long time.

From these perspectives, we can gain a better understanding of how misinformation and negative information spread and persist within a network. However, these are merely abstract considerations, and concrete numerical models and simulations are needed to analyze the actual information diffusion mechanisms and network dynamics. Other factors, such as the accuracy and influence of information, would also need to be taken into account.

## 21. Conclusion:Consideration of Inference of News Sources

Here, we introduce trace anomalies and metrics as an approach to quantitatively analyze cases where the risk of spreading negative information, rumors, fake news, and other filter bubbles is likely to be exposed. A trace anomaly refers to the spread of information in an unexpected way, usually indicating anomalous behavior or hidden dynamics in the system. Metric variation indicates how diffusion characteristics, such as the rate and pattern of information diffusion, change over time.

In the analysis approach of identifying the sources of socially negative news or fake news, modeling diffusion routes on the network, and conducting simulations, we aim to graphically output the results. Here, we introduce the trace anomaly coefficient $a$, the curvature scalar $R$, and the variation in the metric $h$ to discuss the magnitude of diffusion errors from the trace anomaly.

$a$ represents the trace anomaly coefficient, $R$ stands for the curvature scalar, and $h$ denotes the variation in the metric from a flat metric $\eta_{\mu\nu}$. We introduce this variation to calculate the change from the expectation value of $T_\mu^\mu$ using the Ward identity and design it to discuss the magnitude of diffusion errors from the trace anomaly coefficient.

Let $X$ be a product of general local fields, and consider the proof and calculation process when a trace anomaly occurs and depends on $h$ (variation in the metric). Below are the steps involved:

## Definition of Trace Anomaly

1. Definition of Trace Anomaly The trace anomaly is an anomaly related to the trace $(T_\mu^\mu)$ of the energymomentum tensor $T(z)$ in the Ward identity. The trace anomaly is generally expressed as follows:

$$\langle T_\mu^\mu \rangle = a \cdot R$$

Here, $a$ is the trace anomaly coefficient, and $R$ is the curvature scalar.

## Variation in the Metric

2. Variation in the Metric $h$ The variation in the metric $h$ represents the change from a flat metric $\eta_{\mu\nu}$.

$$g_{\mu\nu} = \eta_{\mu\nu} + h_{\mu\nu}$$

Here, $g_{\mu\nu}$ is the metric tensor, $\eta_{\mu\nu}$ is the flat metric, and $h_{\mu\nu}$ is the varied metric.

## Form of Trace Anomaly Depending on $h$

3. Form of Trace Anomaly Depending on $h$ We assume that the trace anomaly coefficient $a$ depends on the varied metric $h_{\mu\nu}$.

$$a = a(h)$$

## Calculation of Trace Anomaly

4. Calculation of Trace Anomaly To calculate the trace anomaly, we use the Ward identity to compute the expectation value of $T_\mu^\mu$.

$$\langle T_\mu^\mu \rangle = \sum_i \left( \frac{\partial S}{\partial h_{\mu\nu_i}} \right) \langle T_\mu^\mu \rangle_i$$

Here, $S$ is the action in conformal field theory, $h_{\mu\nu_i}$ represents parameters representing part of the variation in the metric, and $\langle T_\mu^\mu \rangle_i$ is the expectation value of the trace for the corresponding case.

5. Determining the Dependency of Trace Anomaly Coefficient To find the dependence of the trace anomaly coefficient $a$ on $h$, we proceed with the calculation.

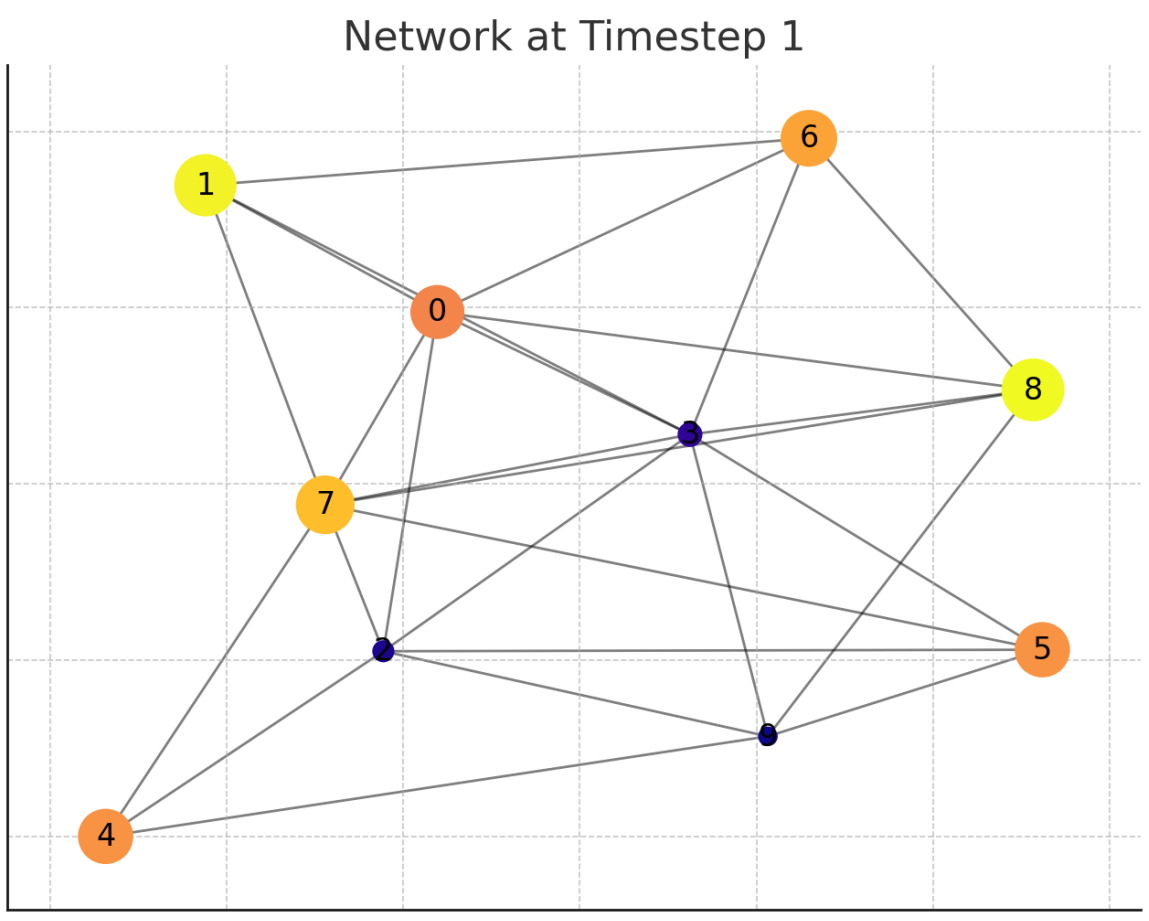

Fig. 30: Network at Timestep, N=1

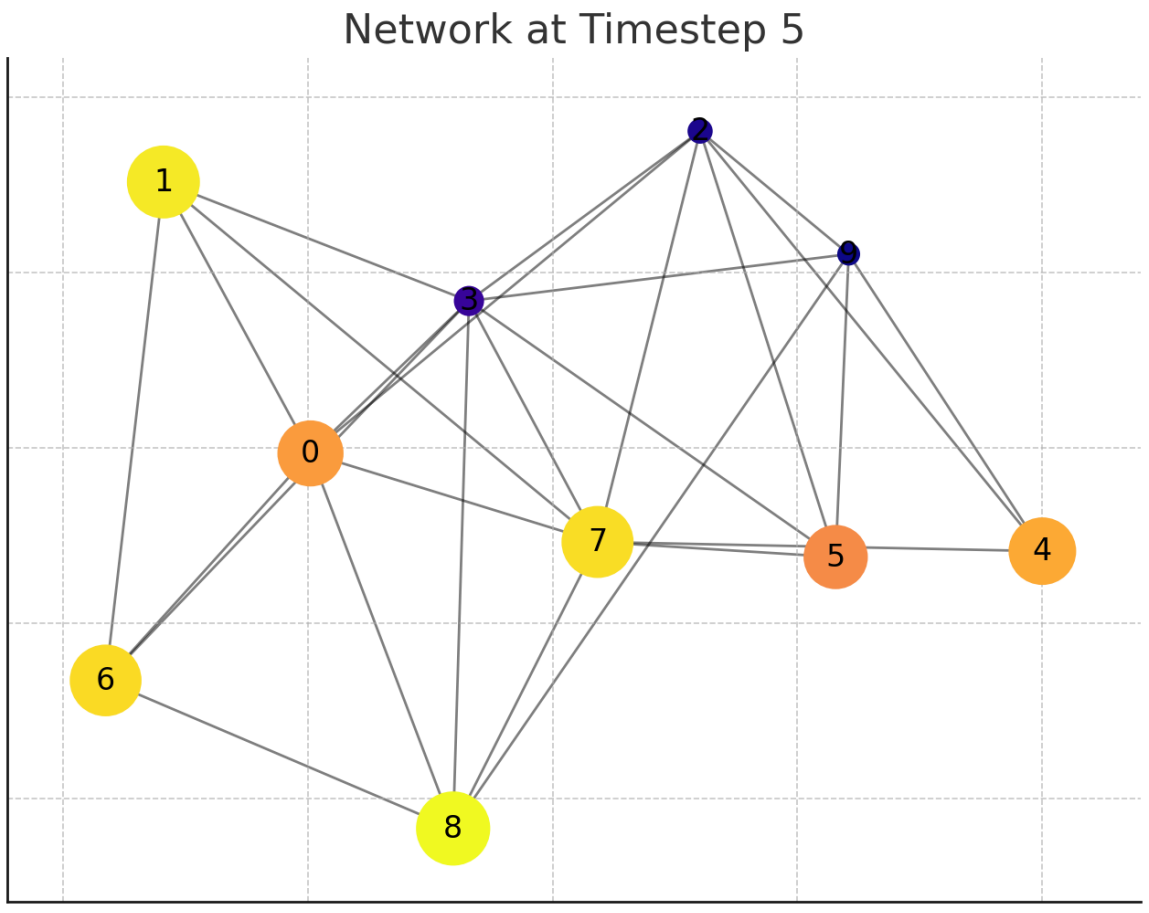

Fig. 31: Network at Timestep, N=5

Considering the variation in the metric $h$ up to first order, we expand $a(h)$ and identify the coefficients with respect to $h$.

6. Result for Trace Anomaly Coefficient The dependence of the trace anomaly coefficient $a(h)$ is determined.

The results of the network diagram show the states at time steps 1 and 5, with the color of the node representing the level of information at that point in time. In information diffusion simulations, a trace anomaly refers to information diffusing in an unexpected way, usually indicating anomalous behavior or hidden dynamics in the system. Metric variation indicates how diffusion characteristics, such as the rate and pattern of information diffusion, change over time.

## Network at Time Step 1

Light-colored nodes indicate a high level of information; these nodes may be information sources. Dark colored nodes indicate a low level of information, indicating that the infor-mation has not yet arrived or has very limited information. Node 0 is connected to many other nodes and appears to be acting as a central hub of information.

## Network at time step 5

The number of light-colored nodes is increasing, which indicates that information is spreading from time step 1 to 5. The connection pattern between nodes appears unchanged, but there are some changes in the location of nodes with higher levels of information. In particular, the high information levels at nodes 1, 6, and 8 are maintained.

## Considerations in terms of information diffusion

If the positions of nodes with high information levels do not change between time steps, this indicates little variation in the metric and may mean that information is diffusing in a constant pattern. If the information level of a node changes over time, this indicates that there is variability in the metric and that the information is spread non-uniformly or irregularly.

To identify trace anomalies, the difference between the expected pattern of information diffusion and the actual pattern should be evaluated. For example, if a node has a higher or lower level of information than expected, there may be an anomaly present. This anomaly may suggest a hidden influential node in the network or a node that is susceptible to external influences.

The term "information" in this context can be interpreted as negative information, rumors, fake news, and other information that is at high risk of being spread by the filter bubble. By analyzing trace anomalies and metric variation, we can understand how such information spreads within a network and develop strategies to control or stop its spread.

## Trace Anomalies

If information is spreading in unexpected ways, this may indicate anomalies in communication patterns between nodes. For example, a situation where some nodes have unusually high information levels or information is spreading rapidly among nodes in unexpected ways. Comparing time steps 1 and 5, certain nodes (e.g., node 1 and node 6) remain consistently brightly colored and maintain high information levels. This indicates that these nodes are either information sources or important centers where information tends to take root.

## Metric Variation

From time steps 1 to 5, the extent of information diffusion in the network appears to be expanding. This is evidenced by the fact that node 3 and node 7 change from dark to light colors at time step 5. Through metric fluctuations, the speed and pattern of information diffusion can be analyzed.

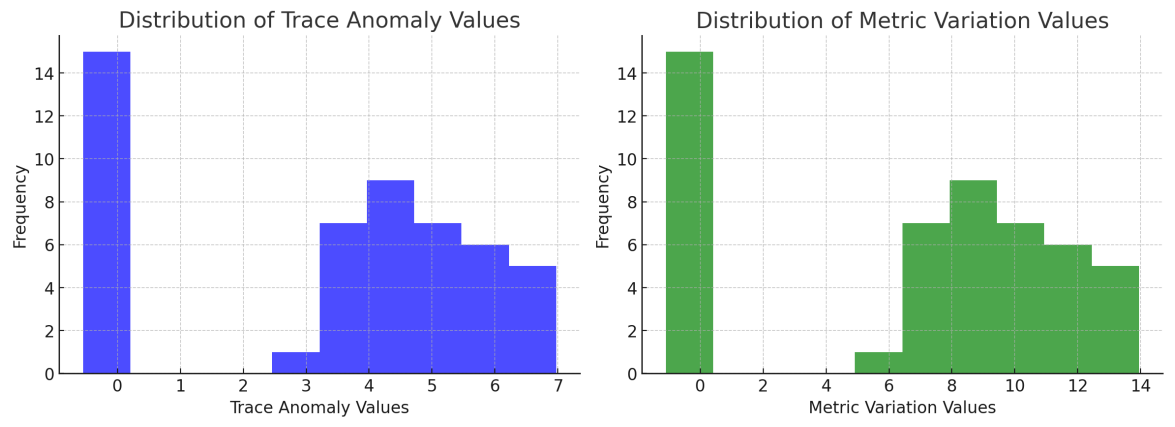

Fig. 32: Trace Anomaly and Metric Variation Values

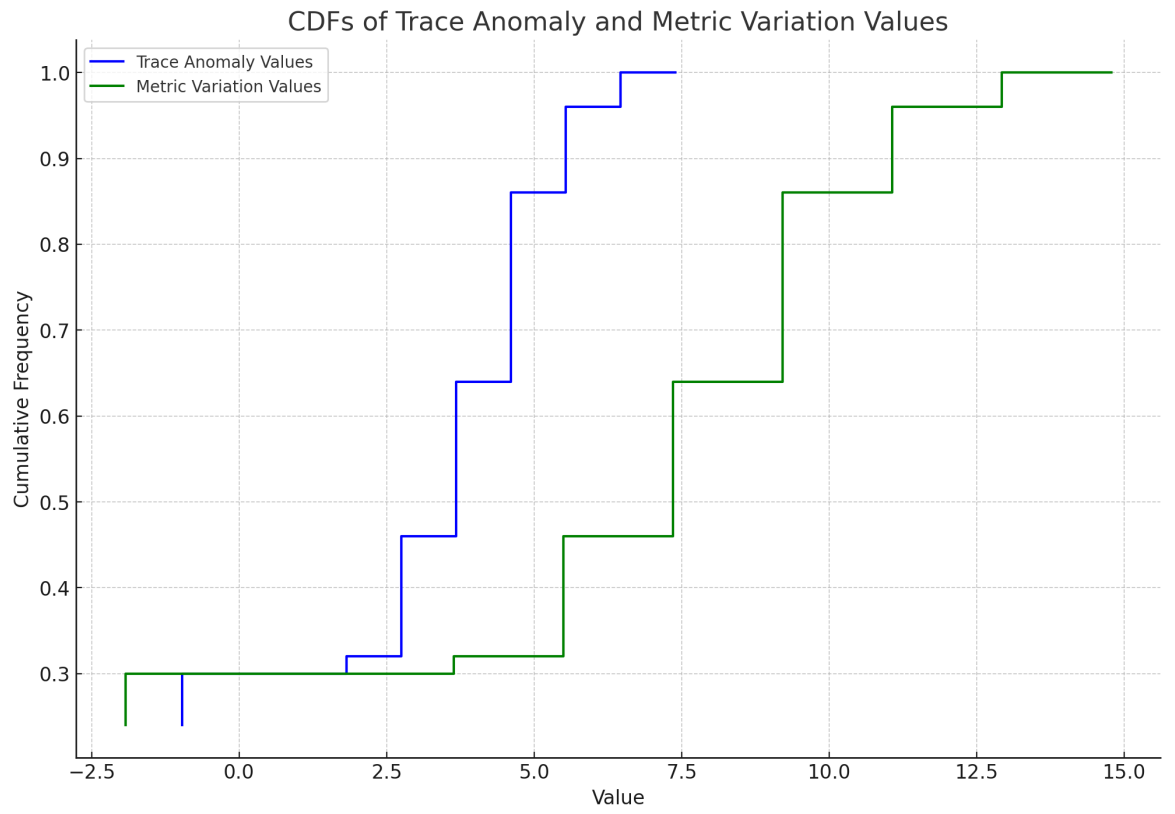

Fig. 33: CDFs of Trace Anomaly and Metric Variation Values

For example, it is possible to observe the speed at which nodes accept new information and the pattern of information movement within the network.

## Quantitative Analysis of Information Diffusion

Detecting and analyzing these metrics and anomalies is essential to curb the spread of negative information and fake news. It is important to understand how information is spread within the network and what role specific nodes and links play in the spread. This allows for intervention strategies such as filtering information to specific nodes or reinforcing information from trusted sources.

Overall, through network analysis, we will gain a better understanding of how negative information and fake news spreads and forms filter bubbles. This will provide important insights to mitigate the problem of social misinformation.

Two key statistics are presented in the results. These are the value of the trace anomaly and the variation value of the metric. These cumulative distribution functions (CDFs) and distributions will help to quantitatively capture important characteristics associated with information diffusion.

## CDF and Distribution of Trace Anomalies

The CDF of a trace anomaly shows the cumulative probability that a particular value will be observed; locations where the CDF shows a steep step indicate a concentration of trace anomaly values, indicating that there have been significant changes around a particular trace anomaly value. In the distribution plots, the trace anomaly values peak on the low side, which indicates that the expected diffusion pattern is maintained in many cases. However, the long tails toward higher values suggest that there are a certain number of cases where unexpected or unusual diffusion patterns are occurring.

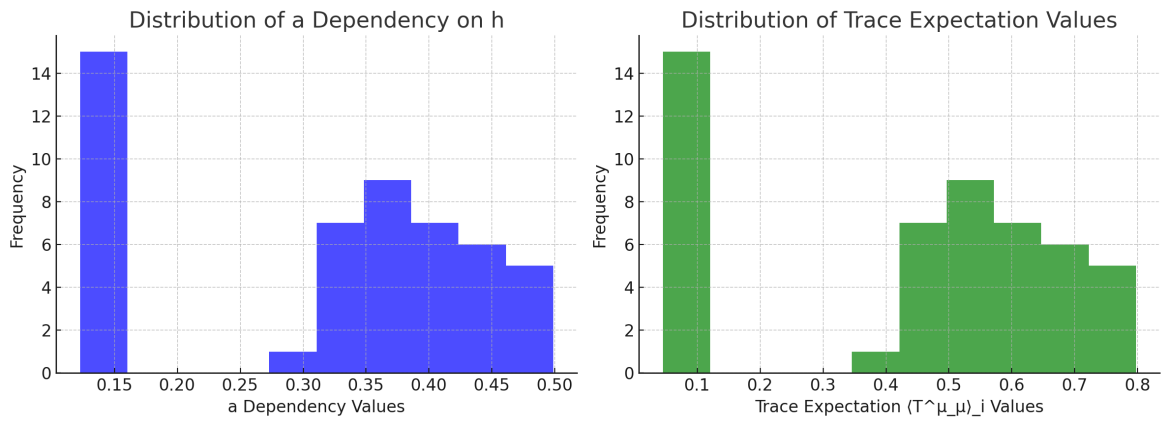

Fig. 34: Dependency on $h$ and Trace Expectation Values

## CDF and Distribution of Metric Variability

It can be seen that the CDF of the metric variation increases smoothly over a wider range of values. This indicates that metric variability can take on a variety of values, suggesting that the characteristics of information diffusion are changing over time. The distribution of metric variation shows the highest frequency around the median, but has tails at both lower and higher values. This indicates that while there are many standard cases in the rate and pattern of diffusion, there is considerable variability.

## Analysis of Information Diffusion

The analysis of trace anomaly and metric variation shows that the dynamics of information diffusion are not constant and exhibit a variety of diffusion patterns. If the trace anomaly is high, information may be spreading in an erratic or unintended pattern, which can help identify regions or nodes that are susceptible to filter bubbles or fake news. If metric variation is high, there may be an event that is causing information to spread rapidly, which may be a sign of social impact or a need for risk management intervention.

These statistics can be used to understand the mechanisms of information diffusion and develop strategies to prevent the formation of filter bubbles and the spread of fake news. In particular, areas that exhibit unusual diffusion patterns or large metric variations are important targets for intervention and education.

## Consideration of the Relationship between Trace Anomaly Coefficient $a$ and Trace Expectation $\langle T^{\mu}_{\mu} \rangle_i$

In this section, we will discuss the relationship between the trace anomaly coefficient $a$ and the trace expectation $\langle T^{\mu}_{\mu} \rangle_i$ based on the results. These values are essential physical quantities in the context of Conformal Field Theory (CFT) and are used as indicators of abnormal behavior, hidden dynamics, and diffusion characteristics of information in the context of information diffusion.

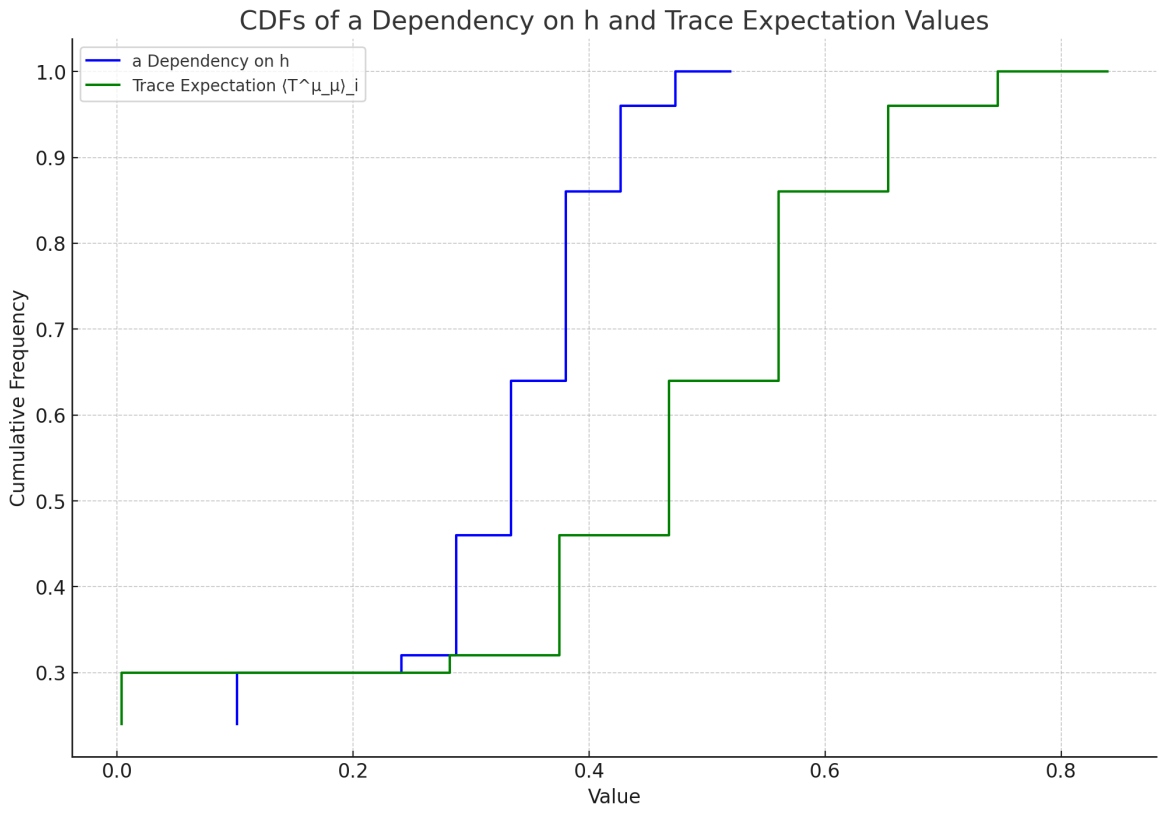

Fig. 35: CDFs of a Dependency on $h$ and Trace Expectation Values

## Analysis of CDF

The Cumulative Distribution Function (CDF) represents the cumulative distribution of data points below a specific value. Points where the CDF curve rises steeply indicate the concentration of data points at that value, reflecting high frequencies of trace anomaly coefficients and trace expectations.

The CDF concerning the dependency of $a$ on $h$ gradually increases with the increase of $h$. This suggests that lower values of $a$ are more common for small values of $h$, and there is a tendency for $a$ to increase with the increase of $h$.

The CDF of the expectation value $\langle T^{\mu}_{\mu} \rangle_i$ shows a more gradual increase, indicating that the values of $\langle T^{\mu}_{\mu} \rangle_i$ are concentrated in a specific range.

## Analysis of Distributions

The distribution of the dependency of $a$ shows high-frequency values within a specific range of $h$, suggesting that specific values of $h$ have a significant impact on the trace anomaly.

The distribution of the expectation value $\langle T^{\mu}_{\mu} \rangle_i$ shows a broader range of values. This suggests that the trace expectation varies across different values of $h$, indicating diverse diffusion characteristics of information.

## Applications to Information Diffusion

The trace anomaly coefficient $a$ and the trace expectation $\langle T^{\mu}_{\mu} \rangle_i$ are essential parameters in the context of information diffusion. They can enhance our understanding of how fake news and negative information spread and help predict diffusion patterns.

The increase in $a$ at specific values of $h$ may indicate a higher risk of information spreading unexpectedly. This can be managed through system monitoring and intervention.

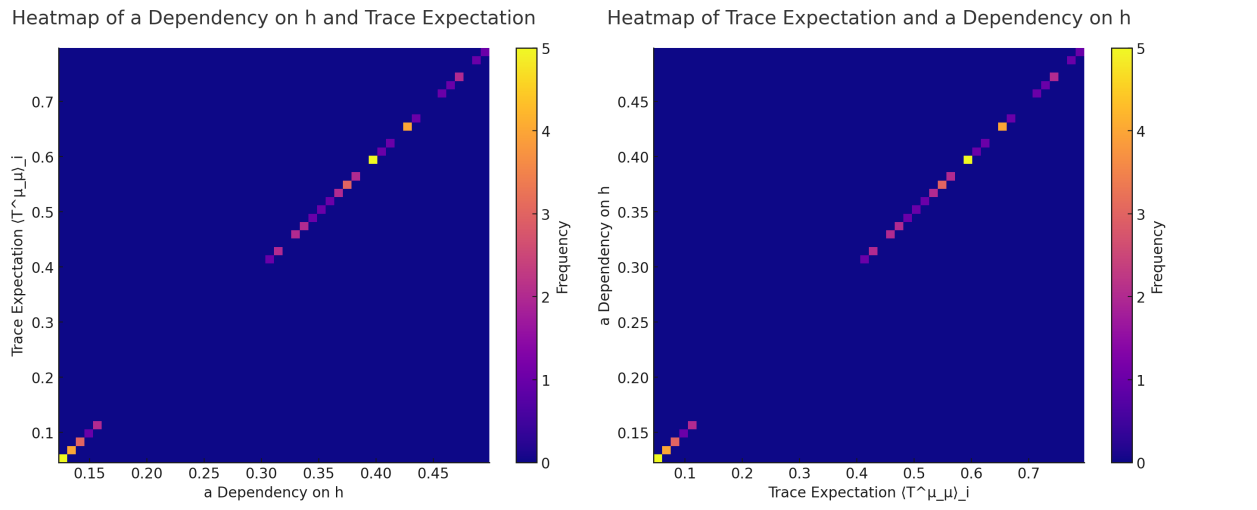

Fig. 36: Heatmap of a Dependency on $h$ and Trace Expectation Values

The width of the distribution of the expectation value $\langle T^{\mu}_{\mu} \rangle_i$ reflects how information diffusion patterns change over time. Analyzing this can lead to a better understanding of the flow of information and help implement appropriate measures.

## Analysis Using Heatmaps

The results' heatmap visually represents the dependence of the trace anomaly coefficient $a$ on $h$ and the relationship with the trace expectation $\langle T^{\mu}_{\mu} \rangle_i$. These values are used for analyzing physical quantities and diffusion characteristics of information in the context of Conformal Field Theory (CFT).

## Heatmap Analysis

The heatmap is color-coded based on $a$'s dependency on $h$ or the values of $\langle T^{\mu}_{\mu} \rangle_i$, where warmer colors indicate higher frequencies in that region. - Both heatmaps exhibit clear diagonal lines, indicating a strong positive correlation between $a$ and $h$ or $\langle T^{\mu}_{\mu} \rangle_i$. This suggests that an increase in one value tends to lead to an increase in the other. This relationship may indicate conditions where trace anomalies become more prominent for specific values of $h$ or conditions where trace expectations are higher. Particularly in a physical context, this implies distinctive behavior in specific states or phases of the system.

## Applications to Information Diffusion

In the context of information diffusion, the values of $h$ may be related to the influence of information sources or the speed of information propagation. For example, nodes with large $h$ values may indicate that information spreads quickly from those nodes or that the information has a significant impact on other nodes. The trace anomaly coefficient $a$ suggests the potential for information to spread nonlinearly or unpredictably. This is crucial for understanding the mechanisms behind the diffusion of fake news or misinformation. This analysis provides valuable insights into understanding the dynamics of information diffusion and formulating strategies to control it. Understanding how information spreads within a network is particularly important in addressing the issue of social misinformation. The insights gained from these heatmaps can be used to identify filter bubbles and counteract them. For instance, monitoring nodes with high $h$ values and improving the quality of information from those nodes can help suppress the diffusion of misinformation.

# 22. Conclusion:Analyzing Diffusion Patterns and Paths in the Context of Filter Bubbles Using Conformal Field Theory with Feynman's Green Function and Radial Ordering

In order to analyze diffusion patterns and diffusion paths with filter bubble risks, such as negative news and fake news, we consider an approach based on Conformal Field Theory (CFT) using Feynman's Green functions and radial ordering. Here, we present the mathematical equations, parameters, and the process of calculation for this approach.

## Mathematical Equations and Calculation Process

## Definition of Feynman's Green Function

The Feynman's Green function $G(x_1, x_2)$ between two fields $\phi(x_1)$ and $\phi(x_2)$ is defined as follows:

$$G(x_1, x_2) = \langle T[\phi(x_1)\phi(x_2)] \rangle$$

Here, $T$ represents the time-ordering operator, assuming that the time at $x_1$ is in the past compared to the time at $x_2$.

## Application of Radial Ordering

Radial ordering $R$ is an operation that arranges Feynman's Green functions from past to future. The radial order operator $R(\phi_1, \phi_2)$ represents the product of two fields $\phi_1$ and $\phi_2$ ordered from past to future.

## Calculation of Vacuum Expectation Values

The vacuum expectation value $\langle R(\phi_1, \phi_2) \rangle$ of the radial order operator is obtained by calculating the expectation value of the radial order operator. In the vacuum state $|0\rangle$, the vacuum expectation value of the radial order operator $R$ is represented as follows:

$$\langle R(\phi_1, \phi_2) \rangle = \langle 0 | R(\phi_1, \phi_2) | 0 \rangle$$

## Analysis of Diffusion Patterns and Paths

To analyze specific diffusion patterns, we model the interactions between nodes on the network as $\phi(x_1)$ and $\phi(x_2)$ and calculate correlation functions based on these fields' radial orders. Using these correlation functions, we analyze the distribution of information propagation and impact, particularly identifying diffusion paths for negative news.

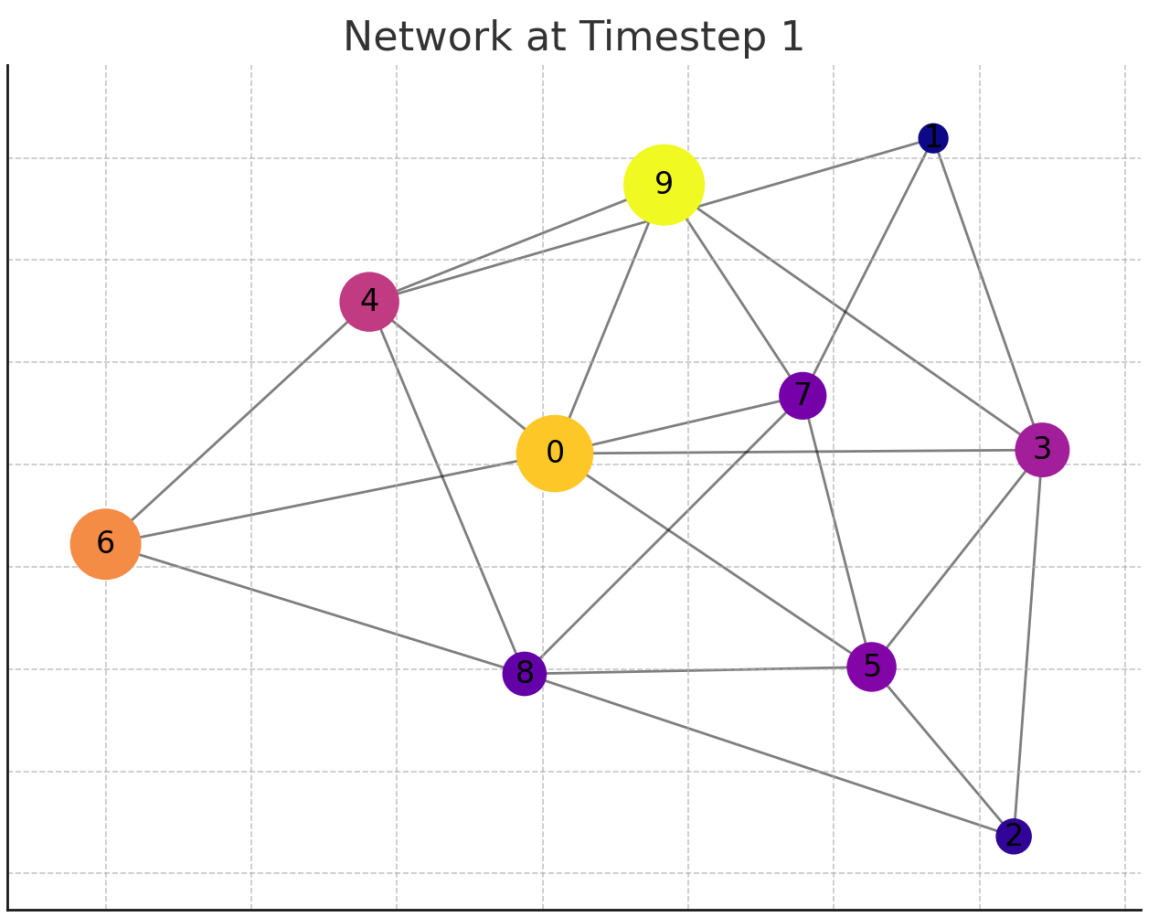

Fig. 37: Network at Timestep, N=1

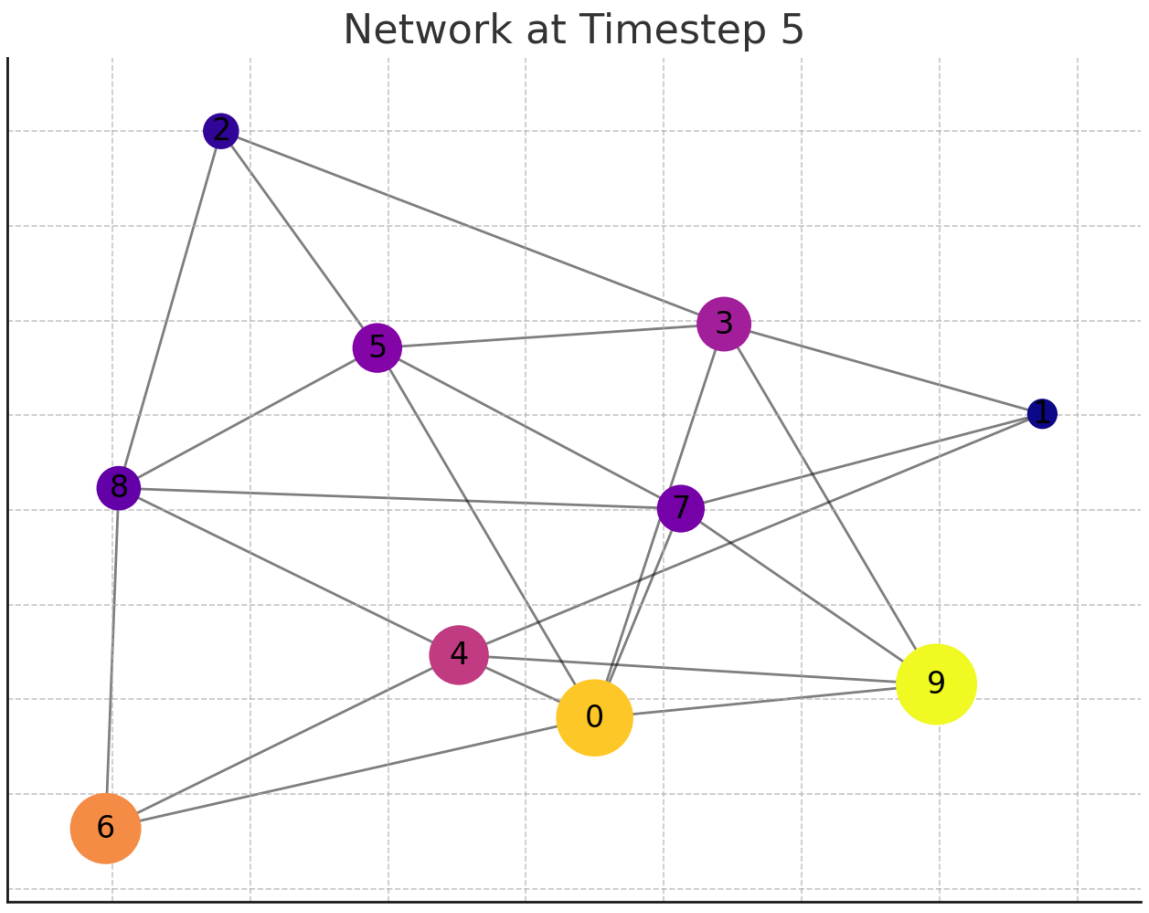

Fig. 38: Network at Timestep, N=5

## Simulation Approach

1. Define fields $\phi(x)$ to represent interactions between nodes on the network and calculate correlation functions based on the radial order of these fields. 2. Calculate vacuum expectation values of radial orders at each time step and simulate information diffusion patterns. 3. Analyze the temporal changes in vacuum expectation values and identify diffusion paths for negative news.

In this simulation, the information level in the network is calculated based on the initial density matrix assigned to each node and the value of the radial ordinal operator $R$. Traces of the radial ordinal operator based on conformal field theory appear to be used to quantify the information level at each node.

## Time Step 1

Light-colored nodes (e.g. nodes 0, 6, and 9) are shown as having a high information level, and these may represent the nodes with the highest information density at the beginning of the simulation. Dark-colored nodes (e.g. nodes 1, 2, and 3) are shown as having a low information level; these may indicate nodes where the information is not yet very diffuse or where the information density is low.

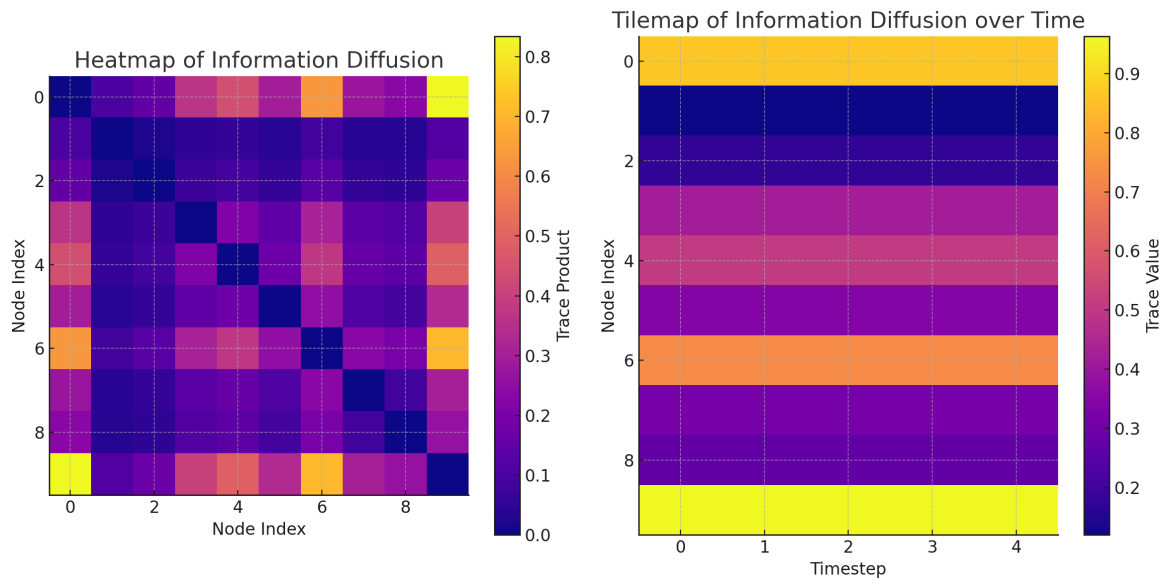

Fig. 39: Radial ordinal operator $R(\phi_1, \phi_2)$Density matrix $\rho$heatmap, tilemap

## Time Step 5

As the time step progresses, we observe that the placement of the light-colored nodes changes (e.g. nodes 4, 7, and 9). This indicates that information is spreading in the network and that there may be an active exchange of information between certain nodes. Nodes in darker colors are also changing and may indicate areas where information has not yet diffused sufficiently or where new information has decreased.

## Information Level Considerations

These network states help us understand how information is diffusing and which nodes are information centers. The value of the dynamic ordinal operator $R$ may define how information moves between nodes or which node is the source of information diffusion. Nodes with high values of $R$ may be the source or major diffusion point of information. This information can help us understand how fake or negative news spreads and develop strategies to curb them.

## Conclusion

This simulation provides a theoretical framework for understanding how fake news and negative information spreads within a network. It allows us to analyze the flow of information and develop strategies to improve the quality of information.

These heat maps and tile maps show the results of simulations of information diffusion using traces of dynamic ordinal operators. These visualizations quantify how information diffuses over time within a given network.

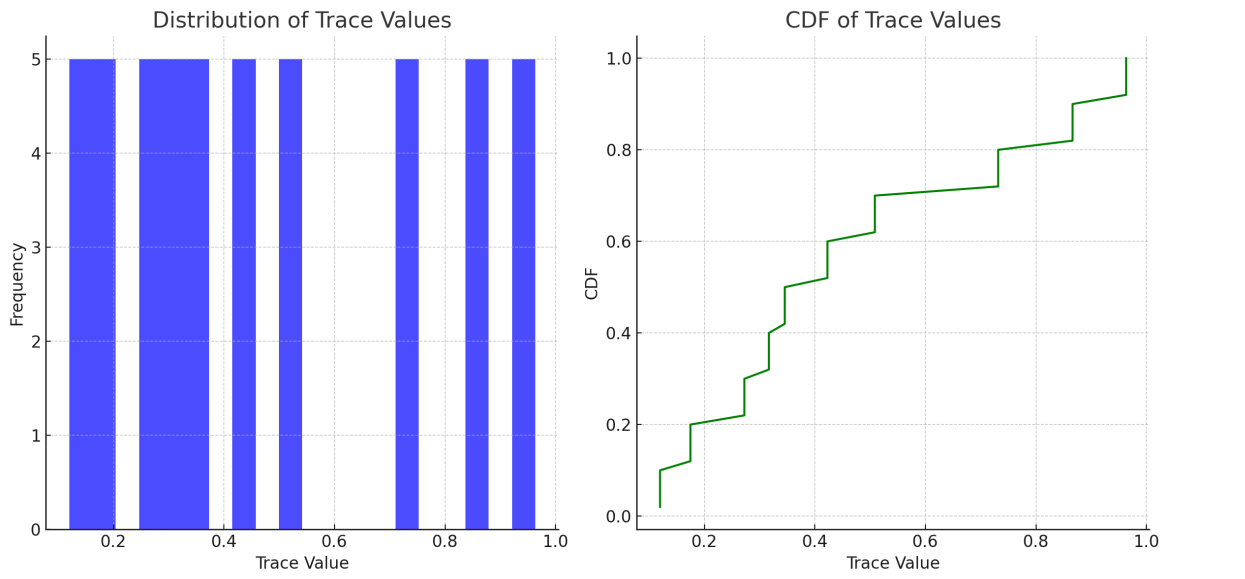

Fig. 40: Radial ordinal operator $R(\phi_1, \phi_2)$Density matrix $\rho$, Density, CDF

## Heat Maps

Heat maps show the intensity of the level of information (product of traces) between nodes at a particular point in time. Regions where the color is warmer (yellow or orange) indicate a higher level of information and represent nodes where information is more strongly diffused. In contrast, areas where the color is closer to cool (blue or purple) indicate a lower level of information, representing nodes where the diffusion of information is weaker or has not yet reached the node.

## Tile Map

The tile map shows how information is diffused through the time step. The intensity of the colors changes with each time step, showing how the information changes over time. This tile map is useful for showing from which node the information first began to diffuse and which node is the center of the information over time.

## Distribution of Trace Values and CDF

The distribution of trace values shows the variation in the level of information at each node in the simulation. From this distribution, we can see that most nodes have a medium information level, while some have very high or very low information levels. The CDF (cumulative distribution function) shows how the trace values are distributed overall and represents the percentage of nodes with information levels above a certain value. These visualizations help us understand how information levels vary in certain parts of the network. They can identify patterns in the spread of fake or negative news and how this information tends to be spread. The simulation is useful for monitoring the flow of information and understanding how specific nodes or connections affect the spread of information. This may help in developing strategies to control information diffusion.

The following calculations are introduced into the above process Calculating the vacuum expectation The vacuum expectation value $\langle R(\phi_1, \phi_2) \rangle$ is obtained by calculating the expectation value of the radial ordinal operator. In the vacuum state $|0\rangle$, the vacuum expectation of the radial ordinal operator $R$ is expressed as follows

$$\langle R(\phi_1, \phi_2) \rangle = \langle 0 | R(\phi_1, \phi_2) | 0 \rangle$$

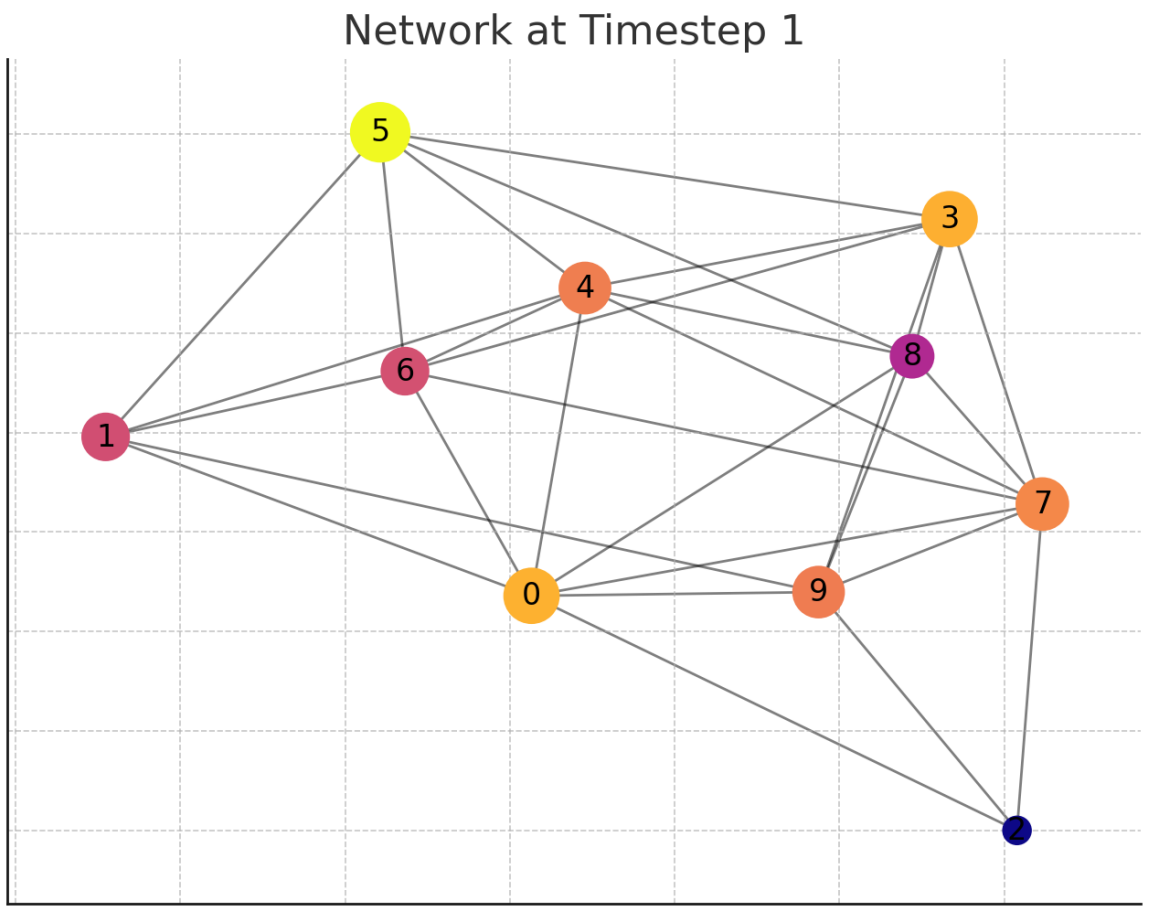

Fig. 41: Network at Timestep, N=1

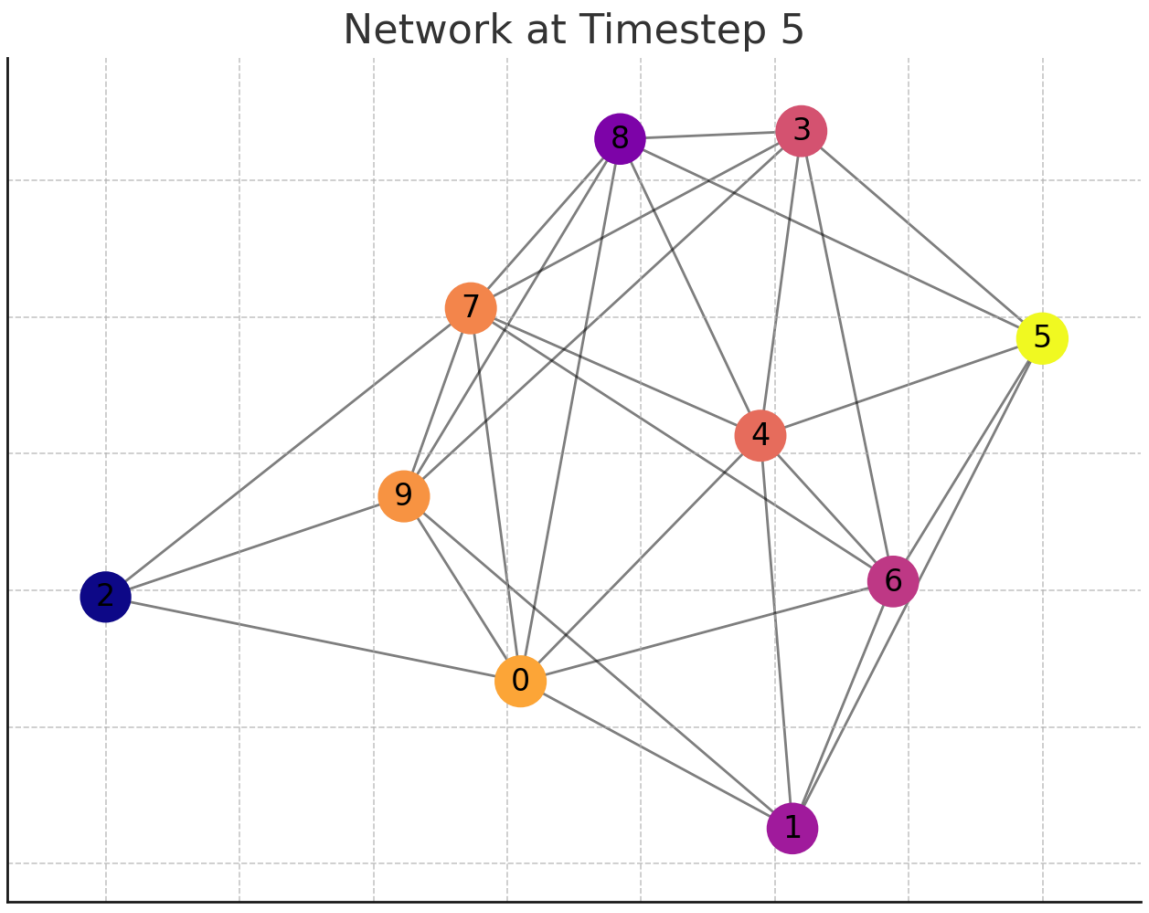

Fig. 42: Network at Timestep, N=5

Resuls seem to represent the states of a simulated information diffusion process at two different timesteps (Timestep 5 and Timestep 1) using concepts from quantum field theory, particularly Feynman's Green's functions and radial ordering.

In quantum field theory, Green's functions describe the propagation of particles or fields between two points, and radial ordering is used to handle the time evolution of operators in conformal field theory. The radial order places operators at later times in front of those at earlier times when calculating their contributions to correlation functions. In the context of information diffusion, this could represent the influence that the timing and sequence of information dissemination have on the overall spread within a network.

## Traces of Radial Order Operators

In theoretical physics, the trace of an operator often represents a sum over all possible states or a sum over the diagonal elements of a matrix representing the operator. For information diffusion, a trace over radial order operators might represent a sum over all possible diffusion paths or the influence of various nodes based on their temporal order of information reception.

## Influence of Past Time on Diffusion Patterns

By examining the state of the network at different timesteps, we can infer the temporal dynamics of information spread. Nodes that become highly connected or intensely colored at later times may have been influenced by nodes that reached those states at earlier times. This suggests a pattern where certain 'influencer' nodes or clusters play a critical role in disseminating information.

## Change Over Time

Comparing the two timesteps, we can observe how information (or in this analogy, 'particles' or 'fields') propagates through the network. Nodes that were less colored at Timestep 1 but became more colored by Timestep 5 likely represent points where the information has been recently acquired or has become more influential. Conversely, nodes that remain consistently colored across timesteps may represent steadystate spreaders of information, maintaining their influence over time.

## Diffusion Patterns

If we treat the nodes as operators and the edges as propagators, the diffusion pattern can be viewed as a superposition of multiple paths of information flow. The trace of the radial order operators would then give us a measure of the total influence exerted by all nodes up to a given time, accounting for the 'history' of information spread within the network.

To further analyze the past influence and diffusion patterns, we would ideally look for changes in the network's connectivity over time, shifts in the centrality of certain nodes, and possibly the emergence or disappearance of influential nodes or clusters. This would require a time series analysis of the network states to identify patterns and influences.

Results representing the simulation results of information diffusion over time, likely modeled by Feynman's Green's functions and radial ordering concepts.

## Tilemap of Information Diffusion over Time

The first heatmap depicts how the information level at each node changes over different timesteps. The vertical axis lists node indices, while the horizontal axis represents time. Each tile's color intensity indicates the level of information at that node at a given time, with warmer colors representing higher levels of information.

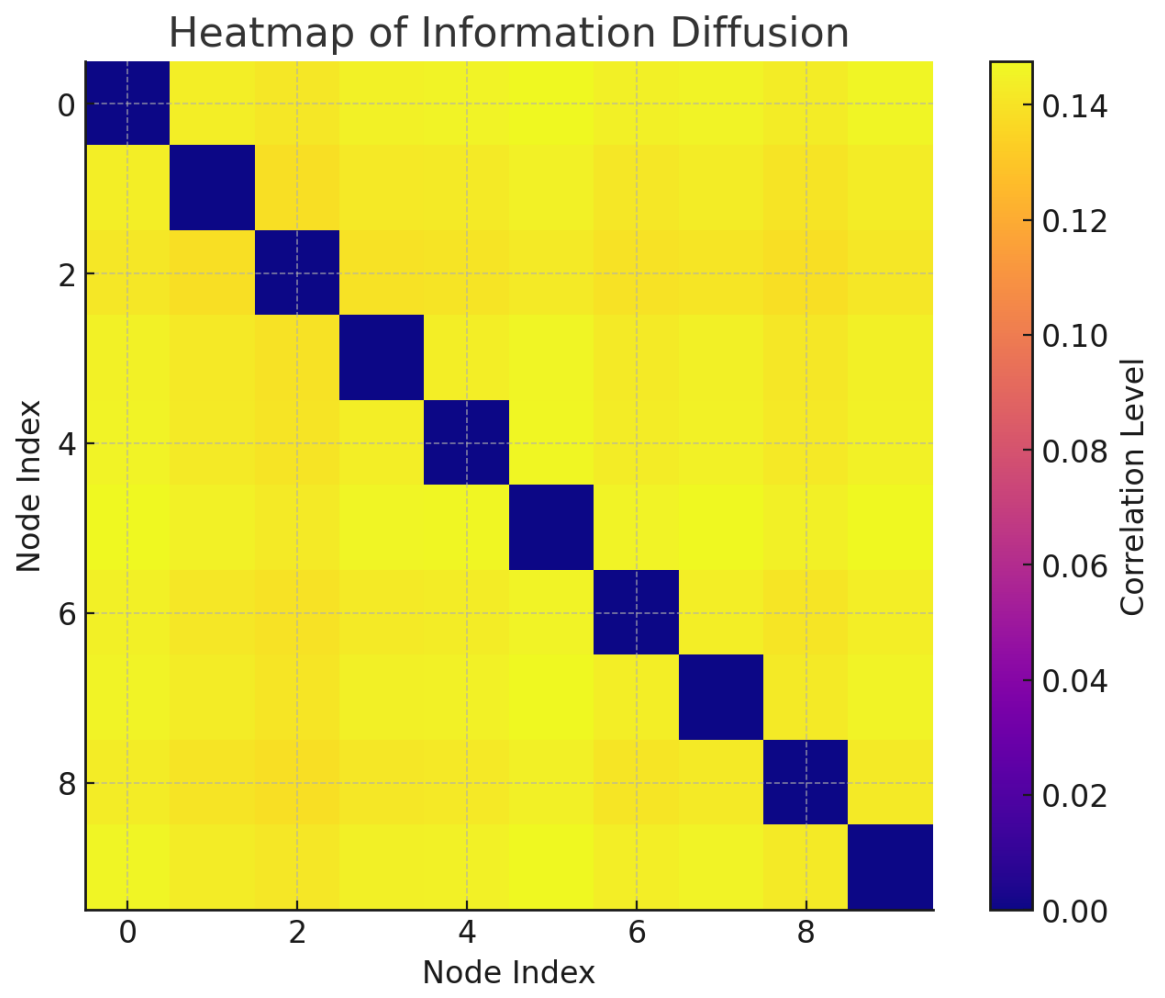

Fig. 43: Information Diffusion heatmap

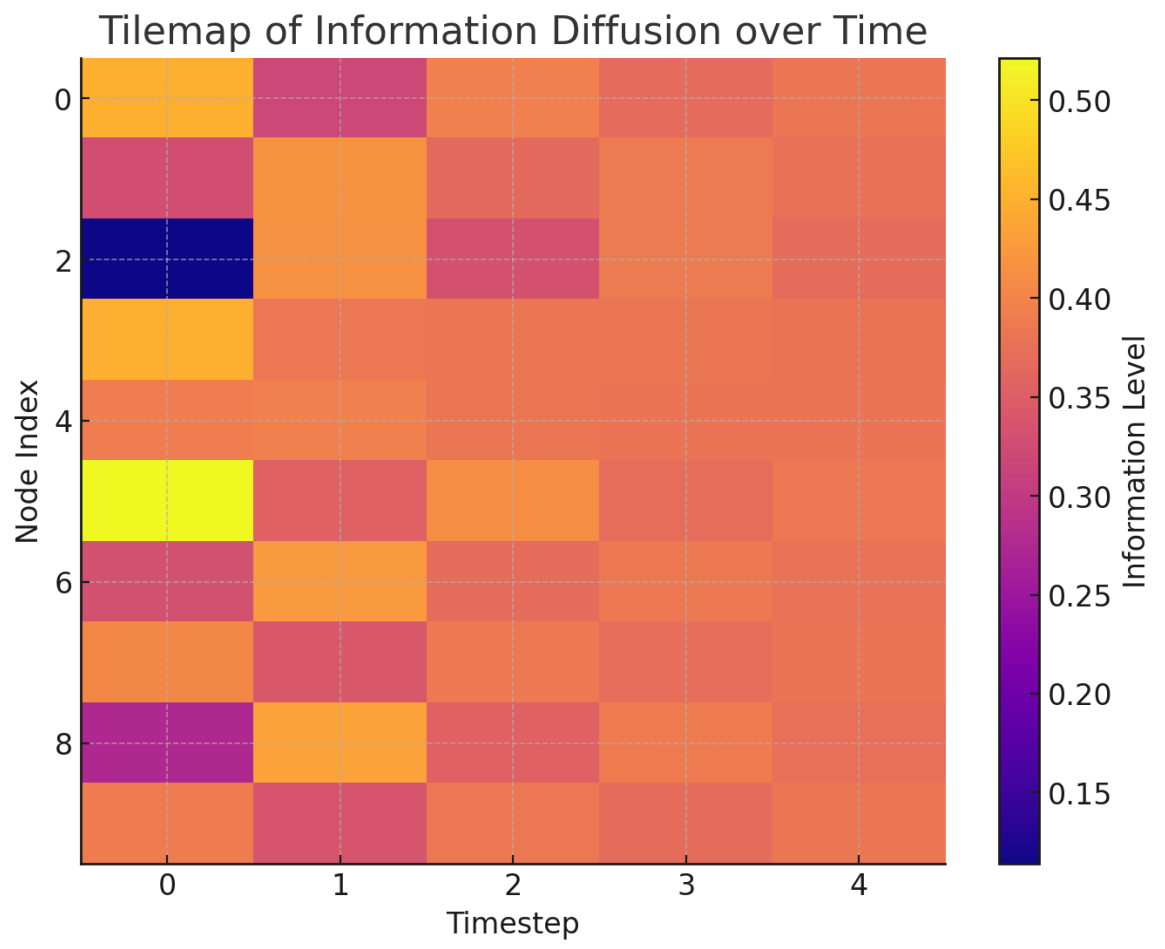

Fig. 44: Information Diffusion tilemap

## From this tilemap, we can observe the following patterns

Certain nodes exhibit a significant change in information level at particular timesteps, which might indicate moments when these nodes were particularly active in spreading or receiving information.The progression of colors across the timesteps for a single node may indicate the pattern of information diffusion for that node – whether it's a steady increase, fluctuation, or a sudden spike.

## Heatmap of Information Diffusion

The second heatmap looks like a correlation matrix, where both axes represent node indices, and the color intensity in each cell indicates the level of correlation in information diffusion between pairs of nodes. Darker colors on the diagonal are normal since they represent selfcorrelation (i.e., each node has a perfect correlation with itself).

## Key observations from this heatmap include

Offdiagonal dark squares indicate pairs of nodes that have a strong correlation in their information levels, suggesting that these nodes may influence each other or be influenced by common sources within the network. The overall yellow background suggests low correlation among most pairs of nodes, which could mean that the network has a complex structure with many independent pathways for information spread.

## Radial Ordering and Trace Analysis

Radial ordering in quantum field theory deals with the temporal sequence of events – an important factor in the spread of information. In this context, the "trace" of the radial ordering operator might be interpreted as the cumulative effect of all information diffusion events up to that point in time.

## To analyze the influence of past times on diffusion patterns

Look for patterns in the tilemap to identify if there is a 'cascade' effect, where the spread of information from certain nodes leads to subsequent spread in others.Analyze the correlation heatmap to discern if certain nodes consistently show a high correlation, which may suggest a repeated pattern or a direct influence from one to another.Comparing the two heatmaps may reveal the nodes that are central to information spread at different times, as indicated by changes in their information levels and their correlation with other nodes. By tracing the radial order operator, one could theoretically sum over all these diffusion patterns to get a sense of the overall spread dynamics. This could highlight how earlier timesteps influenced the later spread of information, revealing key nodes or moments critical to the diffusion process.

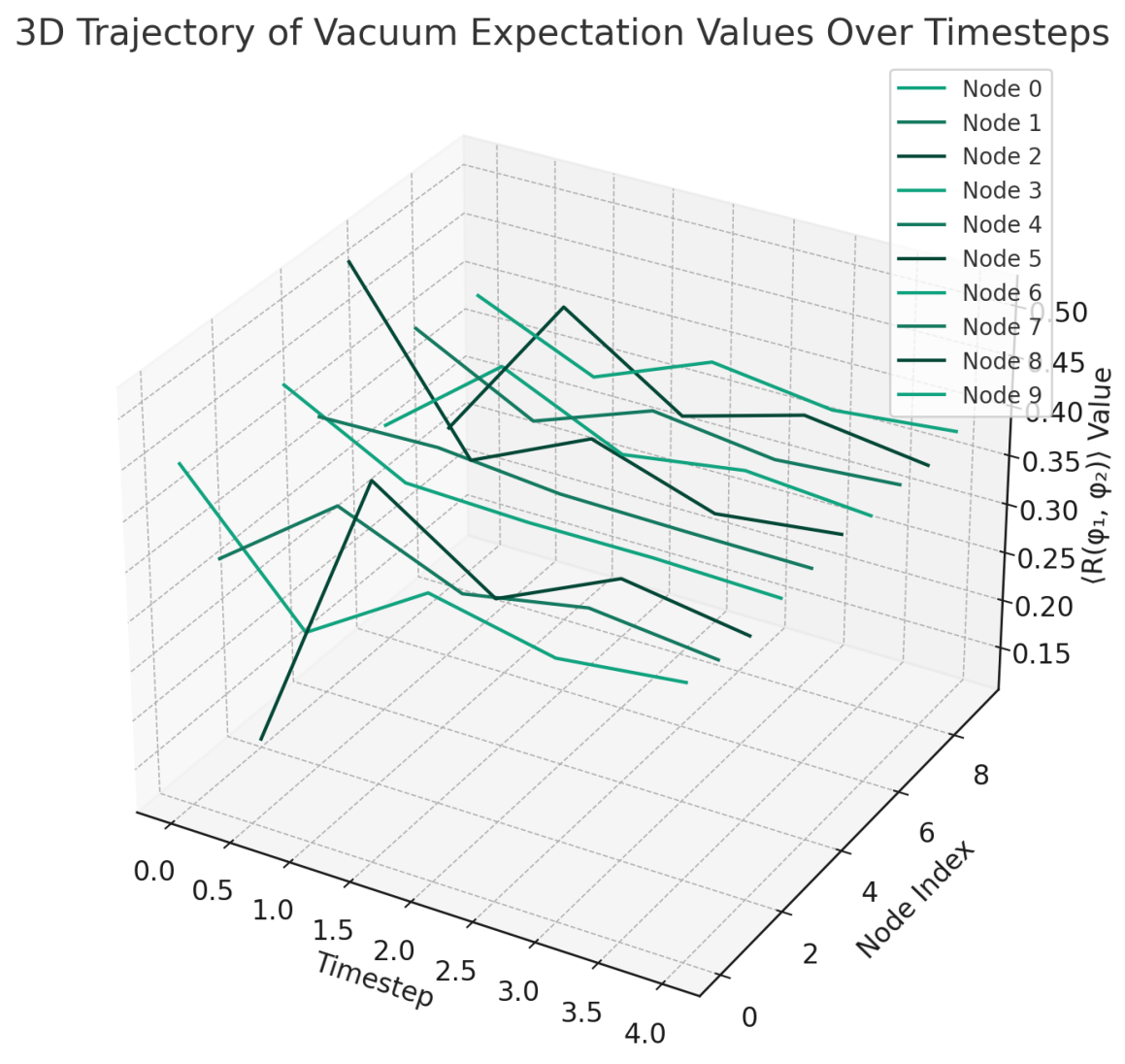

Fig. 45: 3D Trajectory of Vacuum Expectation Values Over Timesteps

Results seems to represent the trajectories of vacuum expectation values $\langle R(\phi_1, \phi_2)\rangle$ over time for different nodes on a cylindrical distribution, likely in the context of a quantum field theory analogy for information diffusion.

In quantum field theory, the vacuum expectation value (VEV) of an operator is its average value in the vacuum state, which is the state of lowest energy with no particles present. In the context of your simulation, $\langle R(\phi_1, \phi_2)\rangle$ could represent some measure of the state or influence of a node regarding information diffusion, where $\phi_1$ and $\phi_2$ could be variables or parameters defining the state of the node or the information itself.

Analysis of Past Time Influence and Diffusion Patterns Trajectories Over Time Each line in the graph tracks the VEV for a node across different timesteps. The trajectory's shape can give insights into how the influence or state of each node changes over time. For example, a trajectory with a pronounced peak suggests a moment when the node had a significant impact on the diffusion process, while a trough might indicate a period of less influence. Temporal Patterns By observing the trajectories, we can identify patterns such as nodes that consistently rise in influence over time, those that have cyclical patterns, and those that may have sudden spikes or drops. This information can be crucial for understanding how information is likely to flow through the network in the future based on past behavior. Comparing Node Behaviors By examining the trajectories in conjunction with each other, we can discern whether certain nodes behave similarly over time, indicating possible correlations or dependencies in their roles in the diffusion process. Trace of Radial Order Operators If we consider each node's trajectory as being influenced by radial order operators that dictate the 'spread' of information, then taking a trace of these operators would involve integrating over all nodes and all timesteps to provide a holistic measure of the information diffusion throughout the network. This could show how earlier states of nodes influence the later states, as the VEVs at each timestep would be affected by all previous timesteps. Diffusion Patterns on the Cylinder Given that the trajectories are plotted on a cylindrical distribution, the angular component (which might be represented implicitly in the 3D trajectory as the 'circular' aspect around the cylinder) could show the spread of information in a less linear and more interconnected fashion, akin to how rumors or news might spread in a real social network.

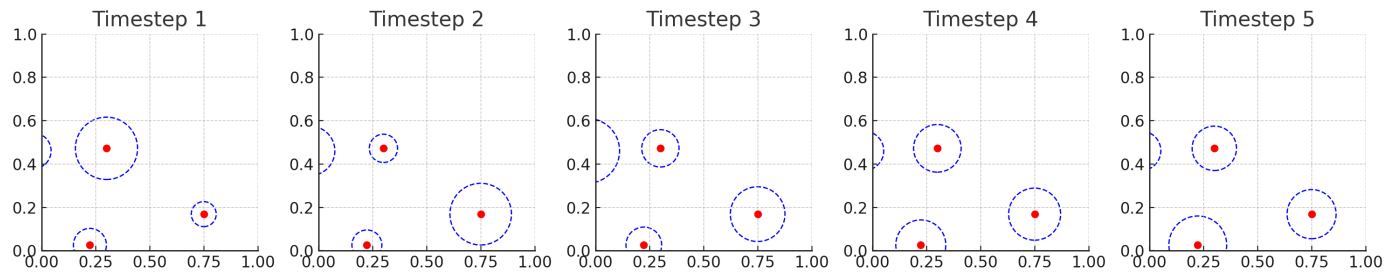

Fig. 46: Integral path $C_i$, behavior around the insertion point of the primary field $\phi_i$

The overall analysis would benefit from identifying key nodes and times that are critical to the diffusion process, understanding the interplay between different nodes, and determining how past states influence future states. This could help in predicting future patterns of information spread and identifying strategic points for intervention in the case of misinformation.

Results show a series of snapshots at different timesteps of a simulated information diffusion process. Each snapshot illustrates nodes (as red dots) and their associated integration paths (as blue dashed circles) on a 2D plane. The radii of the circles, which change randomly, represent the residues (information levels) at each node. This setup seems to be a graphical representation of a radial ordering operation within a conformal field theory framework applied to the spread of information, potentially fake news or negative news.

### Node Positions and Integration Paths

The red dots (nodes) appear fixed in position across timesteps, suggesting that the network structure is static, while the blue dashed circles (integration paths) change in size, implying variability in how information is being processed or its influence at different nodes over time.

### Temporal Evolution of Information Levels

The changing sizes of the circles from Timestep 1 to Timestep 5 could indicate fluctuations in the nodes' information levels. Nodes with increasing circle radii may suggest an accumulation or amplification of information, while decreasing radii could suggest a dissipation or reduction of influence.

### Residues and Information Spread

In the context of conformal field theory, a residue often refers to the coefficient of the pole in a Laurent series and can have a physical interpretation. Here, it could be analogous to the impact or 'weight' of the information at a node. Larger circles may indicate nodes with greater 'residue', possibly representing key spreaders of information or influential nodes in the network.

### Diffusion Patterns for Fake or Negative News

The simulation results could help understand the propagation patterns of fake or negative news. Nodes with consistently large radii across timesteps could be hotspots for spreading such news, while nodes with smaller or shrinking radii may be less influenced or more resistant to the spread.

### Radial Ordering Operators and Vacuum Expectation Values

Radial ordering imposes a specific arrangement of operators based on the 'time' coordinate, which in this simulation is analogous to the timesteps of the diffusion process. The integration paths around the primary fields ($\Phi_i$) indicate the influence region for each node. The vacuum expectation values in this context likely measure the average level of information (or misinformation) at each node, factoring in the influence from all other nodes according to the radial ordering.

By analyzing these graphics, we can potentially identify patterns such as which nodes are central to the diffusion at different times, how the information impact changes over time, and what the overall dynamics of the network might be. This insight could be critical for developing strategies to counteract the spread of misinformation by targeting influential nodes or times in the spread.

## 23. Conclusion:Modeling Diffusion Risk of Negative News in Conformal Field Theory (CFT)

Let's consider the mathematical approach when considering the modeling of diffusion risk of negative news, including primary fields, infinitely small conformal transformations, the definition of the Virasoro algebra's time evolution, stress-energy tensor, conformal weights, and Weyl scaling factors.

## Key Concepts and Mathematical Formulations

### Primary Fields and Infinitely Small Conformal Transformations

Infinitely small conformal transformations for primary fields are fundamental in conformal field theory and can be expressed as follows:

$$\delta\Phi(z,\bar{z}) = -\epsilon(z)\partial\Phi(z,\bar{z}) - \bar{\epsilon}(\bar{z})\bar{\partial}\Phi(z,\bar{z})$$

Here, $\Phi(z,\bar{z})$ represents a primary field, and $\epsilon(z)$ and $\bar{\epsilon}(\bar{z})$ are the parameters of the infinitesimal transformation.

### Time Evolution of the Virasoro Algebra

The time evolution of the Virasoro algebra is related to the energy-momentum tensor $T(z)$ and its mode expansion. The mode expansion can be expressed as follows:

$$T(z) = \sum_{n=-\infty}^{\infty} L_n z^{-n-2}$$

Here, $L_n$ represents the generators of the Virasoro algebra.

### Stress-Energy Tensor and Conformal Weights

The stress-energy tensor $T^{\mu\nu}$ represents the distribution of physical energy and momentum. Conformal weights are crucial parameters that indicate the scaling behavior of primary fields.

### Weyl Scaling Factors

Weyl scaling factors capture the impact of scaling in conformal transformations. The Weyl scaling factor $\Omega(x)$ is defined as follows:

$$\Omega(x) = |f'(x)|^{\Delta}$$

Here, $\Delta$ is the conformal weight, and $f'(x)$ is the derivative of the conformal transformation.

### Lagrangian Density

The Lagrangian density $\mathcal{L}$ is a function that describes the dynamic behavior of physical systems. In conformal field theory, $\mathcal{L}$ defines the dynamics of fields.

Using these concepts, you can define the necessary equations and parameters for modeling the diffusion risk of fake news and negative news. This modeling can be useful for analyzing information propagation patterns and assessing risks in specific scenarios.

## Application of Radial Ordering Operator and Vacuum Expectation Value in the Context of Negative News Diffusion Risk Analysis

In the context of analyzing the diffusion risk of negative news, let's consider the definitions of the radial ordering operator $R(\phi_1,\phi_2)$ and the vacuum expectation value $\langle R(\phi_1,\phi_2)\rangle$ in conformal field theory (CFT).

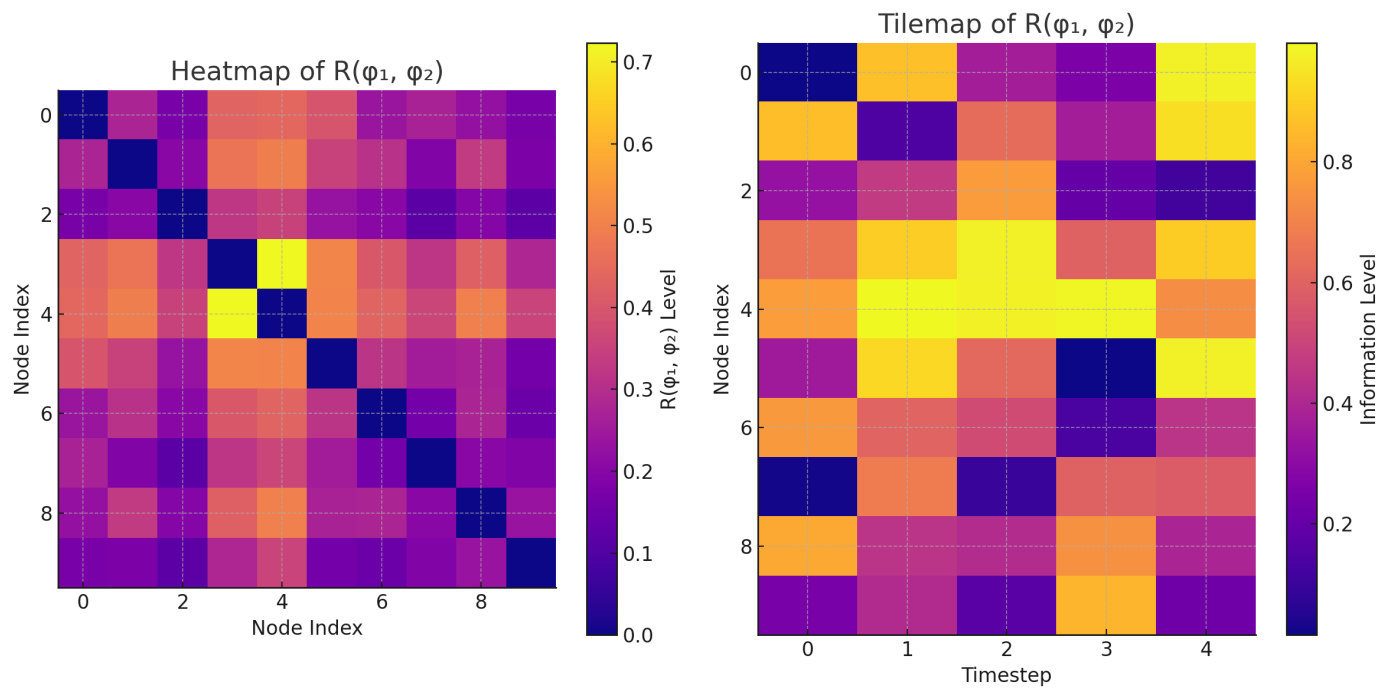

Fig. 47: Tilemap of $R(\phi_1,\phi_2)$

### Radial Ordering Operator $R(\phi_1,\phi_2)$

The radial ordering operator is an operator used in conformal field theory to arrange the product of two fields in a specific order. It is often referred to as the "time ordering operator" in standard terminology, but the concept of "radial" becomes significant in the context of conformal field theory. The definition of this operator is as follows:

$$R(\phi_1,\phi_2) = \begin{cases} \phi_1(x_1)\phi_2(x_2) & \text{if } x_1 \text{ is earlier than } x_2 \\ \phi_2(x_2)\phi_1(x_1) & \text{if } x_2 \text{ is earlier than } x_1 \end{cases}$$

Here, $x_1$ and $x_2$ represent the coordinates (time and spatial position) of the respective fields.

### Vacuum Expectation Value $\langle R(\phi_1,\phi_2)\rangle$

The vacuum expectation value is obtained by calculating the expectation value of the radial ordering operator. It represents the field correlations in the ground state (vacuum state). The formula is as follows:

$$\langle R(\phi_1,\phi_2)\rangle = \langle 0|R(\phi_1(x_1),\phi_2(x_2))|0\rangle$$

Here, $|0\rangle$ represents the vacuum state (the lowest energy state).

### Application to Negative News Diffusion Risk Analysis

When applying this theory to the analysis of the diffusion risk of negative news, $\phi(x)$ can represent the state of propagation of specific information or news. $R(\phi_1,\phi_2)$ represents interactions between different sources of information or nodes, and $\langle R(\phi_1,\phi_2)\rangle$ indicates the average or overall tendencies of these interactions. This approach can be used to model the flow of information and influence diffusion on a network and may be particularly useful in understanding how negative news spreads and affects society.

Results associated with the radial ordering operator $R(\phi_1,\phi_2)$, which seem to represent the spread of social negative news across a network. In this context, the radial ordering operator could be a mathematical tool used to model the dynamics of how information, particularly negative news, spreads through a social network. The heatmap likely shows the pairwise relationships or correlations between nodes at a single point in time, while the tilemap shows the evolution of information levels at individual nodes over time.

When considering primary fields and the risk of negative news spreading infinitely, we're essentially looking at the theoretical framework where these primary fields are points in a network that are particularly influential. The infinitesimal conformal transformation and its associated Virasoro algebra can describe the evolution of these points over time.

## StressEnergy Tensor and Conformal Weight

In conformal field theory (CFT), the stressenergy tensor $T$ is crucial. It encodes the dynamics of the theory, including how fields respond to deformations of spacetime (or in our adapted context, changes in the network structure). The conformal weight (h) of a primary field is a measure of how it transforms under scaling, and the Virasoro algebra defines the time evolution of these fields.

## Weyl Scaling Factor

The Weyl scaling factor would adjust how the network model responds to scale transformations, which in the social network context might represent the ability of a node to either amplify or dampen the information passing through it.

## Analysis Based on Lagrangian Density

Incorporating these concepts into a Lagrangian density for the network, we would be looking at an equation that governs the dynamics of information spread. The Lagrangian density $\mathcal{L}$ would include terms that represent the interaction between nodes (the primary fields), their intrinsic influence (conformal weights), and their response to changes in the 'shape' of the network (modeled by the Weyl scaling factor).

## Considerations for the Graphs

Heatmap Analysis In the heatmap, highintensity areas (yellow) suggest nodes with a high level of interaction or correlation in terms of negative news spreading. These could be points where the risk of negative news spreading is significant. Conversely, lower intensity areas (purple) indicate less interaction or lower risk.Tilemap Analysis The tilemap shows how the risk at each node evolves over time. Nodes that consistently exhibit higher information levels across timesteps could be those where the risk of negative news becoming entrenched is higher. Fluctuations over time might suggest nodes that are more dynamic in their influence on the spread of news.

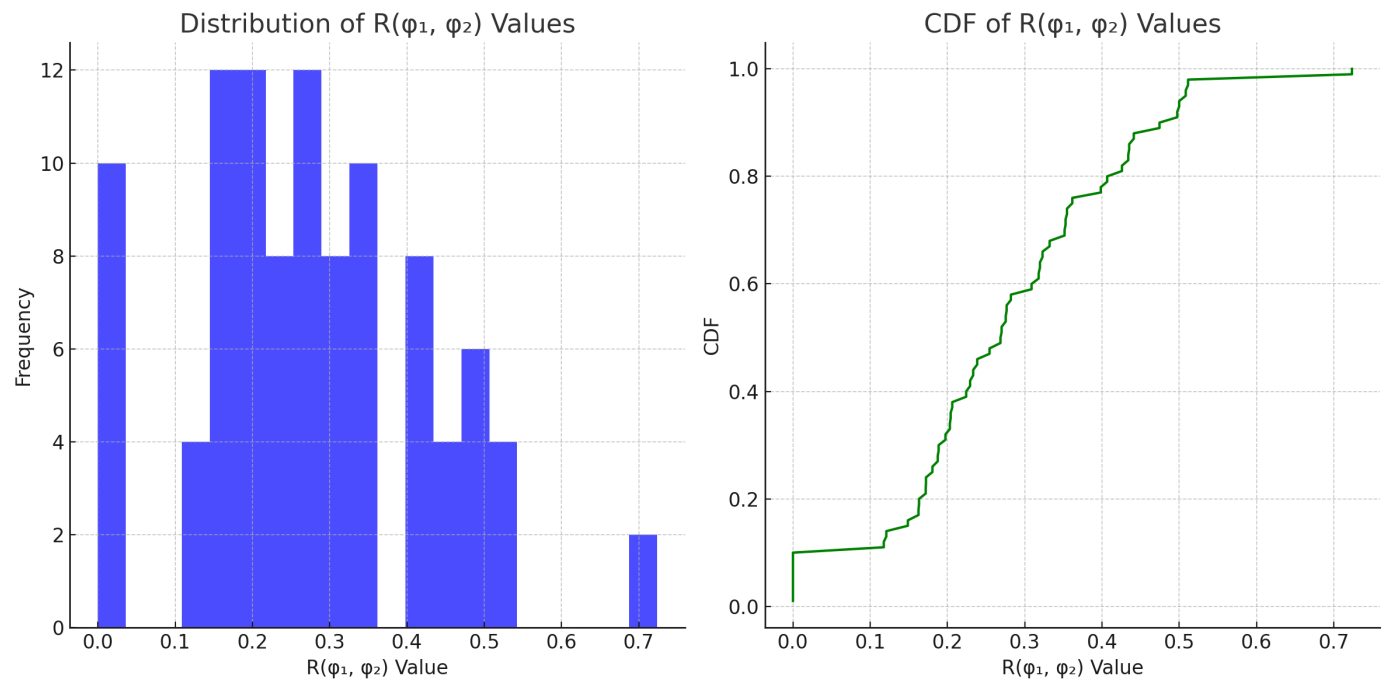

Fig. 48: CDF, Distribution of $R(\phi_1, \phi_2)$

## Identifying Risk Areas

Risk Emergence Nodes that start with lower information levels but show a rapid increase might be places where the risk of negative news spreading is emerging.Infinite Growth Nodes with a high and increasing information level might represent areas of the network susceptible to unbounded growth in the spread of negative news, akin to a field experiencing an infinite conformal transformation.

## Calculation Process and Formulas

To perform a quantitative analysis, we would need the specific form of the Lagrangian density and the associated equations of motion for the fields. These would provide us with the mathematical framework to calculate the change in information levels at each node, considering the influence of all the aforementioned factors.

In summary, the provided graphs, alongside the theoretical constructs from CFT, could offer a powerful model for understanding and predicting the dynamics of negative news spread within a social network. The key to this analysis is identifying influential nodes (primary fields), understanding their interactions (modeled by $R(\phi_1, \phi_2)$), and observing how these interactions evolve over time, as influenced by the theoretical constructs of CFT.

Results appear to show the distribution and cumulative distribution function (CDF) for values of a radial ordering operator $R(\phi_1, \phi_2)$, which might be used to model the spread of negative news in a social network within the framework of conformal field theory (CFT).

## Histogram and CDF Analysis

The histogram represents the frequency distribution of $R(\phi_1, \phi_2)$ values across the network or over multiple instances. It shows how often each range of values occurs. The CDF graph, on the other hand, shows the probability that a value of $R(\phi_1, \phi_2)$ will be less than or equal to a particular value. It provides a sense of the overall spread of values and can help identify thresholds or critical points.

## Considering CFT Concepts

The Virasoro algebra, stressenergy tensor $T$, conformal weights, and Weyl scaling factors are all part of the theoretical machinery of CFT. In the context of social networks, they can be adapted to describe how information (or misinformation) spreads and evolves over time.

## Virasoro Algebra and Time Evolution

The Virasoro algebra can provide a way to understand how the influence of a node changes over time. In particular, it describes how the stressenergy tensor evolves, which would be related to the change in information levels in the network.

## StressEnergy Tensor

The stressenergy tensor $T$ in a CFT context measures the flow of energy and momentum. Adapted to social networks, it could represent the flow and intensity of information at a point in the network.

## Conformal Weight

The conformal weight of a primary field in CFT measures its scaling dimension. In a network, it might represent the relative importance or influence of a node. Nodes with a higher conformal weight could have a more significant role in spreading information.

## Weyl Scaling Factor

This factor in CFT represents how a system responds to scaling transformations. In a social network, it might represent the sensitivity of nodes to the amplification of information they receive or transmit.

## Lagrangian Density and Dynamics

The Lagrangian density in CFT is a scalar quantity that describes the dynamics of the fields within the theory. In the network context, it would represent the 'rules' or dynamics governing how information spreads, which might include factors like the network topology, node influence, and the nature of the information itself.

## Insights from the Graphs, Distribution of $R(\phi_1, \phi_2)$ Values

The histogram shows a range of values for $R(\phi_1, \phi_2)$, with some values occurring more frequently. This could point to common levels of node influence or typical patterns of information spread.

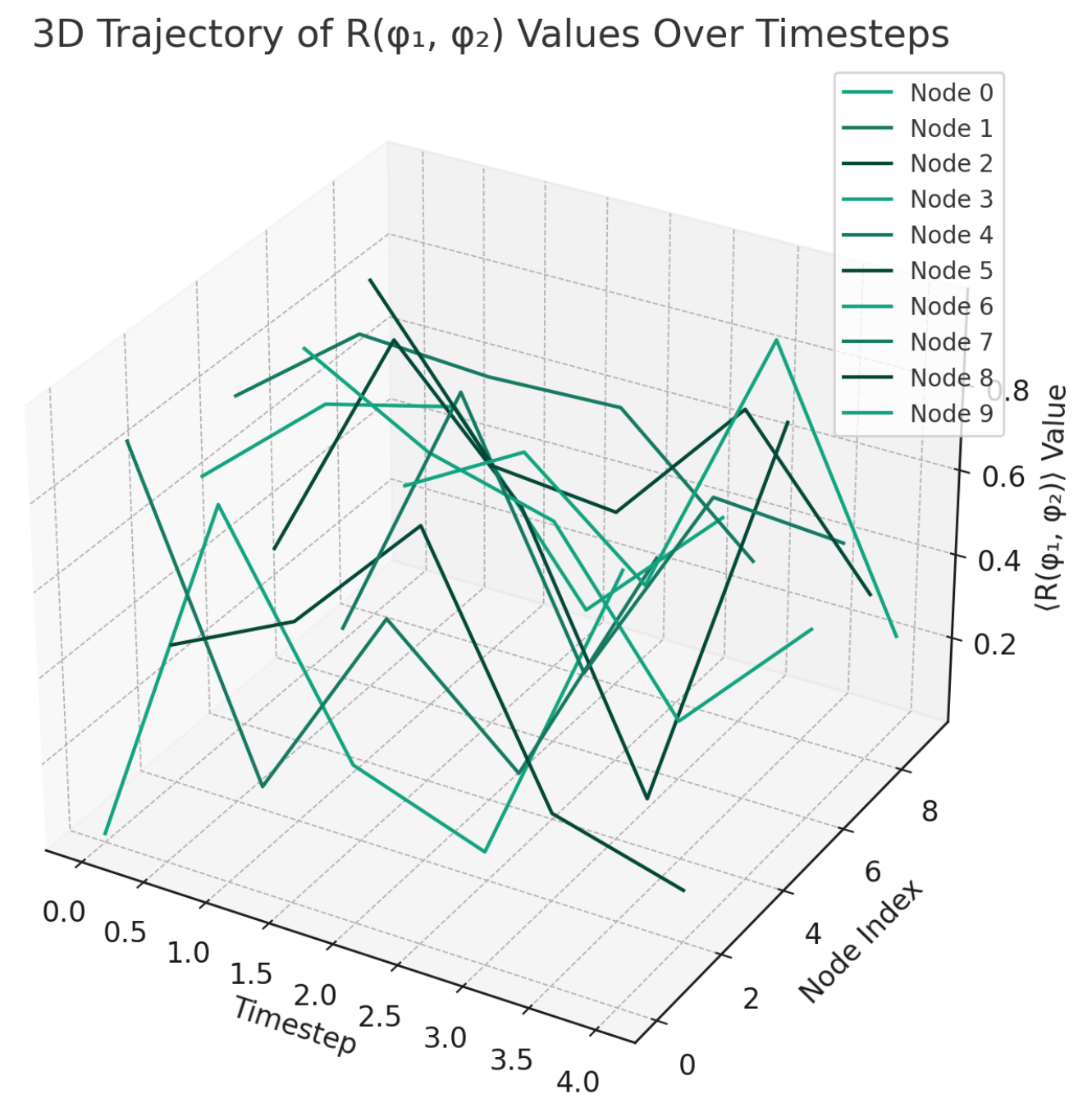

Fig. 49: 3Dmap of $R(\phi_1, \phi_2)$

## CDF of $R(\phi_1, \phi_2)$ Values

The CDF indicates that there are thresholds beyond which the probability of higher $R(\phi_1, \phi_2)$ values increases more rapidly. This might be indicative of a point where the spread of negative news accelerates or becomes more likely.

## Identifying Risk Locations

Nodes or pairs of nodes corresponding to higher $R(\phi_1, \phi_2)$ values, especially those at the higher end of the CDF, may represent parts of the network where negative news is more likely to spread or where its impact is more severe. These could be areas of focus for interventions to mitigate the spread of negative news.

To move from a qualitative to a quantitative analysis, we would need the exact form of the CFT adapted for the social network context. With that, we could use the distribution of $R(\phi_1, \phi_2)$ values along with the CFTderived equations of motion to predict how the spread of information might evolve and identify potential intervention points to mitigate the spread of negative news.

Results shows the values of the radial ordering operator $R(\phi_1, \phi_2)$ for different nodes over various timesteps. This visualization can give us insights into the dynamic behavior of each node with respect to the spread of negative news in a social network modeled by conformal field theory (CFT) concepts.

## Interpretation of the 3D Trajectory Graph

Node Axis Each line in the graph represents a different node in the network. Timestep Axis This axis indicates the time evolution of the system. Value Axis The $R(\phi_1, \phi_2)$ values, which could represent the level of influence or 'information potential' at each node.

## Analysis Considering CFT Constructs

Virasoro Algebra The Virasoro algebra in CFT defines the time evolution of the conformal fields. In this case, it might define how the influence of each node evolves over time due to the spread of negative news. StressEnergy Tensor (T) This tensor quantifies the density and flux of energy and momentum in the field theory. For our network, it might represent the strength and directionality of the information flow between nodes. Conformal Weight (h) This is a measure of a field's response to scaling transformations. Nodes with higher conformal weights might represent more influential individuals or groups within the network, who can significantly alter the spread pattern of the news. Weyl Scaling Factor Reflects how the scale of the system affects the nodes. In the context of social networks, it might represent the resilience or susceptibility of nodes to the spread of negative news.

## Insights from the 3D Trajectory Graph

Fluctuations The graph shows fluctuations in $R(\phi_1, \phi_2)$ values over time, which may indicate the variable nature of each node's influence. Some nodes might be more consistent in their influence levels, while others fluctuate more dramatically.Node Comparison By comparing the trajectories of different nodes, we can identify which ones have more significant roles in the spread of negative news. Nodes whose trajectories tend to stay at higher levels or exhibit upward trends might be of particular interest. Critical Points Sharp changes in the trajectory might indicate events or times when the node's influence changed markedly, which could be critical points for intervention to mitigate the spread of negative news.

## Further Considerations

Lagrangian Density In CFT, the Lagrangian density $\mathcal{L}$ describes the dynamics of the fields. To thoroughly analyze the spread of negative news, we would need to construct a Lagrangian that includes terms for the interaction between nodes, the influence of external factors, and the internal dynamics of the information spread.Time Evolution and Intervention Understanding how $R(\phi_1, \phi_2)$ values change over time is essential for predicting future states of the network and designing timely interventions. To proceed with a deeper analysis, we would need the explicit functional form of $R(\phi_1, \phi_2)$, the equations of motion derived from the Lagrangian density, and any additional parameters or constraints specific to the social network model. With these details, we could potentially forecast the spread of negative news and identify strategic points for targeted interventions to control the spread.

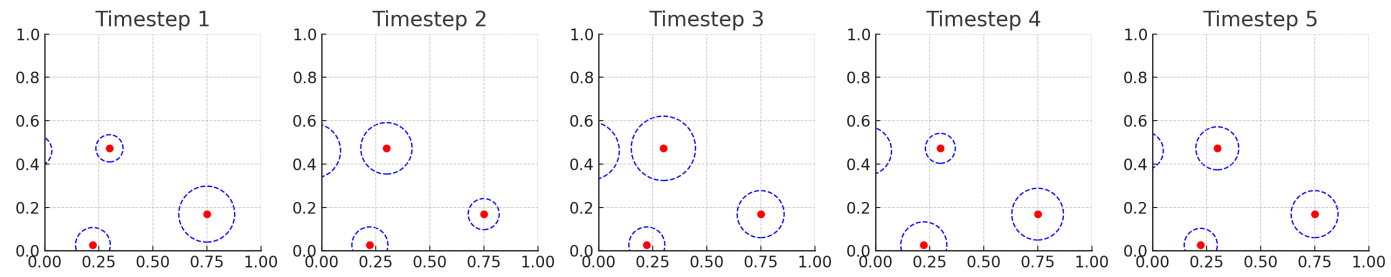

Fig. 50: The deformation of the integral path $C_i$

Result is a sequence of graphs for timesteps 1 through 5, depicting the behavior around insertion points of primary fields $\Phi_i$ in the network, represented by red dots. The blue dashed circles around each node indicate integration paths, which have varying radii across different timesteps. This variability can be interpreted as changes in the influence or 'information potential' of the nodes over time within the context of the diffusion of social negative news.

## Integration Paths and Primary Fields

Red Dots (Primary Fields $\Phi_i$): These represent nodes in the network where information (negative news) is being inserted or from which it is spreading.Blue Dashed Circles (Integration Paths): The varying sizes of the circles at each timestep suggest changes in the influence radius or the 'reach' of each node's information spread.

## Conformal Transformations and Virasoro Algebra

Infinitesimal Conformal Transformations: These would represent small changes in the network's structure or the way information is spread, potentially leading to significant shifts in the dynamics of the network over time.

## Virasoro Algebra

This provides a mathematical framework to understand the time evolution of the nodes' influence.

## StressEnergy Tensor

In CFT, the stressenergy tensor encodes the flow of energy and momentum. In this network context, it might represent the intensity and direction of information flow.

## Conformal Weight (h)

This measures how primary fields scale under a dilation. For nodes in a network, it might represent their relative influ-ence on the spread of information; nodes with higher weights would have a larger impact.

## Weyl Scaling Factor

This would adjust how information spread scales in different parts of the network, potentially modeling nodes' susceptibility or resilience to the spread of negative news.

## Timestep Evolution

By observing the changes in the integration paths over the timesteps, one can identify which nodes are becoming more or less influential over time.

## Risks of Negative News Spread

Nodes with expanding integration paths (larger circles) over consecutive timesteps might be areas where the risk of negative news spreading is increasing. Conversely, shrinking paths might indicate a decrease in influence or containment of information spread.

## Lagrangian Density

A Lagrangian density for this system would include terms representing the nodes' interactions, the spread dynamics, and external influences. It would allow for the derivation of equations of motion to model the spread of negative news quantitatively.

## Strategic Insights

Nodes with rapidly growing integration paths could be key targets for interventions to prevent the spread of negative news.

## Modeling Spread Dynamics

The changes in integration paths could be modeled mathematically to predict future behavior and identify the critical moments for potential intervention.

To conduct a detailed and predictive analysis, one would need to define a functional form of the Lagrangian density that captures the dynamics of information spread in the network. The equations of motion derived from this Lagrangian could then be solved, potentially using numerical methods, to forecast the spread of negative news and develop strategies to mitigate it.

Results show the distribution and cumulative distribution function (CDF) of Virasoro algebra values, which are relevant in the context of conformal field theory (CFT) applied to a system, potentially modeling the spread of social negative news.

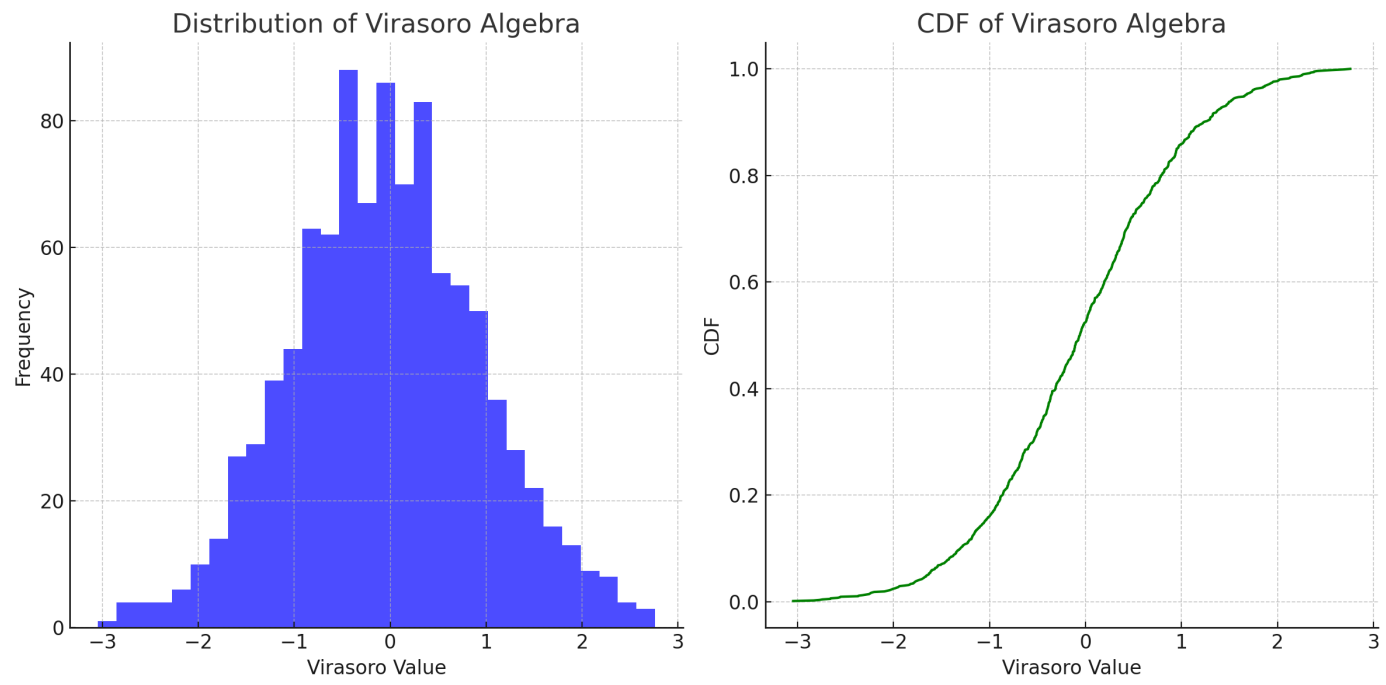

Fig. 51: CDF, Distribution of Virasoro Algebra

## Virasoro Algebra Distribution

The histogram displays the frequency of Virasoro values, which could represent various states or magnitudes of influence within the social network. The bellshaped distribution suggests that most values cluster around a central point, with fewer occurrences as you move away from the center, indicating that extreme influences are less common.

## CDF of Virasoro Algebra

The CDF graph shows the probability that a Virasoro value is less than or equal to a certain value. It provides a cumulative perspective of the distribution, where the steepness of the curve indicates how quickly the accumulated probability increases. The sigmoid shape of the CDF indicates that there are thresholds or points at which the probability of the Virasoro values increases more significantly.

## Interpretation in the Context of Social Networks

## Virasoro Values

In the analogy to social networks, these values could represent the intensity or effectiveness of the nodes' influence in spreading negative news.

## Central Tendency

The central clustering of values suggests that most nodes have a moderate level of influence, while fewer nodes have extremely high or low influence.

## Critical Points

The points where the CDF curve becomes steeper might represent critical thresholds in the network dynamics. Crossing these thresholds could lead to a rapid increase in the spread of negative news.

## Social Network Dynamics with CFT Constructs

## Infinitesimal Conformal Transformations

These transformations might represent slight changes in the social network that can lead to largescale effects on the spread of news.

## Temporal Change and Virasoro Algebra

The time evolution of the Virasoro values can indicate how the influence of certain nodes or groups of nodes evolves, potentially identifying nodes that become more critical in spreading information as time progresses.

## StressEnergy Tensor

This tensor represents the flow of information in the network. Highintensity values in the distribution could correlate with nodes that are central to the flow and spread of negative news.

## Conformal Weights and Weyl Scaling

These factors would affect the scaling and the impact of the nodes. Nodes with high conformal weights might be particularly influential, and their behavior could be crucial in controlling the spread of negative news.

## Application to Mitigating Negative News Spread

## Modeling Spread Dynamics

The distribution of Virasoro values, paired with the network's equivalent of the stressenergy tensor and conformal weights, could help model how negative news might spread through the network.

## Strategic Interventions

By understanding the CFTbased dynamics, interventions could be strategically placed at nodes identified by the Virasoro algebra as having significant influence on the network's behavior.

For a comprehensive analysis, one would require a detailed mathematical model that encapsulates these CFT concepts within the framework of the social network. This model would allow for the simulation of interventions and their potential effects on controlling the spread of negative news.

Results depict the distribution and cumulative distribution function (CDF) for two different sets of values: those related to the stressenergy tensor and those related to the conformal weight within the context of a conformal field theory (CFT) model applied to social dynamics, specifically the spread of negative news.

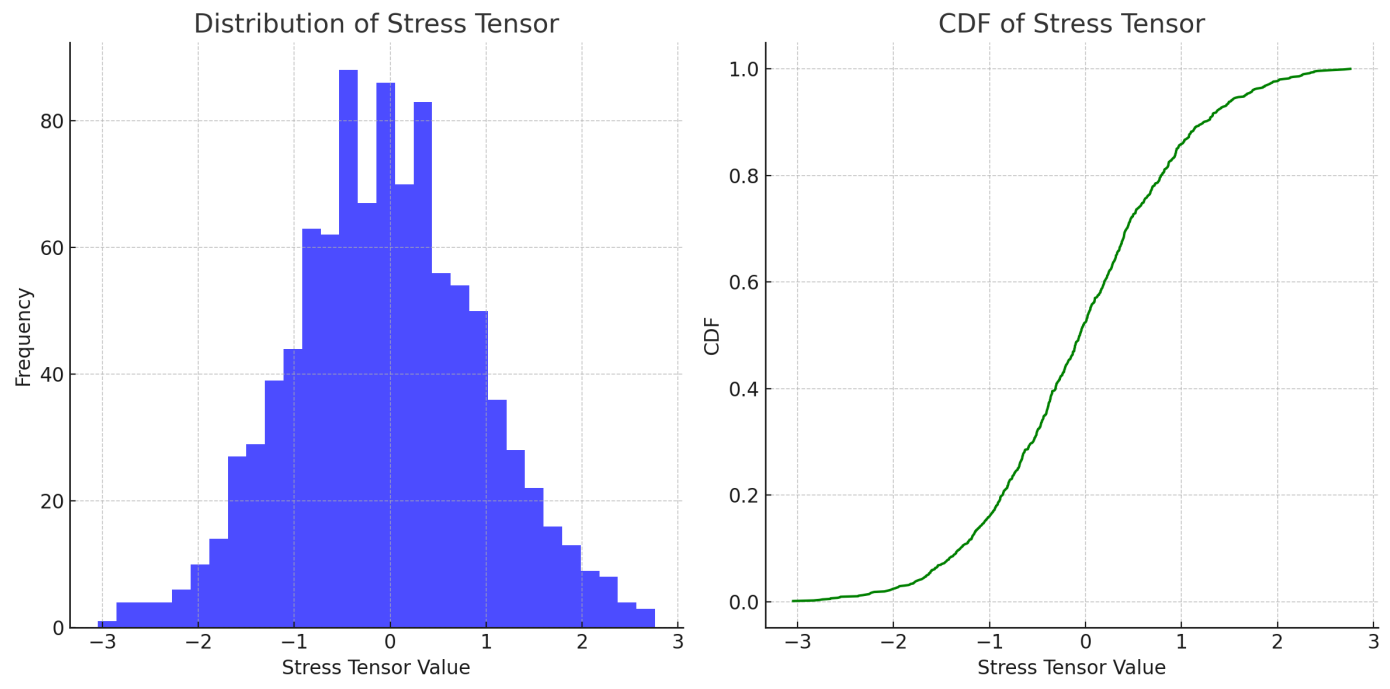

Fig. 52: CDF, Distribution of Stress Tensor

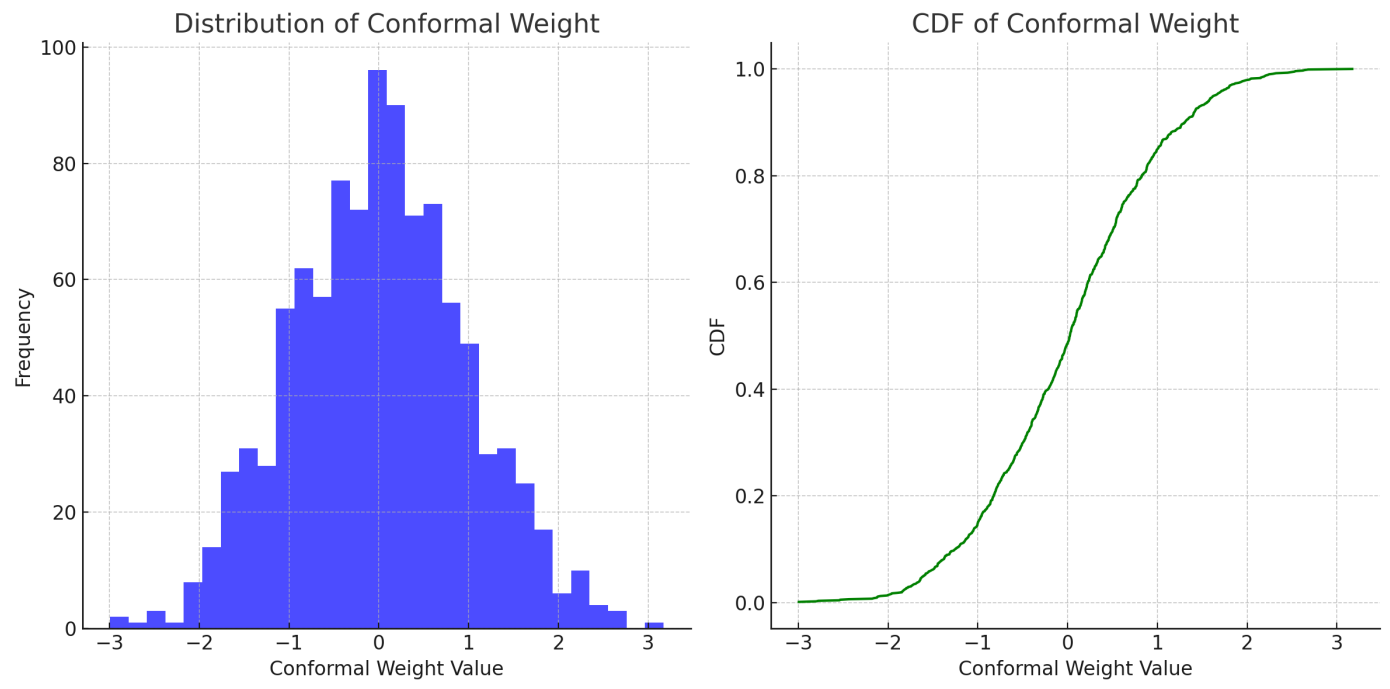

Fig. 53: CDF, Distribution of Conformal Weight

## StressEnergy Tensor Distribution and CDF

The stressenergy tensor in CFT quantifies the density and flux of energy and momentum. Adapted to a social network model, it could symbolize the intensity and direction of the flow of information. In the provided histogram and CDF, the values likely reflect the distribution of 'informational stress' within the network. The normallike distribution implies that most nodes or links have a moderate level of stress, with fewer experiencing extreme levels. The CDF shows that there are certain thresholds where the probability of experiencing higher stress levels jumps significantly.

## Conformal Weight Distribution and CDF

Conformal weight in CFT measures the scaling dimension of a field and its transformation properties under scaling transformations. In a social network context, it might correspond to the influence or importance of a node in terms of its role in disseminating information. The histogram again shows a normallike distribution, suggesting that most nodes have an average level of influence, with extremes being less common. The CDF indicates that the likelihood of a node having a higher conformal weight increases steadily, with no sudden jumps, implying a relatively balanced distribution of influence among the nodes.

## Analysis and Interpretation Locations of Negative News Spread

Nodes associated with higher values in the stress tensor distribution might be areas where information is heavily concentrated or rapidly flowing, indicating potential hotspots for the spread of negative news.

## Infinite Conformal Transformations and Virasoro Algebra

These would be related to changes in the network that can lead to substantial differences in the spread patterns of news. Nodes with significant temporal changes in their associated Virasoro values might be undergoing transitions that could amplify the spread of negative news.

## Role of Conformal Weights

Nodes with higher conformal weights would be critical influencers within the network. Their actions could disproportionately affect the spread of information, making them key targets for monitoring or intervention.

## Weyl Scaling Factor

This factor would adjust the impact of scaling transformations on the nodes. Nodes that are highly sensitive to Weyl scaling might be more susceptible to changes in the overall network dynamics, potentially becoming catalysts for the spread of negative news.

## Lagrangian Density

A Lagrangian describing this system would include terms accounting for the interactions (modeled by the stressenergy tensor) and transformations (conformal weights and scaling factors) of the network. It would allow for the derivation of equations of motion, which could be used to simulate the spread of negative news and devise strategies to mitigate it.

By analyzing the distributions and CDFs of these values, one could potentially forecast where and how negative news might spread within the network and identify critical points for intervention. This would involve constructing a dynamic model using the principles of CFT, calibrated to the specific characteristics of the social network in question.

Results relationship between the Weyl scaling factor and Lagrangian density. These concepts can be translated from their use in conformal field theory (CFT) to the analysis of social networks, specifically in understanding the diffusion of negative news.

## Weyl Scaling Factor and Lagrangian Density in CFT Context

The Weyl scaling factor is associated with how a field theory responds to changes in scale, which could correspond to the sensitivity of different nodes or connections in a social network to the spread of information or influence.The Lagrangian density describes the dynamics of a field theory. In a social network, it could be thought of as encapsulating the 'rules' or mechanisms that govern how information spreads through the network.

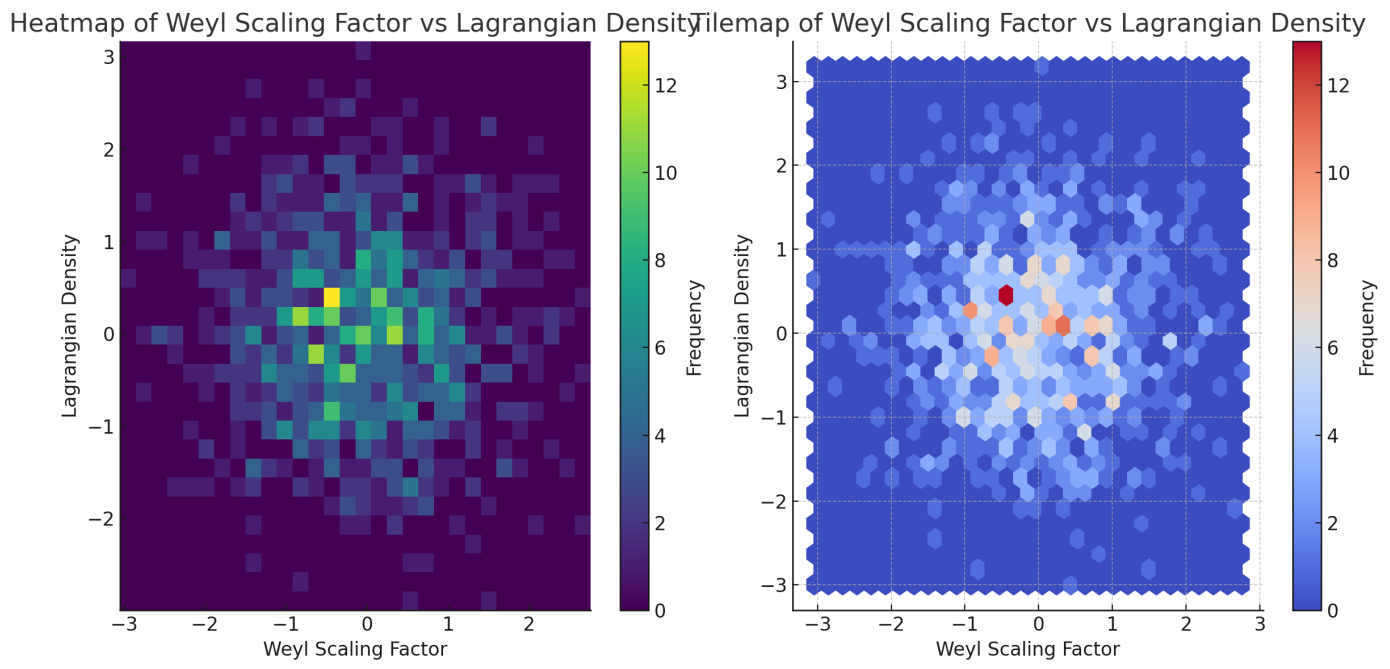

Fig. 54: Heatmap and Tilemap of Weyl Scaling Factor vs Lagrangian Density

## Interpretation of the Heatmaps

## Heatmap of Weyl Scaling Factor vs. Lagrangian Density

Shows the frequency of occurrences for different combinations of Weyl scaling factors and Lagrangian densities. Regions with higher frequency (yellow and green) indicate more common combinations, while less common combinations are shown in purple.

## Tilemap of Weyl Scaling Factor vs. Lagrangian Density

This representation could be showing the same relationship as the heatmap but possibly over time or under different conditions.

## Insights from Heatmap Analysis

## Clusters and Outliers

The concentration of points in certain areas of the heatmap can indicate 'normal' operating conditions within the network. Outliers or regions with high intensity could represent abnormal or critical states where the network is particularly susceptible or resistant to the spread of negative news.

## Correlation Between Factors

If certain values of the Weyl scaling factor are frequently associated with specific ranges of Lagrangian densities, this could suggest a direct relationship in how nodes influence the dynamics of news spread. For example, nodes with a high

Weyl scaling factor might consistently exhibit a particular pattern of influence (as indicated by the Lagrangian density).

### Risk Assessment

In terms of risk related to the spread of negative news, regions in the heatmap with higher intensity might be areas where the network structure or dynamics are more conducive to rapid and widespread dissemination of information. These could be targeted for monitoring or intervention.

### Applying the Analysis

### Primary Field Locations

In CFT, primary fields are points where the conformal symmetry is welldefined. In a social network, these could be influential nodes or hubs. Nodes corresponding to highintensity areas in the heatmap might be these primary fields.

### Time Evolution and Virasoro Algebra

The Virasoro algebra's role in the time evolution of the system could be reflected in how the distribution of points in the heatmap changes over time or under different conditions.

### Stress Tensor Values

The stressenergy tensor's values, if translated to this context, could represent the stress or pressure on different nodes due to the information flow. This could be used to anticipate which areas of the network might become stressed next.

### Conformal Weights

These could be indicative of the importance or 'weight' of nodes in terms of their role in spreading information.

To provide a detailed analysis or simulation, we would need the actual mathematical model or formulas that link these CFT concepts to the social network dynamics. With that information, one could potentially calculate the spread of negative news and develop strategies to mitigate it, possibly by altering the network dynamics (as represented by the Lagrangian density) or by adjusting the influence of certain nodes (reflected by the Weyl scaling factor).

## 24. Future Implications

The concepts of conformal field theory and momentum tensor can be explained by replacing them with the process of fake news, including social diffusion of information, information loops and wasteful diffusion, such as filter bubbles, to better understand information diffusion and its effects.

For example, considering the process of spreading fake news, information loops and wasteful diffusion of

### Momentum of information

From a tensor perspective, information in fake news, information loops and wasteful diffusion has "information momentum" depending on its content and impact. This momentum affects the speed and extent to which information diffuses.

### Law of Conservation of Information

The spread of fake news, loops of information and wasteful diffusion follow the law of conservation of information. That is, information does not increase or decrease during the spreading process, but maintains a constant amount. This can be difficult once fake news is spread and the spread can be inhibited.

### Specificity of information

When fake news, information loops or futile spreading is specific to a particular social context or people's beliefs, the information behaves like a singularity. Singularity of information affects a specific group of people or community, and the proliferation proceeds.

### Energy density of information

Depending on the content and priority of fake news, information loops and useless diffusion, the energy density of information varies. Important information has a high energy density and spreads quickly.

### Correlation function of information

In social information diffusion, there is also a correlation function between information. That is, as one piece of information spreads, information related to it also tends to spread.

Thus, by replacing the concepts of conformal field theory and the momentum tensor with the process of information diffusion, we can deepen our understanding of patterns and influencing factors related to information diffusion and come up with ideas useful for effective management and control of information.

## Aknowlegement

The author is grateful for discussion with Prof. Serge Galam and Prof.Akira Ishii.